\pdfoutput=1
\documentclass[pre,superscriptaddress,twocolumn,amsmath,amssymb,floatfix]{revtex4-1}

\usepackage{graphicx}
\usepackage{hyperref}
\usepackage{comment}
\usepackage{esvect}
\usepackage{makecell}
\usepackage{cleveref}
\usepackage{amsmath}
\usepackage[position=top,caption=false]{subfig}

\newcommand{\labelarial}{\fontsize{12pt}{0pt}\usefont{T1}{phv}{m}{n}}

\begin{document}

\title{A common trajectory recapitulated by urban economies}

\author{Inho Hong}
\affiliation{Department of Physics, Pohang University of Science and Technology, Pohang 37673, Republic of Korea}
\author{Morgan R. Frank}
\affiliation{Media Laboratory, Massachusetts Institute of Technology, Cambridge, MA, 02139, USA}
\author{Iyad Rahwan} 
\affiliation{Media Laboratory, Massachusetts Institute of Technology, Cambridge, MA, 02139, USA}
\affiliation{Institute for Data, Systems, and Society, Massachusetts Institute of Technology, Cambridge, MA, 02139, USA}
\author{Woo-Sung Jung} 
\affiliation{Department of Physics, Pohang University of Science and Technology, Pohang 37673, Republic of Korea}
\affiliation{Department of Industrial and Management Engineering, Pohang University of Science and Technology, Pohang 37673, Republic of Korea}
\affiliation{Asia Pacific Center for Theoretical Physics, Pohang 37673, Republic of Korea}
\author{Hyejin Youn}
\email{hyejin.youn@kellogg.northwestern.edu}
\affiliation{Kellogg School of Management at Northwestern University, Evanston, IL 60208, USA}
\affiliation{Northwestern Institute on Complex Systems, Evanston, IL 60208, USA}
\affiliation{London Mathematical Lab, London WC2N 6DF, UK }
\affiliation{Santa Fe Institute, 1399 Hyde Park Road, Santa Fe, NM 87501, USA}

\begin{abstract} 
Is there a general economic pathway recapitulated by individual cities over and over? 
Identifying such evolution structure, if any, would inform models for the assessment, 
maintenance, and forecasting of urban sustainability and economic success as a quantitative baseline. 
This premise seems to contradict the existing body of empirical evidences for 
path-dependent growth shaping the unique history of individual cities. 
And yet, recent empirical evidences and theoretical models 
have amounted to the universal patterns, mostly size-dependent, thereby
expressing many of urban quantities as a set of simple scaling laws. 
Here, we provide a mathematical framework to integrate repeated cross-sectional data, 
each of which freezes in time dimension, into a frame of reference for longitudinal 
evolution of individual cities in time. 
Using data of over 100 millions employment 
in thousand business categories between 1998 and 2013, we decompose 
each city's evolution into a pre-factor and relative changes 
to eliminate national and global effects. In this way, we show
the longitudinal dynamics of individual cities 
recapitulate the observed cross-sectional regularity. Larger cities are not 
only scaled-up versions of their smaller peers but also of their past.
In addition, our model shows that both specialization and diversification
are attributed to the distribution of industry's scaling exponents, resulting 
a critical population of 1.2 million at which 
a city makes an industrial transition into innovative economies.
\end{abstract}

\date{\today}

\maketitle

\begin{figure*}[!ht]
	\centering
	\includegraphics[width=0.99\textwidth]{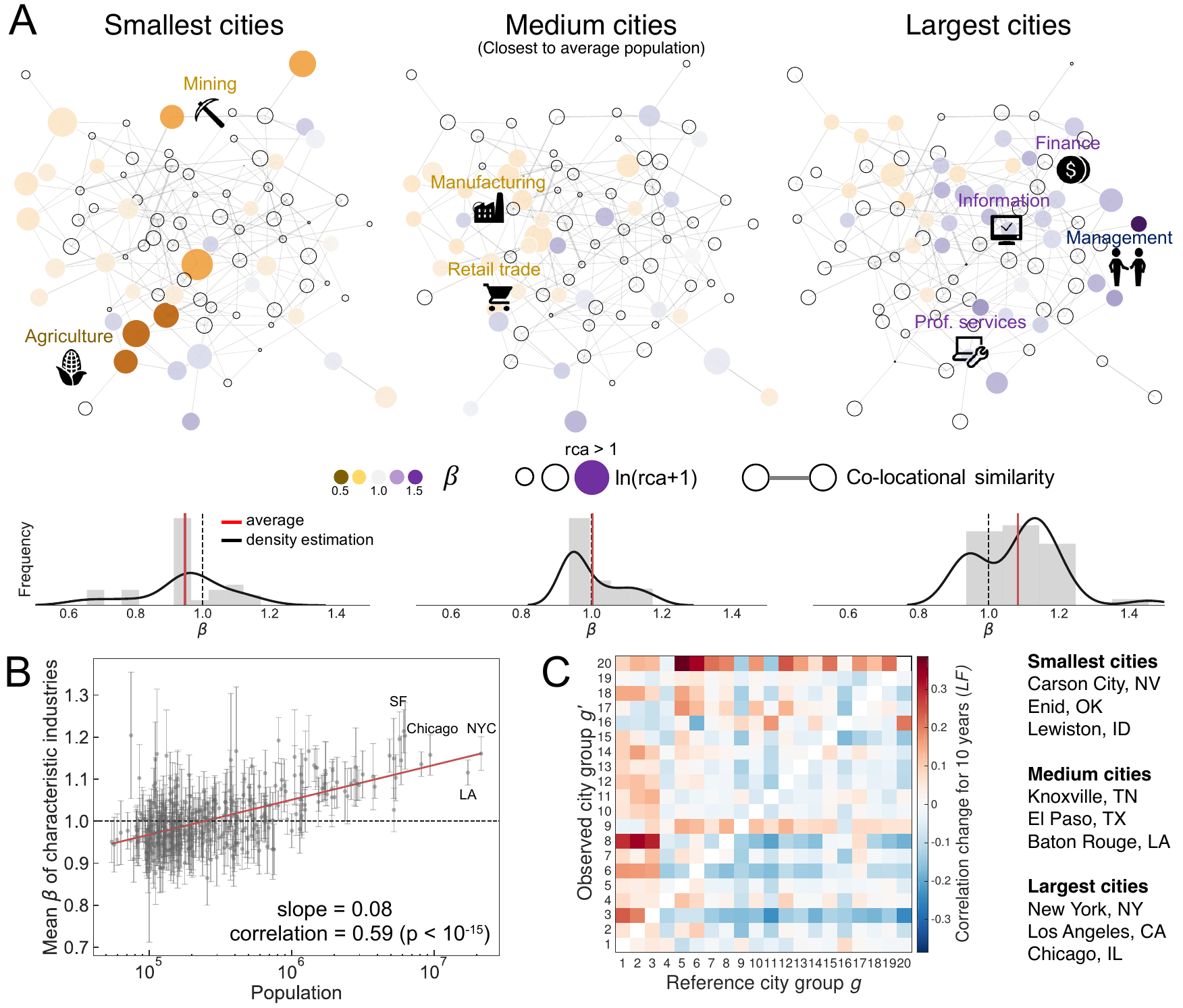}
	\caption{
    The characteristic industries of a city are predicted by city size via scaling exponent. 
	(A) The characteristic industries in small, medium, and large cities.
	Each industry (node) is sized by its logarithmic Revealed Comparative Advantages (RCA), and colored by the scaling exponent ($\beta$). 
	Empty nodes are non-characteristic industries ($rca < 1$). 
	Two industries $i$ and $j$ are connected when their co-locational similarity $\psi(i,j)$ is larger than 0.15. 
	Every value is averaged for 16 years time span.
	(B) The average scaling exponent of characteristic industries in each city (y-axis) compared to city size (x-axis).
	Error bars represent standard errors.
	(C) The lead-follow matrix demonstrates increases (red) or decreases (blue) in industrial similarity between cities ranked and grouped by size. 
	Each cell represents the similarity change over 10 year of an observed city group (y-axis) with respect to a reference group (x-axis). 
	The positive upper triangle means that smaller cities in the future become more similar to larger cities at the present.
	}
	\label{fig:fig1}
\end{figure*}

For societies worldwide, large cities play an important role in economic productivity and innovation~\cite{glaeser1992growth,quigley1998urban}.
The resulting economic and social opportunities drive worker migration to cities, thus producing an era of accelerated urban growth~\cite{migrationReport,rozenblat2007firm} and the advent of ``megacities''~\cite{megacity}.
Therefore, policy makers in growing cities must identify and leverage any available insight into the sustainability and future development of their local economies.

Specialization, a centerpiece of urban growth and change, has long been the virtue of urban development, 
shaping culturally and economically every city's idiosyncratic path-dependent trajectory. 
The prominence of urban economies are characterized, and hence understood by distinctive socio-economic activities,
mainly in a form of specialization~\cite{henderson1991urban, duranton2000diversity}.
This specialized urban development is considered as a result of a bundle of available natural resources~\cite{ellison1997geographic,ellison1999geographic}, a cluster of geographic advantages~\cite{porter1997new,delgado2014clusters,glaeser2005urban}, a sequence of strategic decisions~\cite{markusen1996interaction,evans2009creative,storper2015rise}, or a series of historical contingencies~\cite{henderson1991urban,martin2006path,glaeser2005reinventing}.
Therefore, studies often choose one or two cities of interest, and describe and explain their distinctive trajectories of economic development focusing one or more of the factors above (e.g. New York~\cite{glaeser2005urban}, Boston~\cite{glaeser2005reinventing}, Los Angeles and San Francisco~\cite{storper2015rise}).
These individual trajectories are observed to diverge rather than
converge to a single path~\cite{markusen2006distinctive,turok2009distinctive}, and such details are difficult to perform for all US cities in a comparative way. 

In the presence of path-dependent specialization, 
the very existence of general patterns across cities is puzzling.
These regularities include the famous Zipf's law, urban scaling, 
the number-size rule, and the universal distribution and hierarchical structure of business, 
which are largely expressed as a function of city size~\cite{bettencourt_growth_2007,batty2008size,batty2013new,barthelemy2011spatial, bettencourt2010urban,gomez-lievano_explaining_2016,schiller2016urban,zipf1949human, fujita_evolution_1999, mori2008number, bettencourt_origins_2013,youn2016scaling}.  
Indeed, population size has been a great indicator of many of urban properties, 
among which we are particularly interested in the industrial composition.
For example, small cities heavily rely on manufacturing-based labor while large cities on cognitive labor~\cite{henderson1986efficiency, florida2004rise,michaels2013task,youn2016scaling, frank2018small}.

These regularities have served as an excellent reference point for urban growth model, 
but they are nonetheless cross-sectional, freeze in time. 
To what extent do these cross-sectional regularities embody the longitudinal trajectories of individual cities?
It is conceivable that the common hierarchical structure, inside and across cities, 
results self-similar growth in time, also manifested as cross-sectional regularities ~\cite{christaller1933zentralen,fujita_evolution_1999, Lee2017}. 
The general economic development, largely observed at the national level, may be \emph{recapitulated} 
by individual cities ~\cite{hidalgo_product_2007, hidalgo_building_2009,thurner_schumpeterian_2010,klimek_empirical_2012,shutters2016constrained, sachs1995economic}.
Identifying such underlying structure will make a better predictive model for growing cities, and thus help urban policy makers to assess developmental opportunities.

In this study, we analyze the economic activities in U.S. cities between 1998 and 2013. 
First, we demonstrate that population size determines the industrial composition in a city. 
We then quantify to what extent this cross-sectional regularity is recapitulated by individual cities’ longitudinal dynamics. 
Finally, we show how cities make transition between the manufacturing economy to the productive and innovative economy, and
analytically derive the transition population.

\section*{Results}

\subsection*{Characteristic Industries Determined by City Size}
We characterize the economies of 350 U.S. cities between 1998 and 2013 by industries' relative size using the North American Industry Classification System (NAICS) (see \hyperref[method:data]{Materials and Methods} and SI.~\ref{SI-sec:data} for full description on data). 
Revealed Comparative Advantage (RCA), also known as location quotient, is
a widely used measure for a region's industrial specialization ~\cite{hidalgo_product_2007,isserman1977location,glaeser1992growth,markusen2006distinctive}. 
It is the city's relative employment share to the national average for the industry $i$, 
and we define \emph{characteristic} industry if its share in city $c$ is greater than national level at time $t$, 
$rca(c, i, t) > 1$ (solid color in Fig.~\ref{fig:fig1}A) ~\cite{balassa1965trade,hidalgo_product_2007}.


Scaling theory remains as a powerful framework for size-dependent nature of urban characteristics~\cite{bettencourt_growth_2007,youn2016scaling,bettencourt2016urban,gomez-lievano_explaining_2016}
and dynamics~\cite{bettencourt2010urban,alves2015scale}
despite its statistical caveat~\cite{arcaute_constructing_2015,leitao2016scaling}. 
Urban quantities are expressed as
\begin{equation}
    Y(c,i,t) \approx Y_0(i,t)\cdot N(c,t)^{\beta(i,t)}
    \label{eqn:scaling}
\end{equation}
where pre-factor $Y_{0}(i,t)$ and scaling exponent $\beta(i,t)$ are fitted given empirical observations of employment $Y(c,i,t)$ and population $N(c,t)$. 
We find the Eq.~\ref{eqn:scaling} is mostly in a good agreement with our dataset ($R^2 > 0.65$, see SI.~\ref{SI-sec:scaling}). 
Then, $\beta(i)$ characterizes the size-dependency of industry $i$ (sublinear, linear and superlinear). 

Figure~\ref{fig:fig1} summarizes our analyses in a network representation where characteristic industries are 
colored by $\beta$ and sized by RCA. 
Two industry $i$ and $j$ are connected by co-occurrence similarity, $\psi (i,j)$ across US cities (see \hyperref[method:rca]{Materials and Methods})
~\cite{hidalgo_product_2007,neffke_revealed_2008,muneepeerakul_urban_2013,shutters2016constrained}.
As population grows, where the city's characteristic industries occupying in the network moves (from agriculture and mining to managerial and technology-based industries), and this transition seems to be predicted by scaling exponents ~\cite{youn2016scaling}.

\begin{figure*}[!t]
	\centering
	\includegraphics[width=0.99\textwidth]{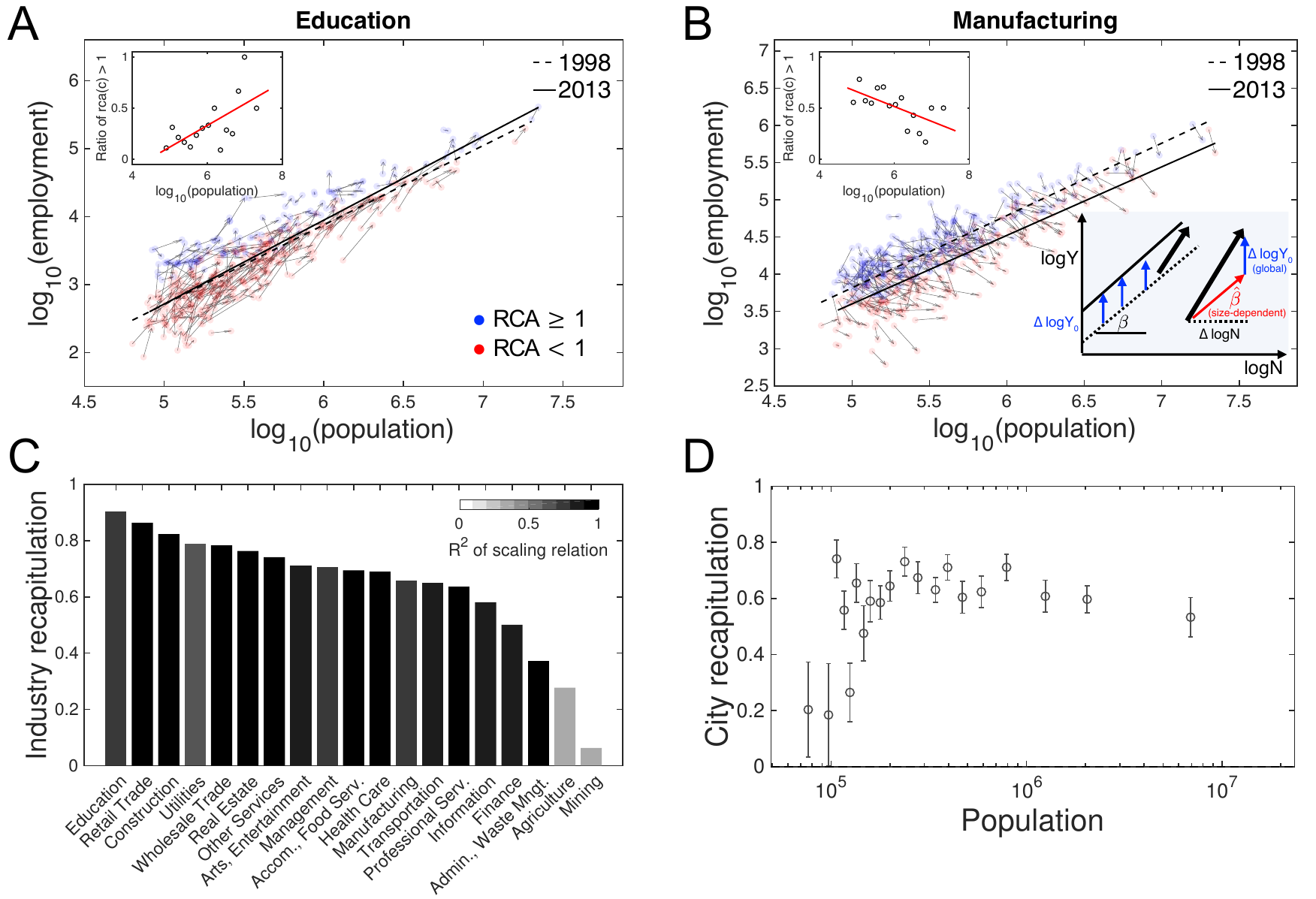}
	\caption{
	Individual industries and entire urban economies recapitulate the employment of larger cities.
	(A) The trajectory of each city on the population-industry plane of ``educational services'' industry. 
	Arrows depict the change of population and industry size of each city from 1998 to 2013. 
    Cities with $rca\geq1$ are indicated with blue arrows while cities with $rca<1$ have red arrows. 1998 (dashed) and in 2013 (solid).
	The ratio of characteristic cities in 2013 is depicted as a function of log-binned population in the inset.
	(B) The trajectory of each city by employment in the Manufacturing industry as in (A).
    We provide a schematic for decomposing changing employment as an additional inset.
	A city's trajectory (black arrow) can be decomposed into size-dependent growth (red arrow) and time-dependent global growth (blue arrow). 
	(C) The recapitulation score for each industry in the 2-digit NAICS classification. 
	Bar colors indicate the performance of the urban scaling model for that industry (i.e. $R^2$ of model fit according to Eq.~\ref{eqn:scaling}.
    For industries that are well-described by urban scaling (i.e. $R^2>0.65$), the average industrial recapitulation score of 0.70 is reasonably high.
	(D) The recapitulation score for city groups using industries with strong scaling relationships.
	}
	\label{fig:fig2}
\end{figure*}
 
City size determines what are the characteristic industries and their directions are manifested as scaling exponents.
For example, small cities are characterized by sublinear industries ($\beta(i)<1$), such as agriculture and mining, while large cities are by superlinear industries ($\beta(i)>1$) such as management and professional services. 
This result is consistent with existing observations that small cities are relying more on manual labor~\cite{henderson1986efficiency} while large cities on cognitive labor~\cite{michaels2013task,florida2004rise,youn2016scaling}, suggesting an industrial transition according to 
$\beta(i)$ from sublinear to superlinear sectors (see Fig.~\ref{fig:fig1}A). 
Fig.~\ref{fig:fig1}B confirms this overall trend at the level of individual U.S. cities. The average scaling exponent $\beta(i)$ for the characteristic industries in each city is strongly correlated with population size (Pearson correlation $\rho=0.59$).

\subsection*{Lead-Follow Matrix}

So far, our analysis uses a static snapshot of urban economies, but how do these cross-sectional relationships relate to temporal change in cities?
Here, we explore urban recapitulation through the industrial composition of different sized cities over time.
Cities are grouped into 20 equally-populated city groups (denoted by $g$) according to city size, and we represent the industrial characteristic of group $g$ as industry vector $\vec{G}(g,t)$ composed of logarithmic RCA values averaged for cities in group $g$ (see \hyperref[method:lead-follow]{Materials and Methods}).
Then, the Pearson correlations $\phi(g,g',t,\tau)$ of $\vec{G}(g,t)$ and $\vec{G}(g',t+\tau)$ describes how the industrial compositions of city groups $g$ and $g'$ relate over different time periods $t$ and $t+\tau$.
A lead-follow score aggregates these industrial changes over each starting year in our data set, according to
\begin{equation}
    LF(g,g') = \frac{\sum_{\tau=1}^{10}10\tau\cdot\langle\phi(g,g',t,\tau)-\phi(g,g',t,0) \rangle_t}{\sum_{\tau=1}^{10}\tau^2}.
    \label{industryChange}
\end{equation}
This score is basically the average similarity change between reference group $g$ and observed group $g'$ over a 10 year period as default, and a lead-follow matrix for the scores of all pairs summarizes the overall trend (see SI.~\ref{SI-sec:lead-follow} for other time periods).

The resulting lead-follow matrix provides temporal evidence for urban recapitulation (see Fig.~\ref{fig:fig1}C).
Groups of small cities become more similar to groups of large cities over a ten year period while groups of large cities become increasingly dissimilar to groups of small cities.
This observation suggests that there exists a general pathway explored by large cities and followed by small cities over time.
We interpret the lead-follow matrix as strong evidence for urban recapitulation of industrial compositions, and we endeavor to explain the phenomenon further.
City size determines both characteristic industries therein and the future growth. 

\subsection*{Explaining Urban Recapitulation}

How does city size characterize urban economic structures and their dynamics?
Characterized industries are determined by city size and scaling exponents. 
First, revealed comparative advantage of city is expressed as an industry's relative size 
to a national level. The industry's size is explained by scaling relation Eq.~\ref{eqn:scaling}. 
Therefore, we can express RCA as a function of scaling exponents as following: 
\begin{equation}
    rca(c,i,t) \propto \frac{N(c,t)^{\beta(i)-1}}{\sum_{c'\in C}N(c',t)^{\beta(i)}}  \\ 
    \label{proof1}
\end{equation}

The equation can be further approximated by Zipf's law $P(N(c)=N)\propto N^{-2}$ \cite{rosen1980size} (see SI.~\ref{SI-sec:integrated_framework}):  
\begin{equation}
    rca(c,i,t) \propto \frac{(\beta(i)-1)\cdot N(c,t)^{\beta(i)-1}}{N_{max}(t)^{\beta(i)-1}-N_{min}(t)^{\beta(i)-1}}
    \label{proof2}
\end{equation}
where $N_{max}(t)$ and $N_{min}(t)$ are size of the largest and smallest city at time $t$, respectively (see SI.~\ref{SI-sec:integrated_framework} for a complete derivation).

This framework determines the characteristic industries and 
predicts which industry will be characteristic given scaling relations. 
When cities grow, superlinear industries, $\beta(i)>1$, will be characteristic 
industries as a result of increasing RCA value, as shown in Fig.~\ref{fig:fig1}A\&B.
We exemplify this observation empirically by examining the growth of individual U.S. cities from 1998 to 2013 compared to their RCA values for the Education industry (see Fig.~\ref{fig:fig2}A), which has $\beta(i)=1.21$, and the Manufacturing industry (see Fig.~\ref{fig:fig2}B), which has $\beta(i)=0.94$.

Both our analytic framework and empirical observations imply the existence of urban recapitulation by city size.
Now, how can we measure the degree of recapitulation for individual industries and cities?
The size-dependency $\beta(i)$ estimates the employment growth when the population grows in time. 
We compare the cross-sectional ratio $\beta(i)$ and the observed longitudinal growth ratio $\hat{\beta}(i)$ as a measure of recapitulation.
From Eq.~\ref{eqn:scaling}, we have
\begin{equation}
    \Delta \log(Y(c,i)) \approx \Delta \log(Y_0(i)) + \beta(i)\cdot\Delta \log(N(c))
    \label{scalingDeriv}
\end{equation}
where $\Delta\log(Y(c,i))=\log(Y(c,i,2013) - \log(Y(c,i,1998))$, and the same for $\log{Y_{0}}$ and $\log{N}$.
This calculation decomposes employment growth into each city's size-dependent growth (i.e. $\beta(i)\cdot\Delta\log(N(c))$) and global time-dependent growth (i.e. $\Delta\log(Y_0(i))$. See the schematic in Fig.~\ref{fig:fig2}B).
By fitting Eq.~\ref{scalingDeriv} to empirical employment data, we derive the longitudinal size-dependency $\hat{\beta}(i)$ and the global trend $\Delta\log(\hat{Y}_0(t))$ across all cities between 1998 and 2013, and we compare $\hat{\beta}(i)$ with the cross-sectional size-dependency $\beta(i)$.
If an individual industry is perfectly recapitulated by growing cities, then cross-sectional and longitudinal size-dependencies should be equal ($\hat{\beta}(i)=\beta(i)$).
Therefore, we employ an industry recapitulation score according to
\begin{equation}
    S(i) = 1-\Big\lvert\frac{\hat{\beta}(i)-\beta(i)}{\beta(i)}\Big\rvert.
    \label{recapScore}
\end{equation}
The industry recapitulation score becomes one when every city perfectly recapitulates, and equals to zero when employment growth has no longitudinal size-dependency.

\begin{figure*}
	\includegraphics[width=\textwidth]{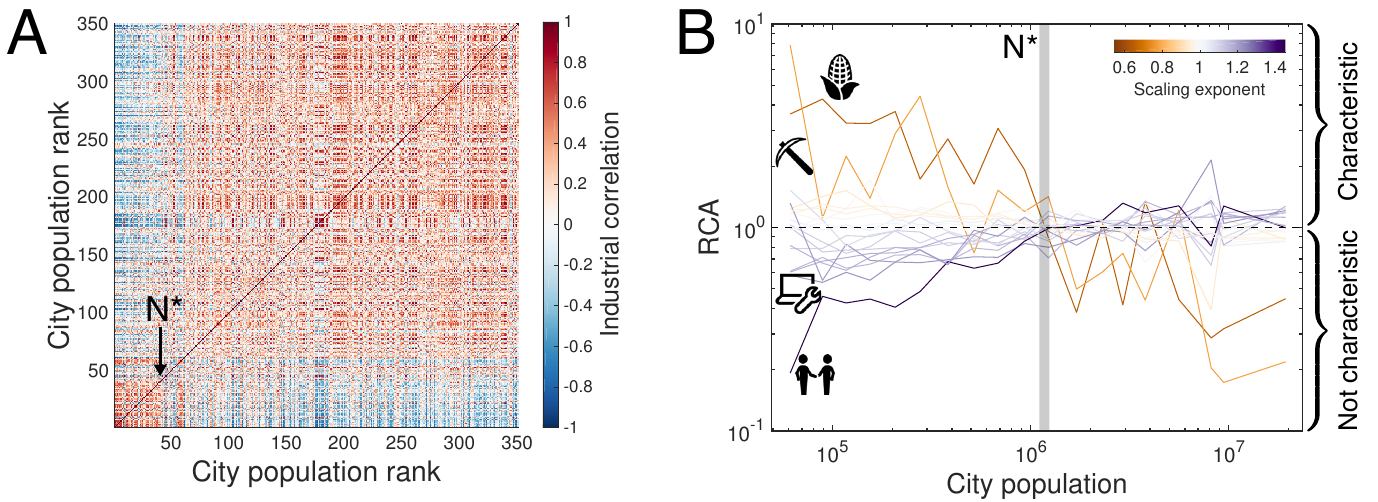}
	\caption{
    Small cities transition into an innovative large-city economy after a critical size.
	(A) After ordering US cities in decreasing size (x- and y-axis), we measure the correlation of logarithmic RCA scores between city pairs at the 2-digit NAICS classification (color).
	We use $N^{*}$ to denote a critical size rank that separates cities based on industrial composition.
	(B) We compare the prominence of industries (y-axis) with different cross-sectional scaling exponents (color) across cities of different sizes (x-axis).
	Again, we observe a critical city size ($N^{*}$) that divides small city economies from innovative large city economies. 
	}
	\label{fig:city_clustering}
\end{figure*}

We observe high recapitulation scores (i.e. $S(i)>0.50$) for industries that are well-described by urban scaling (i.e. $R^2>0.65$).
This strong longitudinal size-dependency is a sufficient condition of the constant deviation from a common trajectory~\cite{bettencourt2010urban}.
In contrast to the recent research on traffic congestion~\cite{depersin2018global}, the common trajectory is well explained by city size. 
The most significant difference is teasing out the global trend, $\Delta\log{(Y_0)}$, which makes common upward or downward drifts as shown in manufacturing in Fig.~\ref{fig:fig2}B.
Its recapitulation score around $0.7$ shows that decomposition is essential to properly measure the size-dependent growth.

Analogous to industries, do cities recapitulate with the same strength?
We generalize the decomposition in Eq.~\ref{scalingDeriv} for individual cities, and measure the recapitulation scores of city groups, accordingly.
We extract the size-dependent growth of each city by subtracting the global growth $\Delta\log(\hat{Y}_0(t))$ estimated from Eq.~\ref{scalingDeriv}.
Then, we measure the longitudinal size-dependency $\hat{B}(g,i)$ averaged for the cities in city group $g$ and industry $i$ similar to $\hat{\beta}(i)$ (see SI.~\ref{SI-sec:recapitulation} for the details).
\begin{equation}
	\hat{B}(g,i) = \frac{\sum_{c \in g}\left(\Delta\log{Y}(c,i) - \Delta\log{\hat{Y}_{0}(i)}\right)\cdot\Delta\log{N}(c)}{\sum_{c \in g} \left( \Delta\log{N}(c) \right)^{2}}	
\end{equation}
where $c \in g$ is the set of cities in city group $g$, and the groups are determined in the order of populations.
Finally, we measure a city group recapitulation score according to 
\begin{equation}
    S(g) = 1-\frac{1}{|I|}\sum_{i\in I}\Big\lvert\frac{\hat{B}(i,g)-\beta(i)}{\beta(i)}\Big\rvert.
    \label{cityRecap}
\end{equation}
where $i \in I$ is the set of all industries. Similar to before, larger values of $S(g)$ mean large tendency of recapitulation.

In addition to industries, a strong tendency of recapitulation is also observed for city groups.
Fig.~\ref{fig:fig2}D demonstrates that recapitulation occurs across all city groups as $\langle S(g)\rangle_{g} \approx 0.70$.
Most city groups exhibit a high strength of recapitulation, except for a few groups of very small cities. 
From the measured recapitulation scores of industries and cities, we conclude that cities recapitulate a general pathway ruled by city size, in agreement with the observed lead-follow matrix and our framework connecting RCA and scaling relation.

\subsection*{Transition into Innovative Economies}
The observed \emph{recapitulation} predicts the structural change of urban economy, represented by characteristic industries, as city size grows. Can we specify the characteristic size at which a city transitions from a small city economy into the innovative and productive economy of large cities?
Specifically, given the analytical relationship between city size and corresponding characteristic industries in Eq.~\ref{proof2}, at what size do cities begin to leverage innovative industries~\cite{youn2016scaling} with $\beta(i)>1$?
Fig.~\ref{fig:city_clustering}A compares cities according to size and the Pearson correlation of logarithmic RCA values (i.e. $\vec{I}(c,t)=(\dots,\log(rca(c,i,t)+1),\dots)$) for city pairs averaged for entire time span.
We observe a transition at around the 50th largest city indicating that the 50 largest cities are characterized by qualitatively different industries compared to the remaining smaller cities. 

Our framework can explain this observed transition through a fixed point in the $\beta-N$ plane.
From Eq.~\ref{proof2}, we observe a saddle point at $\beta^*(i)=1$ and city size $N^*\approx 1.2$ million, which is around the population of Louisville, KY, the 43rd largest city (see SI.~\ref{SI-sec:integrated_framework} for a complete derivation).
The existence of this fixed point suggests that characteristic industries in growing cities will change based on $\beta(i)$ as cities achieve this critical size.
We expect that small cities with $N(c)<N^*$ will be characterized by industries with $\beta(i)<1$ while large cities with $N(c)>N^*$ will be characterized by industries with $\beta(i)>1$.
Fig.~\ref{fig:city_clustering}B empirically confirms that scaling exponents distinguish the trends of characteristic industries by city size.
The characteristic industries of growing cities transition from industries with $\beta(i)<1$ to industries with $\beta(i)>1$ at precisely the predicted city size $N^*=$ 1.2 millions.

\section*{Discussion}

We provide a comprehensive framework that connects the cross-sectional size-dependency of urban economies to longitudinal industrial composition in US cities in a process that well-described as ``urban recapitulation.''
We provide both analytical and empirical evidence to support the existence of a common trajectory that cities grow along.
Our observations show that urban economies actually recapitulate a common trajectory at an aggregate level, which anticipates economic change as cities grow.

Surprisingly, our recapitulation framework identifies industrial transitions at population size of 1.2 million where the increasing prominence with population of superlinearly-scaled industries crosses the decreasing trend of sublinearly-scaled industries.
It shows good agreement with empirical results that superlinearly-scaled industries are characteristic in large cities while sublinearly-scaled industries are characteristic in small cities.
Scaling exponents are often considered as a proxy of innovation and maturity, as primary and secondary sectors have sublinear scaling, while managerial and professional industries have superlinear scaling~\cite{pumain_cybergeo_2006,youn2016scaling}.
Therefore, an urban economy is expected to become more innovative, creative, and economically desirable as the city grows above some characteristic size.
This result informs policy makers, employers and job-seekers about upcoming structural changes in the local labor market.

At first, our finding seems to contrast the conventional wisdom in urban studies that suggests cities possess unique individual trajectories~\cite{markusen2006distinctive,turok2009distinctive}. 
However, this specialization is also explained by the recapitulation framework.
A specialized city is represented by only a small number of characteristic industries, and we showed that superlinearity and sublinearity distinguish characteristic industries in small and large cities.
Therefore, the specialization trend in size is determined by the number of sublinearly- and superlinearly-scaled industries.
In our case, the small number of sublinearly-scaled industries results in higher specialization in small cities (see Fig.~\ref{fig:n_prominent}).
Specialization is not a random accident or localized feature, but a regular characteristic.
Distinctive pathways of individual cities are also understood by the recapitulation framework as population growth differentiates industrial growth according to their scaling exponents. 
We can regard the deviation from the common trajectory as an \emph{effective} distinctive pathway~\cite{bettencourt2010urban}. 
This enables us to measure distinctiveness more accurately, and hence our analysis may contribute to a well-individualized urban policy~\cite{turok2009distinctive}.
 
In the previous studies on clusters of traded industries~\cite{porter2003economic} and national exports~\cite{hidalgo_product_2007}, economic characteristics and their growth are considered as independent of population size.
Then, why do we observe strong recapitulation by city size?
Most of these studies looked into tradeable sectors and exclude production and consumption in domestic market.
As a result, they overstate their comparative advantages because they sell products in external markets.
By considering both internal external production and consumption, our results show different level of recapitulation by tradeability.
We observe strong recapitulation in the industries that we locally consume, such as education, retail trade, construction, and utility, while weak recapitulation in tradaeble industries, such as agriculture, mining, administrative service, and finance.
Our analysis provides a reliable reference for economic growth by tradeability.

Economic output is often expressed as a product function of labor and capital inputs, which are implicit functions of population~\cite{lobo2013urban}. 
Our approach turns around this convention by bringing this implicit relation to an explicit functional form and isolating industry's size-dependent growth in urban areas from any other factors. 
Individuals make up a city and account for socio-economic activities that constitute labor and result in capital, the product of which becomes the economic output.
Thus, we consider population as a fundamental unit to make a {\it first-order} approach to the economic output. 
Indeed, there are numerous {\it higher-order} attributes unexplained by population: available natural resources, accessibility, and economic intervention~\cite{ellison1997geographic,ellison1999geographic}. 
In addition, not every industry perfectly fits to a scaling function, and yet the functional form serves well to derive a general characteristics of recapitulation and industrial transition at $N^{*}$~\cite{arcaute_constructing_2015,leitao2016scaling}.

Our framework is inspired by the recapitulation theory in biology~\cite{ehrlich1963process} much the same as urban scaling theory~\cite{bettencourt_growth_2007} is motivated from the allometric scaling in biology~\cite{west_general_1997}.
However, there is a conceptual difference between them.
In biological recapitulation, close species grow on a common pathway as they recapitulate a shared evolutionary pathway composed of many common ancestors.
We consider each city as a footprint of evolutionary pathway of the largest city, thus, we regard the scaling relation itself as the common trajectory.
Although cities have followed the general pathway in our short-term data, our framework needs to be evaluated and advanced through further works on long-term data~\cite{pumain_cybergeo_2006}.

Employing the scaling framework, we have discussed several urban economic characteristics: recapitulation, transition, specialization and distinctiveness.
Unveiling the origin of urban scaling will answer how urban economies have evolved with embodying such characteristics.
While there are several reliable explanations using networks~\cite{bettencourt_origins_2013}, necessary complementary factors~\cite{gomez-lievano_explaining_2016}, collective brains~\cite{kline2010population,muthukrishna2016innovation} and social interactions~\cite{yang2017origin}, we still lack a concrete generative mechanism and its empirical confirmation to explain how microlevel individual capabilities aggregate into collective intelligence, and then manifest economic outputs formulated as scaling relations.
Identifying such mechanism will specify insufficient capabilities for growth and enable urban policy makers to build an appropriate policy for sustainable economic success.

\section*{Materials and Methods}
\subsection*{Data}
\label{method:data}
This study uses employment data by North American Industry Classification System (NAICS) industry code and 350 U.S. metropolitan statistical areas (referred to as ``city'' throughout the manuscript) according to the U.S. Bureau of Labor Statistics.
The sizes of industry set are 19, 86, 289, 642 and 978 according to the depth of classification denoted by 2-6 digits.   
Data includes annual employment measurements for the years 1998 through 2013.

\subsection*{Identifying Characteristic Industries with RCA}
\label{method:rca}
We employ revealed comparative advantage (RCA)~\cite{balassa1965trade,hidalgo_product_2007} to identify characteristic industries in each U.S. city.
Let $I$ denote the set of 2-digits NAICS industry codes and $C$ denote the set of U.S. cities, then $Y(c,i,t)$ denotes the employment in industry $i\in I$ in city $c\in C$ at year $t\in[1998,2013]$.
The revealed comparative advantage is given by
\begin{equation}
    rca(c,i,t) = \frac{Y(c,i,t)/\sum_{i'\in I}Y(c,i',t)}{\sum_{c'\in C}Y(c',i,t)/\sum_{c'\in C,i'\in I}Y(c',i',t)},
    \label{rca}
\end{equation}
where $c$, $i$ and $t$ denote city, industry and time, respectively. 

We can characterize industry $i$ as a vector of RCA values of cities as $\vec{I}(i,t) = (\dots,\log(rca(c,i,t)+1),\dots)$.
The time-averaged inter-industry relatedness is defined by the Pearson correlation of two industry vectors $\vec{I}(i,t)$ and $\vec{I}(j,t)$ as 
\begin{equation}
	\psi(i,j) = \langle \rho(\vec{I}(i,t), \vec{I}(j,t))\rangle_{t},
\end{equation}
where $\rho$ is the Pearson correlation function. Figure 1 A shows links  that satisfy $\psi(i,j)>0.15$. 

\subsection*{Lead-Follow Relationship}
\label{method:lead-follow}

In the analysis on lead-follow relationship, we bin cities into 20 equally-populated city groups (denoted by $g$) according to city size, and, for each city group, we calculate the average RCA values of each city according to 
\begin{equation}
	\vec{G}(g,t)=(\dots,\langle \log((rca(c,i,t)+1) \rangle_{c\in g},\dots)_{i\in I},
\end{equation}
where $I$ denotes the set of 2-digit NAICS industry codes.

\begin{acknowledgments}
The authors thank to Ricardo Hausmann, Frank Neffke, Lu{\'\i}s M. A. Bettencourt and C\'esar Hidalgo for useful comments. W.-S. Jung was supported by Basic Science Research Program through the National Research Foundation of Korea (NRF) funded by the Ministry of Education (2016R1D1A1B03932590).
\end{acknowledgments}

\bibliography{references}

\begin{thebibliography}{63}%
\makeatletter
\providecommand \@ifxundefined [1]{%
 \@ifx{#1\undefined}
}%
\providecommand \@ifnum [1]{%
 \ifnum #1\expandafter \@firstoftwo
 \else \expandafter \@secondoftwo
 \fi
}%
\providecommand \@ifx [1]{%
 \ifx #1\expandafter \@firstoftwo
 \else \expandafter \@secondoftwo
 \fi
}%
\providecommand \natexlab [1]{#1}%
\providecommand \enquote  [1]{``#1''}%
\providecommand \bibnamefont  [1]{#1}%
\providecommand \bibfnamefont [1]{#1}%
\providecommand \citenamefont [1]{#1}%
\providecommand \href@noop [0]{\@secondoftwo}%
\providecommand \href [0]{\begingroup \@sanitize@url \@href}%
\providecommand \@href[1]{\@@startlink{#1}\@@href}%
\providecommand \@@href[1]{\endgroup#1\@@endlink}%
\providecommand \@sanitize@url [0]{\catcode `\\12\catcode `\$12\catcode
  `\&12\catcode `\#12\catcode `\^12\catcode `\_12\catcode `\%12\relax}%
\providecommand \@@startlink[1]{}%
\providecommand \@@endlink[0]{}%
\providecommand \url  [0]{\begingroup\@sanitize@url \@url }%
\providecommand \@url [1]{\endgroup\@href {#1}{\urlprefix }}%
\providecommand \urlprefix  [0]{URL }%
\providecommand \Eprint [0]{\href }%
\providecommand \doibase [0]{http://dx.doi.org/}%
\providecommand \selectlanguage [0]{\@gobble}%
\providecommand \bibinfo  [0]{\@secondoftwo}%
\providecommand \bibfield  [0]{\@secondoftwo}%
\providecommand \translation [1]{[#1]}%
\providecommand \BibitemOpen [0]{}%
\providecommand \bibitemStop [0]{}%
\providecommand \bibitemNoStop [0]{.\EOS\space}%
\providecommand \EOS [0]{\spacefactor3000\relax}%
\providecommand \BibitemShut  [1]{\csname bibitem#1\endcsname}%
\let\auto@bib@innerbib\@empty
\bibitem [{\citenamefont {Glaeser}\ \emph {et~al.}(1992)\citenamefont
  {Glaeser}, \citenamefont {Kallal}, \citenamefont {Scheinkman},\ and\
  \citenamefont {Shleifer}}]{glaeser1992growth}%
  \BibitemOpen
  \bibfield  {author} {\bibinfo {author} {\bibfnamefont {E.~L.}\ \bibnamefont
  {Glaeser}}, \bibinfo {author} {\bibfnamefont {H.~D.}\ \bibnamefont {Kallal}},
  \bibinfo {author} {\bibfnamefont {J.~A.}\ \bibnamefont {Scheinkman}}, \ and\
  \bibinfo {author} {\bibfnamefont {A.}~\bibnamefont {Shleifer}},\ }\href@noop
  {} {\bibfield  {journal} {\bibinfo  {journal} {Journal of political economy}\
  }\textbf {\bibinfo {volume} {100}},\ \bibinfo {pages} {1126} (\bibinfo {year}
  {1992})}\BibitemShut {NoStop}%
\bibitem [{\citenamefont {Quigley}(1998)}]{quigley1998urban}%
  \BibitemOpen
  \bibfield  {author} {\bibinfo {author} {\bibfnamefont {J.~M.}\ \bibnamefont
  {Quigley}},\ }\href@noop {} {\bibfield  {journal} {\bibinfo  {journal} {The
  Journal of Economic Perspectives}\ }\textbf {\bibinfo {volume} {12}},\
  \bibinfo {pages} {127} (\bibinfo {year} {1998})}\BibitemShut {NoStop}%
\bibitem [{\citenamefont {{International Organization for
  Migration}}(2015)}]{migrationReport}%
  \BibitemOpen
  \bibfield  {author} {\bibinfo {author} {\bibnamefont {{International
  Organization for Migration}}},\ }\href@noop {} {\emph {\bibinfo {title}
  {World Migration Report 2015 - Migrants and Cities: New Partnerships to
  Manage Mobility}}},\ \bibinfo {type} {Tech. Rep.}\ (\bibinfo {year}
  {2015})\BibitemShut {NoStop}%
\bibitem [{\citenamefont {Rozenblat}\ and\ \citenamefont
  {Pumain}(2007)}]{rozenblat2007firm}%
  \BibitemOpen
  \bibfield  {author} {\bibinfo {author} {\bibfnamefont {C.}~\bibnamefont
  {Rozenblat}}\ and\ \bibinfo {author} {\bibfnamefont {D.}~\bibnamefont
  {Pumain}},\ }in\ \href@noop {} {\emph {\bibinfo {booktitle} {Cities in
  Globalization: Practices, Policies, and Theories}}}\ (\bibinfo  {publisher}
  {Routledge},\ \bibinfo {year} {2007})\ pp.\ \bibinfo {pages}
  {130--156}\BibitemShut {NoStop}%
\bibitem [{\citenamefont {Dobbs}(2010)}]{megacity}%
  \BibitemOpen
  \bibfield  {author} {\bibinfo {author} {\bibfnamefont {R.}~\bibnamefont
  {Dobbs}},\ }\href@noop {} {\bibfield  {journal} {\bibinfo  {journal} {Foreign
  Policy Magazine}\ }\textbf {\bibinfo {volume} {181}},\ \bibinfo {pages} {132}
  (\bibinfo {year} {2010})}\BibitemShut {NoStop}%
\bibitem [{\citenamefont {Henderson}(1991)}]{henderson1991urban}%
  \BibitemOpen
  \bibfield  {author} {\bibinfo {author} {\bibfnamefont {J.~V.}\ \bibnamefont
  {Henderson}},\ }\href
  {https://ideas.repec.org/b/oxp/obooks/9780195069020.html} {\emph {\bibinfo
  {title} {Urban {Development}: {Theory}, {Fact}, and {Illusion}}}},\ \bibinfo
  {type} {{OUP} {Catalogue}}\ (\bibinfo  {institution} {Oxford University
  Press},\ \bibinfo {year} {1991})\BibitemShut {NoStop}%
\bibitem [{\citenamefont {Duranton}\ and\ \citenamefont
  {Puga}(2000)}]{duranton2000diversity}%
  \BibitemOpen
  \bibfield  {author} {\bibinfo {author} {\bibfnamefont {G.}~\bibnamefont
  {Duranton}}\ and\ \bibinfo {author} {\bibfnamefont {D.}~\bibnamefont
  {Puga}},\ }\href@noop {} {\bibfield  {journal} {\bibinfo  {journal} {Urban
  studies}\ }\textbf {\bibinfo {volume} {37}},\ \bibinfo {pages} {533}
  (\bibinfo {year} {2000})}\BibitemShut {NoStop}%
\bibitem [{\citenamefont {Ellison}\ and\ \citenamefont
  {Glaeser}(1997)}]{ellison1997geographic}%
  \BibitemOpen
  \bibfield  {author} {\bibinfo {author} {\bibfnamefont {G.}~\bibnamefont
  {Ellison}}\ and\ \bibinfo {author} {\bibfnamefont {E.~L.}\ \bibnamefont
  {Glaeser}},\ }\href@noop {} {\bibfield  {journal} {\bibinfo  {journal}
  {Journal of political economy}\ }\textbf {\bibinfo {volume} {105}},\ \bibinfo
  {pages} {889} (\bibinfo {year} {1997})}\BibitemShut {NoStop}%
\bibitem [{\citenamefont {Ellison}\ and\ \citenamefont
  {Glaeser}(1999)}]{ellison1999geographic}%
  \BibitemOpen
  \bibfield  {author} {\bibinfo {author} {\bibfnamefont {G.}~\bibnamefont
  {Ellison}}\ and\ \bibinfo {author} {\bibfnamefont {E.~L.}\ \bibnamefont
  {Glaeser}},\ }\href@noop {} {\bibfield  {journal} {\bibinfo  {journal}
  {American Economic Review}\ }\textbf {\bibinfo {volume} {89}},\ \bibinfo
  {pages} {311} (\bibinfo {year} {1999})}\BibitemShut {NoStop}%
\bibitem [{\citenamefont {Porter}(1997)}]{porter1997new}%
  \BibitemOpen
  \bibfield  {author} {\bibinfo {author} {\bibfnamefont {M.~E.}\ \bibnamefont
  {Porter}},\ }\href@noop {} {\bibfield  {journal} {\bibinfo  {journal}
  {Economic Development Quarterly}\ }\textbf {\bibinfo {volume} {11}},\
  \bibinfo {pages} {11} (\bibinfo {year} {1997})}\BibitemShut {NoStop}%
\bibitem [{\citenamefont {Delgado}\ \emph {et~al.}(2014)\citenamefont
  {Delgado}, \citenamefont {Porter},\ and\ \citenamefont
  {Stern}}]{delgado2014clusters}%
  \BibitemOpen
  \bibfield  {author} {\bibinfo {author} {\bibfnamefont {M.}~\bibnamefont
  {Delgado}}, \bibinfo {author} {\bibfnamefont {M.~E.}\ \bibnamefont {Porter}},
  \ and\ \bibinfo {author} {\bibfnamefont {S.}~\bibnamefont {Stern}},\
  }\href@noop {} {\bibfield  {journal} {\bibinfo  {journal} {Research Policy}\
  }\textbf {\bibinfo {volume} {43}},\ \bibinfo {pages} {1785} (\bibinfo {year}
  {2014})}\BibitemShut {NoStop}%
\bibitem [{\citenamefont {Glaeser}(2005{\natexlab{a}})}]{glaeser2005urban}%
  \BibitemOpen
  \bibfield  {author} {\bibinfo {author} {\bibfnamefont {E.~L.}\ \bibnamefont
  {Glaeser}},\ }\href@noop {} {\emph {\bibinfo {title} {Urban colossus: why is
  New York America's largest city?}}},\ \bibinfo {type} {Tech. Rep.}\ (\bibinfo
   {institution} {National Bureau of Economic Research},\ \bibinfo {year}
  {2005})\BibitemShut {NoStop}%
\bibitem [{\citenamefont {Markusen}(1996)}]{markusen1996interaction}%
  \BibitemOpen
  \bibfield  {author} {\bibinfo {author} {\bibfnamefont {A.}~\bibnamefont
  {Markusen}},\ }\href@noop {} {\bibfield  {journal} {\bibinfo  {journal}
  {International Regional Science Review}\ }\textbf {\bibinfo {volume} {19}},\
  \bibinfo {pages} {49} (\bibinfo {year} {1996})}\BibitemShut {NoStop}%
\bibitem [{\citenamefont {Evans}(2009)}]{evans2009creative}%
  \BibitemOpen
  \bibfield  {author} {\bibinfo {author} {\bibfnamefont {G.}~\bibnamefont
  {Evans}},\ }\href@noop {} {\bibfield  {journal} {\bibinfo  {journal} {Urban
  studies}\ }\textbf {\bibinfo {volume} {46}},\ \bibinfo {pages} {1003}
  (\bibinfo {year} {2009})}\BibitemShut {NoStop}%
\bibitem [{\citenamefont {Storper}\ \emph {et~al.}(2015)\citenamefont
  {Storper}, \citenamefont {Kemeny}, \citenamefont {Makarem},\ and\
  \citenamefont {Osman}}]{storper2015rise}%
  \BibitemOpen
  \bibfield  {author} {\bibinfo {author} {\bibfnamefont {M.}~\bibnamefont
  {Storper}}, \bibinfo {author} {\bibfnamefont {T.}~\bibnamefont {Kemeny}},
  \bibinfo {author} {\bibfnamefont {N.}~\bibnamefont {Makarem}}, \ and\
  \bibinfo {author} {\bibfnamefont {T.}~\bibnamefont {Osman}},\ }\href@noop {}
  {\emph {\bibinfo {title} {The rise and fall of urban economies: Lessons from
  San Francisco and Los Angeles}}}\ (\bibinfo  {publisher} {Stanford University
  Press},\ \bibinfo {year} {2015})\BibitemShut {NoStop}%
\bibitem [{\citenamefont {Martin}\ and\ \citenamefont
  {Sunley}(2006)}]{martin2006path}%
  \BibitemOpen
  \bibfield  {author} {\bibinfo {author} {\bibfnamefont {R.}~\bibnamefont
  {Martin}}\ and\ \bibinfo {author} {\bibfnamefont {P.}~\bibnamefont
  {Sunley}},\ }\href@noop {} {\bibfield  {journal} {\bibinfo  {journal}
  {Journal of economic geography}\ }\textbf {\bibinfo {volume} {6}},\ \bibinfo
  {pages} {395} (\bibinfo {year} {2006})}\BibitemShut {NoStop}%
\bibitem [{\citenamefont
  {Glaeser}(2005{\natexlab{b}})}]{glaeser2005reinventing}%
  \BibitemOpen
  \bibfield  {author} {\bibinfo {author} {\bibfnamefont {E.~L.}\ \bibnamefont
  {Glaeser}},\ }\href@noop {} {\bibfield  {journal} {\bibinfo  {journal}
  {Journal of Economic Geography}\ }\textbf {\bibinfo {volume} {5}},\ \bibinfo
  {pages} {119} (\bibinfo {year} {2005}{\natexlab{b}})}\BibitemShut {NoStop}%
\bibitem [{\citenamefont {Markusen}\ and\ \citenamefont
  {Schrock}(2006)}]{markusen2006distinctive}%
  \BibitemOpen
  \bibfield  {author} {\bibinfo {author} {\bibfnamefont {A.}~\bibnamefont
  {Markusen}}\ and\ \bibinfo {author} {\bibfnamefont {G.}~\bibnamefont
  {Schrock}},\ }\href@noop {} {\bibfield  {journal} {\bibinfo  {journal} {Urban
  studies}\ }\textbf {\bibinfo {volume} {43}},\ \bibinfo {pages} {1301}
  (\bibinfo {year} {2006})}\BibitemShut {NoStop}%
\bibitem [{\citenamefont {Turok}(2009)}]{turok2009distinctive}%
  \BibitemOpen
  \bibfield  {author} {\bibinfo {author} {\bibfnamefont {I.}~\bibnamefont
  {Turok}},\ }\href@noop {} {\bibfield  {journal} {\bibinfo  {journal}
  {Environment and planning A}\ }\textbf {\bibinfo {volume} {41}},\ \bibinfo
  {pages} {13} (\bibinfo {year} {2009})}\BibitemShut {NoStop}%
\bibitem [{\citenamefont {Bettencourt}\ \emph {et~al.}(2007)\citenamefont
  {Bettencourt}, \citenamefont {Lobo}, \citenamefont {Helbing}, \citenamefont
  {Kühnert},\ and\ \citenamefont {West}}]{bettencourt_growth_2007}%
  \BibitemOpen
  \bibfield  {author} {\bibinfo {author} {\bibfnamefont {L.~M.~A.}\
  \bibnamefont {Bettencourt}}, \bibinfo {author} {\bibfnamefont
  {J.}~\bibnamefont {Lobo}}, \bibinfo {author} {\bibfnamefont {D.}~\bibnamefont
  {Helbing}}, \bibinfo {author} {\bibfnamefont {C.}~\bibnamefont {Kühnert}}, \
  and\ \bibinfo {author} {\bibfnamefont {G.~B.}\ \bibnamefont {West}},\ }\href
  {\doibase 10.1073/pnas.0610172104} {\bibfield  {journal} {\bibinfo  {journal}
  {PNAS}\ }\textbf {\bibinfo {volume} {104}},\ \bibinfo {pages} {7301}
  (\bibinfo {year} {2007})}\BibitemShut {NoStop}%
\bibitem [{\citenamefont {Batty}(2008)}]{batty2008size}%
  \BibitemOpen
  \bibfield  {author} {\bibinfo {author} {\bibfnamefont {M.}~\bibnamefont
  {Batty}},\ }\href@noop {} {\bibfield  {journal} {\bibinfo  {journal}
  {Science}\ }\textbf {\bibinfo {volume} {319}},\ \bibinfo {pages} {769}
  (\bibinfo {year} {2008})}\BibitemShut {NoStop}%
\bibitem [{\citenamefont {Batty}(2013)}]{batty2013new}%
  \BibitemOpen
  \bibfield  {author} {\bibinfo {author} {\bibfnamefont {M.}~\bibnamefont
  {Batty}},\ }\href@noop {} {\emph {\bibinfo {title} {The new science of
  cities}}}\ (\bibinfo  {publisher} {Mit Press},\ \bibinfo {year}
  {2013})\BibitemShut {NoStop}%
\bibitem [{\citenamefont {Barth{\'e}lemy}(2011)}]{barthelemy2011spatial}%
  \BibitemOpen
  \bibfield  {author} {\bibinfo {author} {\bibfnamefont {M.}~\bibnamefont
  {Barth{\'e}lemy}},\ }\href@noop {} {\bibfield  {journal} {\bibinfo  {journal}
  {Physics Reports}\ }\textbf {\bibinfo {volume} {499}},\ \bibinfo {pages} {1}
  (\bibinfo {year} {2011})}\BibitemShut {NoStop}%
\bibitem [{\citenamefont {Bettencourt}\ \emph {et~al.}(2010)\citenamefont
  {Bettencourt}, \citenamefont {Lobo}, \citenamefont {Strumsky},\ and\
  \citenamefont {West}}]{bettencourt2010urban}%
  \BibitemOpen
  \bibfield  {author} {\bibinfo {author} {\bibfnamefont {L.~M.~A.}\
  \bibnamefont {Bettencourt}}, \bibinfo {author} {\bibfnamefont
  {J.}~\bibnamefont {Lobo}}, \bibinfo {author} {\bibfnamefont {D.}~\bibnamefont
  {Strumsky}}, \ and\ \bibinfo {author} {\bibfnamefont {G.~B.}\ \bibnamefont
  {West}},\ }\href@noop {} {\bibfield  {journal} {\bibinfo  {journal} {PloS
  one}\ }\textbf {\bibinfo {volume} {5}},\ \bibinfo {pages} {e13541} (\bibinfo
  {year} {2010})}\BibitemShut {NoStop}%
\bibitem [{\citenamefont {Gomez-Lievano}\ \emph {et~al.}(2016)\citenamefont
  {Gomez-Lievano}, \citenamefont {Patterson-Lomba},\ and\ \citenamefont
  {Hausmann}}]{gomez-lievano_explaining_2016}%
  \BibitemOpen
  \bibfield  {author} {\bibinfo {author} {\bibfnamefont {A.}~\bibnamefont
  {Gomez-Lievano}}, \bibinfo {author} {\bibfnamefont {O.}~\bibnamefont
  {Patterson-Lomba}}, \ and\ \bibinfo {author} {\bibfnamefont {R.}~\bibnamefont
  {Hausmann}},\ }\href {http://www.nature.com/articles/s41562-016-0012}
  {\bibfield  {journal} {\bibinfo  {journal} {Nature Human Behaviour}\ }\textbf
  {\bibinfo {volume} {1}},\ \bibinfo {pages} {0012} (\bibinfo {year}
  {2016})}\BibitemShut {NoStop}%
\bibitem [{\citenamefont {Schiller}(2016)}]{schiller2016urban}%
  \BibitemOpen
  \bibfield  {author} {\bibinfo {author} {\bibfnamefont {F.}~\bibnamefont
  {Schiller}},\ }\href@noop {} {\bibfield  {journal} {\bibinfo  {journal}
  {Journal of Cleaner Production}\ }\textbf {\bibinfo {volume} {112}},\
  \bibinfo {pages} {4273} (\bibinfo {year} {2016})}\BibitemShut {NoStop}%
\bibitem [{\citenamefont {Zipf}(1949)}]{zipf1949human}%
  \BibitemOpen
  \bibfield  {author} {\bibinfo {author} {\bibfnamefont {G.~K.}\ \bibnamefont
  {Zipf}},\ }\href@noop {} {\emph {\bibinfo {title} {Human Behavior and the
  Principle of Least Effort}}}\ (\bibinfo  {publisher} {Addison-Wesley},\
  \bibinfo {year} {1949})\BibitemShut {NoStop}%
\bibitem [{\citenamefont {Fujita}\ \emph {et~al.}(1999)\citenamefont {Fujita},
  \citenamefont {Krugman},\ and\ \citenamefont {Mori}}]{fujita_evolution_1999}%
  \BibitemOpen
  \bibfield  {author} {\bibinfo {author} {\bibfnamefont {M.}~\bibnamefont
  {Fujita}}, \bibinfo {author} {\bibfnamefont {P.}~\bibnamefont {Krugman}}, \
  and\ \bibinfo {author} {\bibfnamefont {T.}~\bibnamefont {Mori}},\ }\href
  {\doibase 10.1016/S0014-2921(98)00066-X} {\bibfield  {journal} {\bibinfo
  {journal} {European Economic Review}\ }\textbf {\bibinfo {volume} {43}},\
  \bibinfo {pages} {209} (\bibinfo {year} {1999})}\BibitemShut {NoStop}%
\bibitem [{\citenamefont {Mori}\ \emph {et~al.}(2008)\citenamefont {Mori},
  \citenamefont {Nishikimi},\ and\ \citenamefont {Smith}}]{mori2008number}%
  \BibitemOpen
  \bibfield  {author} {\bibinfo {author} {\bibfnamefont {T.}~\bibnamefont
  {Mori}}, \bibinfo {author} {\bibfnamefont {K.}~\bibnamefont {Nishikimi}}, \
  and\ \bibinfo {author} {\bibfnamefont {T.~E.}\ \bibnamefont {Smith}},\
  }\href@noop {} {\bibfield  {journal} {\bibinfo  {journal} {Journal of
  Regional Science}\ }\textbf {\bibinfo {volume} {48}},\ \bibinfo {pages} {165}
  (\bibinfo {year} {2008})}\BibitemShut {NoStop}%
\bibitem [{\citenamefont {Bettencourt}(2013)}]{bettencourt_origins_2013}%
  \BibitemOpen
  \bibfield  {author} {\bibinfo {author} {\bibfnamefont {L.~M.~A.}\
  \bibnamefont {Bettencourt}},\ }\href {\doibase 10.1126/science.1235823}
  {\bibfield  {journal} {\bibinfo  {journal} {Science}\ }\textbf {\bibinfo
  {volume} {340}},\ \bibinfo {pages} {1438} (\bibinfo {year}
  {2013})}\BibitemShut {NoStop}%
\bibitem [{\citenamefont {Youn}\ \emph {et~al.}(2016)\citenamefont {Youn},
  \citenamefont {Bettencourt}, \citenamefont {Lobo}, \citenamefont {Strumsky},
  \citenamefont {Samaniego},\ and\ \citenamefont {West}}]{youn2016scaling}%
  \BibitemOpen
  \bibfield  {author} {\bibinfo {author} {\bibfnamefont {H.}~\bibnamefont
  {Youn}}, \bibinfo {author} {\bibfnamefont {L.~M.~A.}\ \bibnamefont
  {Bettencourt}}, \bibinfo {author} {\bibfnamefont {J.}~\bibnamefont {Lobo}},
  \bibinfo {author} {\bibfnamefont {D.}~\bibnamefont {Strumsky}}, \bibinfo
  {author} {\bibfnamefont {H.}~\bibnamefont {Samaniego}}, \ and\ \bibinfo
  {author} {\bibfnamefont {G.~B.}\ \bibnamefont {West}},\ }\href@noop {}
  {\bibfield  {journal} {\bibinfo  {journal} {Journal of The Royal Society
  Interface}\ }\textbf {\bibinfo {volume} {13}},\ \bibinfo {pages} {20150937}
  (\bibinfo {year} {2016})}\BibitemShut {NoStop}%
\bibitem [{\citenamefont {Henderson}(1986)}]{henderson1986efficiency}%
  \BibitemOpen
  \bibfield  {author} {\bibinfo {author} {\bibfnamefont {J.~V.}\ \bibnamefont
  {Henderson}},\ }\href@noop {} {\bibfield  {journal} {\bibinfo  {journal}
  {Journal of Urban economics}\ }\textbf {\bibinfo {volume} {19}},\ \bibinfo
  {pages} {47} (\bibinfo {year} {1986})}\BibitemShut {NoStop}%
\bibitem [{\citenamefont {Florida}(2004)}]{florida2004rise}%
  \BibitemOpen
  \bibfield  {author} {\bibinfo {author} {\bibfnamefont {R.}~\bibnamefont
  {Florida}},\ }\href@noop {} {\emph {\bibinfo {title} {The rise of the
  creative class}}}\ (\bibinfo  {publisher} {Basic books New York},\ \bibinfo
  {year} {2004})\BibitemShut {NoStop}%
\bibitem [{\citenamefont {Michaels}\ \emph {et~al.}(2013)\citenamefont
  {Michaels}, \citenamefont {Rauch},\ and\ \citenamefont
  {Redding}}]{michaels2013task}%
  \BibitemOpen
  \bibfield  {author} {\bibinfo {author} {\bibfnamefont {G.}~\bibnamefont
  {Michaels}}, \bibinfo {author} {\bibfnamefont {F.}~\bibnamefont {Rauch}}, \
  and\ \bibinfo {author} {\bibfnamefont {S.~J.}\ \bibnamefont {Redding}},\
  }\href@noop {} {\emph {\bibinfo {title} {Task specialization in US cities
  from 1880-2000}}},\ \bibinfo {type} {Tech. Rep.}\ (\bibinfo  {institution}
  {National Bureau of Economic Research},\ \bibinfo {year} {2013})\BibitemShut
  {NoStop}%
\bibitem [{\citenamefont {Frank}\ \emph {et~al.}(2018)\citenamefont {Frank},
  \citenamefont {Sun}, \citenamefont {Cebrian}, \citenamefont {Youn},\ and\
  \citenamefont {Rahwan}}]{frank2018small}%
  \BibitemOpen
  \bibfield  {author} {\bibinfo {author} {\bibfnamefont {M.~R.}\ \bibnamefont
  {Frank}}, \bibinfo {author} {\bibfnamefont {L.}~\bibnamefont {Sun}}, \bibinfo
  {author} {\bibfnamefont {M.}~\bibnamefont {Cebrian}}, \bibinfo {author}
  {\bibfnamefont {H.}~\bibnamefont {Youn}}, \ and\ \bibinfo {author}
  {\bibfnamefont {I.}~\bibnamefont {Rahwan}},\ }\href {\doibase
  10.1098/rsif.2017.0946} {\bibfield  {journal} {\bibinfo  {journal} {Journal
  of The Royal Society Interface}\ }\textbf {\bibinfo {volume} {15}} (\bibinfo
  {year} {2018}),\ 10.1098/rsif.2017.0946}\BibitemShut {NoStop}%
\bibitem [{\citenamefont {Christaller}(1933)}]{christaller1933zentralen}%
  \BibitemOpen
  \bibfield  {author} {\bibinfo {author} {\bibfnamefont {W.}~\bibnamefont
  {Christaller}},\ }\href@noop {} {\enquote {\bibinfo {title} {Die zentralen
  orte in s{\"u}ddeutschland, jena: Gustaf fisher. translated by carlisle w.
  baskin (1966), as central places in southern germany},}\ } (\bibinfo {year}
  {1933})\BibitemShut {NoStop}%
\bibitem [{\citenamefont {Lee}\ \emph {et~al.}(2017)\citenamefont {Lee},
  \citenamefont {Barbosa}, \citenamefont {Youn}, \citenamefont {Holme},\ and\
  \citenamefont {Ghoshal}}]{Lee2017}%
  \BibitemOpen
  \bibfield  {author} {\bibinfo {author} {\bibfnamefont {M.}~\bibnamefont
  {Lee}}, \bibinfo {author} {\bibfnamefont {H.}~\bibnamefont {Barbosa}},
  \bibinfo {author} {\bibfnamefont {H.}~\bibnamefont {Youn}}, \bibinfo {author}
  {\bibfnamefont {P.}~\bibnamefont {Holme}}, \ and\ \bibinfo {author}
  {\bibfnamefont {G.}~\bibnamefont {Ghoshal}},\ }\href {\doibase
  10.1038/s41467-017-02374-7} {\bibfield  {journal} {\bibinfo  {journal}
  {Nature Communications}\ }\textbf {\bibinfo {volume} {8}} (\bibinfo {year}
  {2017}),\ 10.1038/s41467-017-02374-7}\BibitemShut {NoStop}%
\bibitem [{\citenamefont {Hidalgo}\ \emph {et~al.}(2007)\citenamefont
  {Hidalgo}, \citenamefont {Klinger}, \citenamefont {Barabási},\ and\
  \citenamefont {Hausmann}}]{hidalgo_product_2007}%
  \BibitemOpen
  \bibfield  {author} {\bibinfo {author} {\bibfnamefont {C.~A.}\ \bibnamefont
  {Hidalgo}}, \bibinfo {author} {\bibfnamefont {B.}~\bibnamefont {Klinger}},
  \bibinfo {author} {\bibfnamefont {A.-L.}\ \bibnamefont {Barabási}}, \ and\
  \bibinfo {author} {\bibfnamefont {R.}~\bibnamefont {Hausmann}},\ }\href
  {\doibase 10.1126/science.1144581} {\bibfield  {journal} {\bibinfo  {journal}
  {Science}\ }\textbf {\bibinfo {volume} {317}},\ \bibinfo {pages} {482}
  (\bibinfo {year} {2007})}\BibitemShut {NoStop}%
\bibitem [{\citenamefont {Hidalgo}\ and\ \citenamefont
  {Hausmann}(2009)}]{hidalgo_building_2009}%
  \BibitemOpen
  \bibfield  {author} {\bibinfo {author} {\bibfnamefont {C.~A.}\ \bibnamefont
  {Hidalgo}}\ and\ \bibinfo {author} {\bibfnamefont {R.}~\bibnamefont
  {Hausmann}},\ }\href {\doibase 10.1073/pnas.0900943106} {\bibfield  {journal}
  {\bibinfo  {journal} {PNAS}\ }\textbf {\bibinfo {volume} {106}},\ \bibinfo
  {pages} {10570} (\bibinfo {year} {2009})}\BibitemShut {NoStop}%
\bibitem [{\citenamefont {Thurner}\ \emph {et~al.}(2010)\citenamefont
  {Thurner}, \citenamefont {Klimek},\ and\ \citenamefont
  {Hanel}}]{thurner_schumpeterian_2010}%
  \BibitemOpen
  \bibfield  {author} {\bibinfo {author} {\bibfnamefont {S.}~\bibnamefont
  {Thurner}}, \bibinfo {author} {\bibfnamefont {P.}~\bibnamefont {Klimek}}, \
  and\ \bibinfo {author} {\bibfnamefont {R.}~\bibnamefont {Hanel}},\ }\href
  {\doibase 10.1088/1367-2630/12/7/075029} {\bibfield  {journal} {\bibinfo
  {journal} {New Journal of Physics}\ }\textbf {\bibinfo {volume} {12}},\
  \bibinfo {pages} {075029} (\bibinfo {year} {2010})}\BibitemShut {NoStop}%
\bibitem [{\citenamefont {Klimek}\ \emph {et~al.}(2012)\citenamefont {Klimek},
  \citenamefont {Hausmann},\ and\ \citenamefont
  {Thurner}}]{klimek_empirical_2012}%
  \BibitemOpen
  \bibfield  {author} {\bibinfo {author} {\bibfnamefont {P.}~\bibnamefont
  {Klimek}}, \bibinfo {author} {\bibfnamefont {R.}~\bibnamefont {Hausmann}}, \
  and\ \bibinfo {author} {\bibfnamefont {S.}~\bibnamefont {Thurner}},\ }\href
  {http://dx.plos.org/10.1371/journal.pone.0038924} {\bibfield  {journal}
  {\bibinfo  {journal} {PloS one}\ }\textbf {\bibinfo {volume} {7}},\ \bibinfo
  {pages} {e38924} (\bibinfo {year} {2012})}\BibitemShut {NoStop}%
\bibitem [{\citenamefont {Shutters}\ \emph {et~al.}(2016)\citenamefont
  {Shutters}, \citenamefont {Muneepeerakul},\ and\ \citenamefont
  {Lobo}}]{shutters2016constrained}%
  \BibitemOpen
  \bibfield  {author} {\bibinfo {author} {\bibfnamefont {S.~T.}\ \bibnamefont
  {Shutters}}, \bibinfo {author} {\bibfnamefont {R.}~\bibnamefont
  {Muneepeerakul}}, \ and\ \bibinfo {author} {\bibfnamefont {J.}~\bibnamefont
  {Lobo}},\ }\href@noop {} {\bibfield  {journal} {\bibinfo  {journal} {Urban
  Studies}\ }\textbf {\bibinfo {volume} {53}},\ \bibinfo {pages} {3439}
  (\bibinfo {year} {2016})}\BibitemShut {NoStop}%
\bibitem [{\citenamefont {Sachs}\ and\ \citenamefont
  {Warner}(1995)}]{sachs1995economic}%
  \BibitemOpen
  \bibfield  {author} {\bibinfo {author} {\bibfnamefont {J.~D.}\ \bibnamefont
  {Sachs}}\ and\ \bibinfo {author} {\bibfnamefont {A.~M.}\ \bibnamefont
  {Warner}},\ }\href@noop {} {\emph {\bibinfo {title} {Economic convergence and
  economic policies}}},\ \bibinfo {type} {Tech. Rep.}\ (\bibinfo  {institution}
  {National Bureau of Economic Research},\ \bibinfo {year} {1995})\BibitemShut
  {NoStop}%
\bibitem [{\citenamefont {Isserman}(1977)}]{isserman1977location}%
  \BibitemOpen
  \bibfield  {author} {\bibinfo {author} {\bibfnamefont {A.~M.}\ \bibnamefont
  {Isserman}},\ }\href@noop {} {\bibfield  {journal} {\bibinfo  {journal}
  {Journal of the American Institute of Planners}\ }\textbf {\bibinfo {volume}
  {43}},\ \bibinfo {pages} {33} (\bibinfo {year} {1977})}\BibitemShut {NoStop}%
\bibitem [{\citenamefont {Balassa}(1965)}]{balassa1965trade}%
  \BibitemOpen
  \bibfield  {author} {\bibinfo {author} {\bibfnamefont {B.}~\bibnamefont
  {Balassa}},\ }\href@noop {} {\bibfield  {journal} {\bibinfo  {journal} {The
  Manchester School}\ }\textbf {\bibinfo {volume} {33}},\ \bibinfo {pages} {99}
  (\bibinfo {year} {1965})}\BibitemShut {NoStop}%
\bibitem [{\citenamefont {Bettencourt}\ and\ \citenamefont
  {Lobo}(2016)}]{bettencourt2016urban}%
  \BibitemOpen
  \bibfield  {author} {\bibinfo {author} {\bibfnamefont {L.~M.~A.}\
  \bibnamefont {Bettencourt}}\ and\ \bibinfo {author} {\bibfnamefont
  {J.}~\bibnamefont {Lobo}},\ }\href@noop {} {\bibfield  {journal} {\bibinfo
  {journal} {Journal of The Royal Society Interface}\ }\textbf {\bibinfo
  {volume} {13}},\ \bibinfo {pages} {20160005} (\bibinfo {year}
  {2016})}\BibitemShut {NoStop}%
\bibitem [{\citenamefont {Alves}\ \emph {et~al.}(2015)\citenamefont {Alves},
  \citenamefont {Mendes}, \citenamefont {Lenzi},\ and\ \citenamefont
  {Ribeiro}}]{alves2015scale}%
  \BibitemOpen
  \bibfield  {author} {\bibinfo {author} {\bibfnamefont {L.~G.}\ \bibnamefont
  {Alves}}, \bibinfo {author} {\bibfnamefont {R.~S.}\ \bibnamefont {Mendes}},
  \bibinfo {author} {\bibfnamefont {E.~K.}\ \bibnamefont {Lenzi}}, \ and\
  \bibinfo {author} {\bibfnamefont {H.~V.}\ \bibnamefont {Ribeiro}},\
  }\href@noop {} {\bibfield  {journal} {\bibinfo  {journal} {PloS one}\
  }\textbf {\bibinfo {volume} {10}},\ \bibinfo {pages} {e0134862} (\bibinfo
  {year} {2015})}\BibitemShut {NoStop}%
\bibitem [{\citenamefont {Arcaute}\ \emph {et~al.}(2015)\citenamefont
  {Arcaute}, \citenamefont {Hatna}, \citenamefont {Ferguson}, \citenamefont
  {Youn}, \citenamefont {Johansson},\ and\ \citenamefont
  {Batty}}]{arcaute_constructing_2015}%
  \BibitemOpen
  \bibfield  {author} {\bibinfo {author} {\bibfnamefont {E.}~\bibnamefont
  {Arcaute}}, \bibinfo {author} {\bibfnamefont {E.}~\bibnamefont {Hatna}},
  \bibinfo {author} {\bibfnamefont {P.}~\bibnamefont {Ferguson}}, \bibinfo
  {author} {\bibfnamefont {H.}~\bibnamefont {Youn}}, \bibinfo {author}
  {\bibfnamefont {A.}~\bibnamefont {Johansson}}, \ and\ \bibinfo {author}
  {\bibfnamefont {M.}~\bibnamefont {Batty}},\ }\href {\doibase
  10.1098/rsif.2014.0745} {\bibfield  {journal} {\bibinfo  {journal} {Journal
  of The Royal Society Interface}\ }\textbf {\bibinfo {volume} {12}},\ \bibinfo
  {pages} {20140745} (\bibinfo {year} {2015})}\BibitemShut {NoStop}%
\bibitem [{\citenamefont {Leitao}\ \emph {et~al.}(2016)\citenamefont {Leitao},
  \citenamefont {Miotto}, \citenamefont {Gerlach},\ and\ \citenamefont
  {Altmann}}]{leitao2016scaling}%
  \BibitemOpen
  \bibfield  {author} {\bibinfo {author} {\bibfnamefont {J.~C.}\ \bibnamefont
  {Leitao}}, \bibinfo {author} {\bibfnamefont {J.~M.}\ \bibnamefont {Miotto}},
  \bibinfo {author} {\bibfnamefont {M.}~\bibnamefont {Gerlach}}, \ and\
  \bibinfo {author} {\bibfnamefont {E.~G.}\ \bibnamefont {Altmann}},\
  }\href@noop {} {\bibfield  {journal} {\bibinfo  {journal} {Royal Society open
  science}\ }\textbf {\bibinfo {volume} {3}},\ \bibinfo {pages} {150649}
  (\bibinfo {year} {2016})}\BibitemShut {NoStop}%
\bibitem [{\citenamefont {Neffke}\ and\ \citenamefont
  {Svensson~Henning}(2008)}]{neffke_revealed_2008}%
  \BibitemOpen
  \bibfield  {author} {\bibinfo {author} {\bibfnamefont {F.}~\bibnamefont
  {Neffke}}\ and\ \bibinfo {author} {\bibfnamefont {M.}~\bibnamefont
  {Svensson~Henning}},\ }\href
  {http://www2.druid.dk/conferences/viewpaper.php?id=3691&cf=29} {\bibfield
  {journal} {\bibinfo  {journal} {Papers in Evolutionary Economic Geography}\
  }\textbf {\bibinfo {volume} {8}},\ \bibinfo {pages} {19} (\bibinfo {year}
  {2008})}\BibitemShut {NoStop}%
\bibitem [{\citenamefont {Muneepeerakul}\ \emph {et~al.}(2013)\citenamefont
  {Muneepeerakul}, \citenamefont {Lobo}, \citenamefont {Shutters},
  \citenamefont {Goméz-Liévano},\ and\ \citenamefont
  {Qubbaj}}]{muneepeerakul_urban_2013}%
  \BibitemOpen
  \bibfield  {author} {\bibinfo {author} {\bibfnamefont {R.}~\bibnamefont
  {Muneepeerakul}}, \bibinfo {author} {\bibfnamefont {J.}~\bibnamefont {Lobo}},
  \bibinfo {author} {\bibfnamefont {S.~T.}\ \bibnamefont {Shutters}}, \bibinfo
  {author} {\bibfnamefont {A.}~\bibnamefont {Goméz-Liévano}}, \ and\ \bibinfo
  {author} {\bibfnamefont {M.~R.}\ \bibnamefont {Qubbaj}},\ }\href {\doibase
  10.1371/journal.pone.0073676} {\bibfield  {journal} {\bibinfo  {journal}
  {PLoS ONE}\ }\textbf {\bibinfo {volume} {8}},\ \bibinfo {pages} {e73676}
  (\bibinfo {year} {2013})}\BibitemShut {NoStop}%
\bibitem [{\citenamefont {Rosen}\ and\ \citenamefont
  {Resnick}(1980)}]{rosen1980size}%
  \BibitemOpen
  \bibfield  {author} {\bibinfo {author} {\bibfnamefont {K.~T.}\ \bibnamefont
  {Rosen}}\ and\ \bibinfo {author} {\bibfnamefont {M.}~\bibnamefont
  {Resnick}},\ }\href@noop {} {\bibfield  {journal} {\bibinfo  {journal}
  {Journal of Urban Economics}\ }\textbf {\bibinfo {volume} {8}},\ \bibinfo
  {pages} {165} (\bibinfo {year} {1980})}\BibitemShut {NoStop}%
\bibitem [{\citenamefont {Depersin}\ and\ \citenamefont
  {Barthelemy}(2018)}]{depersin2018global}%
  \BibitemOpen
  \bibfield  {author} {\bibinfo {author} {\bibfnamefont {J.}~\bibnamefont
  {Depersin}}\ and\ \bibinfo {author} {\bibfnamefont {M.}~\bibnamefont
  {Barthelemy}},\ }\href@noop {} {\bibfield  {journal} {\bibinfo  {journal}
  {Proceedings of the National Academy of Sciences}\ }\textbf {\bibinfo
  {volume} {115}},\ \bibinfo {pages} {2317} (\bibinfo {year}
  {2018})}\BibitemShut {NoStop}%
\bibitem [{\citenamefont {Pumain}\ \emph {et~al.}(2006)\citenamefont {Pumain},
  \citenamefont {Paulus}, \citenamefont {Vacchiani-Marcuzzo},\ and\
  \citenamefont {Lobo}}]{pumain_cybergeo_2006}%
  \BibitemOpen
  \bibfield  {author} {\bibinfo {author} {\bibfnamefont {D.}~\bibnamefont
  {Pumain}}, \bibinfo {author} {\bibfnamefont {F.}~\bibnamefont {Paulus}},
  \bibinfo {author} {\bibfnamefont {C.}~\bibnamefont {Vacchiani-Marcuzzo}}, \
  and\ \bibinfo {author} {\bibfnamefont {J.}~\bibnamefont {Lobo}},\ }\href
  {\doibase 10.4000/cybergeo.2519} {\bibfield  {journal} {\bibinfo  {journal}
  {Cybergeo : European Journal of Geography}\ } (\bibinfo {year} {2006}),\
  10.4000/cybergeo.2519}\BibitemShut {NoStop}%
\bibitem [{\citenamefont {Porter}(2003)}]{porter2003economic}%
  \BibitemOpen
  \bibfield  {author} {\bibinfo {author} {\bibfnamefont {M.}~\bibnamefont
  {Porter}},\ }\href@noop {} {\bibfield  {journal} {\bibinfo  {journal}
  {Regional studies}\ }\textbf {\bibinfo {volume} {37}},\ \bibinfo {pages}
  {549} (\bibinfo {year} {2003})}\BibitemShut {NoStop}%
\bibitem [{\citenamefont {Lobo}\ \emph {et~al.}(2013)\citenamefont {Lobo},
  \citenamefont {Bettencourt}, \citenamefont {Strumsky},\ and\ \citenamefont
  {West}}]{lobo2013urban}%
  \BibitemOpen
  \bibfield  {author} {\bibinfo {author} {\bibfnamefont {J.}~\bibnamefont
  {Lobo}}, \bibinfo {author} {\bibfnamefont {L.~M.~A.}\ \bibnamefont
  {Bettencourt}}, \bibinfo {author} {\bibfnamefont {D.}~\bibnamefont
  {Strumsky}}, \ and\ \bibinfo {author} {\bibfnamefont {G.~B.}\ \bibnamefont
  {West}},\ }\href@noop {} {\bibfield  {journal} {\bibinfo  {journal} {PLoS
  One}\ }\textbf {\bibinfo {volume} {8}},\ \bibinfo {pages} {e58407} (\bibinfo
  {year} {2013})}\BibitemShut {NoStop}%
\bibitem [{\citenamefont {Ehrlich}\ and\ \citenamefont
  {Holm}(1963)}]{ehrlich1963process}%
  \BibitemOpen
  \bibfield  {author} {\bibinfo {author} {\bibfnamefont {P.~R.}\ \bibnamefont
  {Ehrlich}}\ and\ \bibinfo {author} {\bibfnamefont {R.~W.}\ \bibnamefont
  {Holm}},\ }\href@noop {} {\emph {\bibinfo {title} {The process of
  evolution}}},\ \bibinfo {type} {Tech. Rep.}\ (\bibinfo {year}
  {1963})\BibitemShut {NoStop}%
\bibitem [{\citenamefont {West}\ \emph {et~al.}(1997)\citenamefont {West},
  \citenamefont {Brown},\ and\ \citenamefont {Enquist}}]{west_general_1997}%
  \BibitemOpen
  \bibfield  {author} {\bibinfo {author} {\bibfnamefont {G.~B.}\ \bibnamefont
  {West}}, \bibinfo {author} {\bibfnamefont {J.~H.}\ \bibnamefont {Brown}}, \
  and\ \bibinfo {author} {\bibfnamefont {B.~J.}\ \bibnamefont {Enquist}},\
  }\href {\doibase 10.1126/science.276.5309.122} {\bibfield  {journal}
  {\bibinfo  {journal} {Science}\ }\textbf {\bibinfo {volume} {276}},\ \bibinfo
  {pages} {122} (\bibinfo {year} {1997})}\BibitemShut {NoStop}%
\bibitem [{\citenamefont {Kline}\ and\ \citenamefont
  {Boyd}(2010)}]{kline2010population}%
  \BibitemOpen
  \bibfield  {author} {\bibinfo {author} {\bibfnamefont {M.~A.}\ \bibnamefont
  {Kline}}\ and\ \bibinfo {author} {\bibfnamefont {R.}~\bibnamefont {Boyd}},\
  }\href@noop {} {\bibfield  {journal} {\bibinfo  {journal} {Proceedings of the
  Royal Society of London B: Biological Sciences}\ }\textbf {\bibinfo {volume}
  {277}},\ \bibinfo {pages} {2559} (\bibinfo {year} {2010})}\BibitemShut
  {NoStop}%
\bibitem [{\citenamefont {Muthukrishna}\ and\ \citenamefont
  {Henrich}(2016)}]{muthukrishna2016innovation}%
  \BibitemOpen
  \bibfield  {author} {\bibinfo {author} {\bibfnamefont {M.}~\bibnamefont
  {Muthukrishna}}\ and\ \bibinfo {author} {\bibfnamefont {J.}~\bibnamefont
  {Henrich}},\ }\href@noop {} {\bibfield  {journal} {\bibinfo  {journal} {Phil.
  Trans. R. Soc. B}\ }\textbf {\bibinfo {volume} {371}},\ \bibinfo {pages}
  {20150192} (\bibinfo {year} {2016})}\BibitemShut {NoStop}%
\bibitem [{\citenamefont {Yang}\ \emph {et~al.}(2017)\citenamefont {Yang},
  \citenamefont {Papachristos},\ and\ \citenamefont {Abrams}}]{yang2017origin}%
  \BibitemOpen
  \bibfield  {author} {\bibinfo {author} {\bibfnamefont {V.~C.}\ \bibnamefont
  {Yang}}, \bibinfo {author} {\bibfnamefont {A.~V.}\ \bibnamefont
  {Papachristos}}, \ and\ \bibinfo {author} {\bibfnamefont {D.~M.}\
  \bibnamefont {Abrams}},\ }\href@noop {} {\bibfield  {journal} {\bibinfo
  {journal} {arXiv preprint arXiv:1712.00476}\ } (\bibinfo {year}
  {2017})}\BibitemShut {NoStop}%
\bibitem [{\citenamefont {Vollrath}(1991)}]{vollrath1991theoretical}%
  \BibitemOpen
  \bibfield  {author} {\bibinfo {author} {\bibfnamefont {T.~L.}\ \bibnamefont
  {Vollrath}},\ }\href@noop {} {\bibfield  {journal} {\bibinfo  {journal}
  {Review of World Economics}\ }\textbf {\bibinfo {volume} {127}},\ \bibinfo
  {pages} {265} (\bibinfo {year} {1991})}\BibitemShut {NoStop}%
\bibitem [{\citenamefont {Gao}\ \emph {et~al.}(2017)\citenamefont {Gao},
  \citenamefont {Jun}, \citenamefont {Pentland}, \citenamefont {Zhou},
  \citenamefont {Hidalgo} \emph {et~al.}}]{gao2017collective}%
  \BibitemOpen
  \bibfield  {author} {\bibinfo {author} {\bibfnamefont {J.}~\bibnamefont
  {Gao}}, \bibinfo {author} {\bibfnamefont {B.}~\bibnamefont {Jun}}, \bibinfo
  {author} {\bibfnamefont {A.}~\bibnamefont {Pentland}}, \bibinfo {author}
  {\bibfnamefont {T.}~\bibnamefont {Zhou}}, \bibinfo {author} {\bibfnamefont
  {C.~A.}\ \bibnamefont {Hidalgo}},  \emph {et~al.},\ }\href@noop {} {\bibfield
   {journal} {\bibinfo  {journal} {arXiv preprint arXiv:1703.01369}\ }
  (\bibinfo {year} {2017})}\BibitemShut {NoStop}%
\end{thebibliography}%


\clearpage

\setcounter{tocdepth}{2}
\setcounter{page}{1}
\setcounter{equation}{0}
\setcounter{figure}{0}

\renewcommand{\thesubfigure}{\Alph{subfigure}}
\renewcommand{\theequation}{S\arabic{equation}}
\renewcommand{\thetable}{S\arabic{table}}
\renewcommand{\thefigure}{S\arabic{figure}}

\onecolumngrid
{\centering \large \textbf{Supporting Information: A common trajectory recapitulated by urban economies} \\
\bigskip
\bigskip
}

\twocolumngrid

\section{Datasets}\label{SI-sec:data}

\begin{figure}[!ht]
	\centering
	\labelarial{A}
	\subfloat{\includegraphics[width=0.45\textwidth]{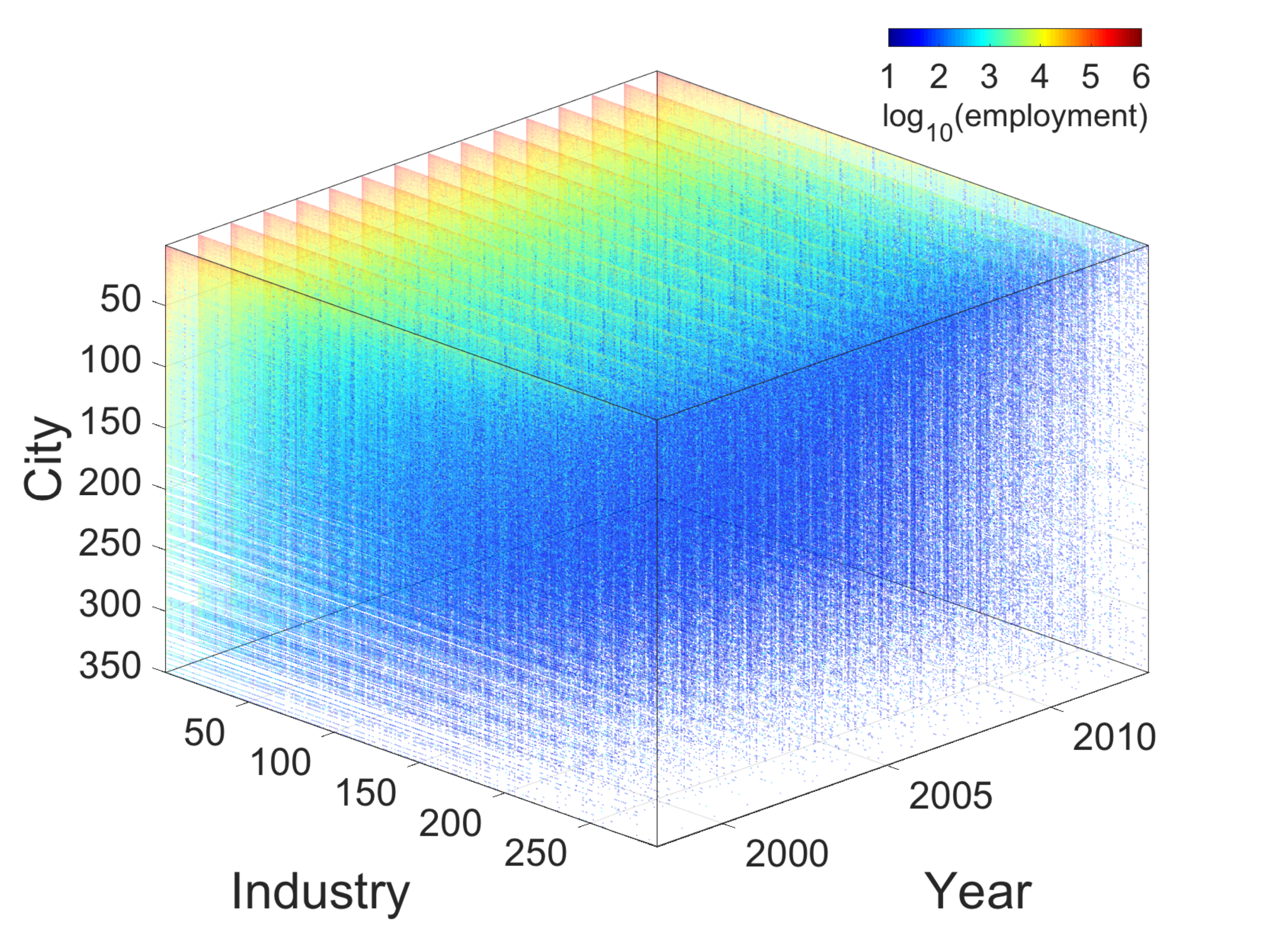}\label{fig:data_3d}}\hspace{0.5cm}
	\\
	\labelarial{B}
	\subfloat{\includegraphics[width=0.45\textwidth]{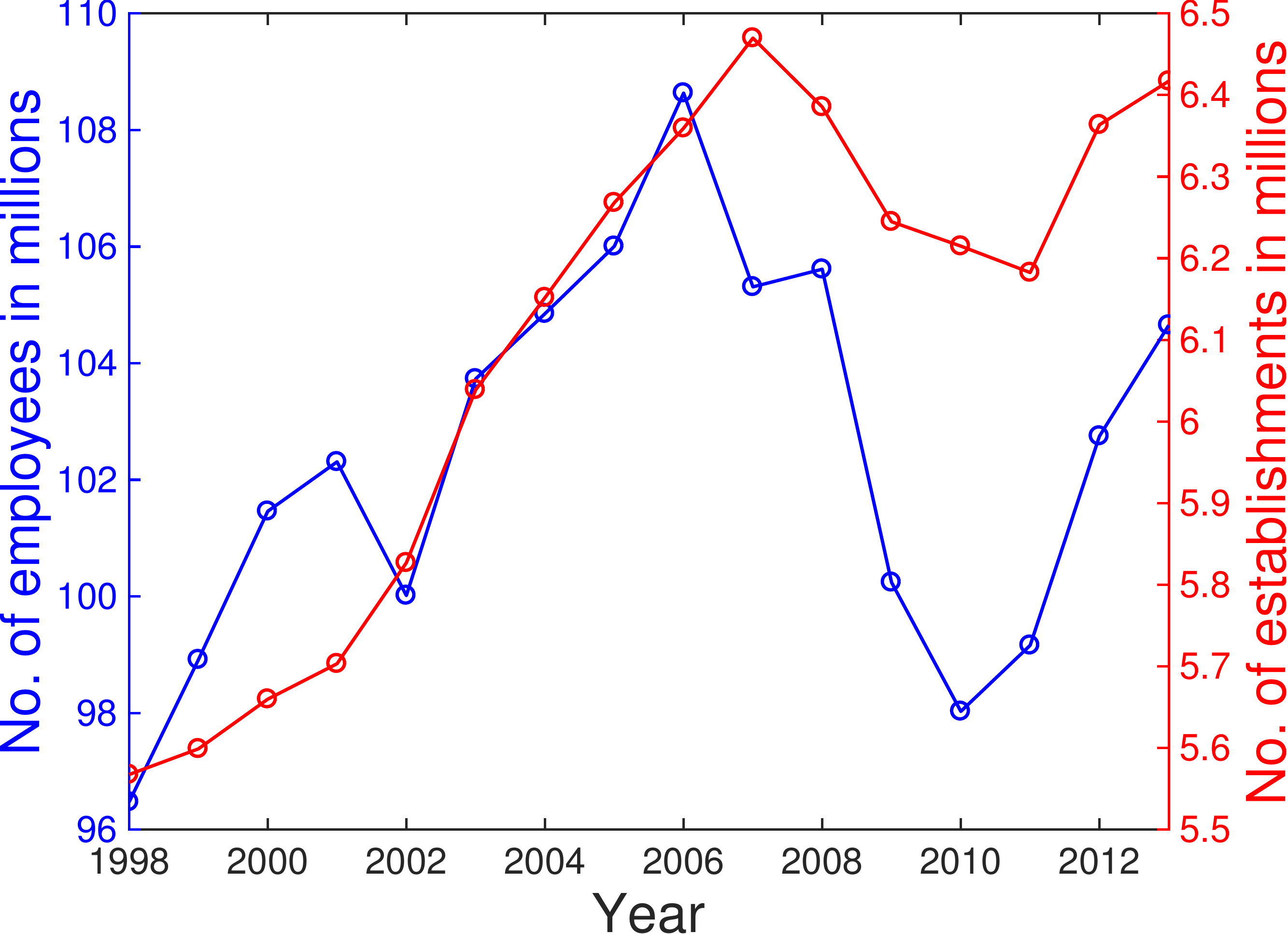}\label{fig:data_timeseries}}
	\caption{
	(A) Visualization of employment data for 4-digits NAICS classifications, Metropolitan Statistical Areas (MSA) and years. The industrial classification is ordered by their weighted sum for the first year (1998) to reform it as a nested matrix. The MSAs are ordered by their populations. 
	(B) Time series of total number of employees (blue) and establishments (red).}
\end{figure}

We study the dynamics and structure of urban industries using 16-years of employment and establishment data from 1998 to 2013 provided by the \href{https://www.census.gov/programs-surveys/cbp.html}{County Business Patterns} (CBP). The number of industry sectors differs for classification levels from 19 to 978 by North American Industry Classification System (NAICS). We use 350 Metropolitan Statistical Areas (MSAs) as a definition of cities with their yearly populations; this data is provided by the \href{https://www.census.gov/programs-surveys/metro-micro/data/tables.html}{U.S. Census Bureau}.
We use $Y(c,i,t)$ to denote the size of industry $i$ in city $c$ at time $t$, and we use $N(c,t)$ to denote the city's population.

An aggregate spatio-temporal view of the complete data set is provided in Figure~\ref{fig:data_3d}. Large cities have a larger number of industries than small cities. Figure~\ref{fig:data_timeseries} depicts the time series of the total number of employees and establishments across all U.S. cities. The temporal changes of employees and establishments show a similar trend which includes a trough near 2010. The number of employees in some industries is denoted as size classes because the data need to avoid disclosure (confidentiality) or they do not meet publication standards. The employee size classes are as follows: A for 0-19, B for 20-99, C for 100-249, E for 250-499, F for 500-999, G for 1,000-2,499, H for 2,500-4,999, I for 5,000-9,999, J for 10,000-24,999, K for 25,000-49,999, L for 50,000-99,999, and M for 100,000 or more. We use the middle value of the range if the data is flagged. In the example of 2013, about 68.5\% of data is flagged due to the small size of employees in specific industry sectors. Although many data points are flagged, they hardly influence the trend because most of the flagged data denote very small industries, thus their contribution is tiny when aggregated into 2- or 3-digit NAICS classifications in the analysis. In the flagged data, about 98.8\% of data are within the range below 1,000 employees (A to F), specifically, 61.8\% for A, 25.5\% for B, 6.9\% for C, 3.0\% for E, and 1.5\% for F. 

The NAICS industry classification system is not static during the period of this study.
Instead, the taxonomy is periodically updated to accommodate new industries.
We accounted summarized these changes into a single taxonomy spanning the entire period of study using NAICS revision data from \href{https://www.census.gov/eos/www/naics}{U.S. Census Bureau}, and MSA revision data from \href{https://www.transportation.org}{American Association of State Highway and Transportation Officials} (AASHTO), \href{https://www.census.gov/programs-surveys/susb.html}{U.S. Census Bureau} and \href{https://en.wikipedia.org/wiki/List_of_Metropolitan_Statistical_Areas}{Wikipedia}.

\subsection{NAICS revisions}
The North American Industry Classification System (NAICS) is used by Federal statistical agencies to classify business establishments for the purpose of collecting, analyzing, and publishing statistical data related to the U.S. business economy. In addition to the temporal evolution of business sectors, this taxonomy itself is periodically revised to account for new industries. These updates are informative of the features that reshape the U.S. economy, including technological change for example. Nevertheless, it is possible that some sectors may remain under the same name in the nomenclature with a completely new content, and we cannot capture innovation in these cases. Instead, we leave them as background technological progress encompassing entire sectors. This ambiguity adds noise to our findings. 

\begin{figure}[t]
	\centering
	\includegraphics[width=0.5\textwidth]{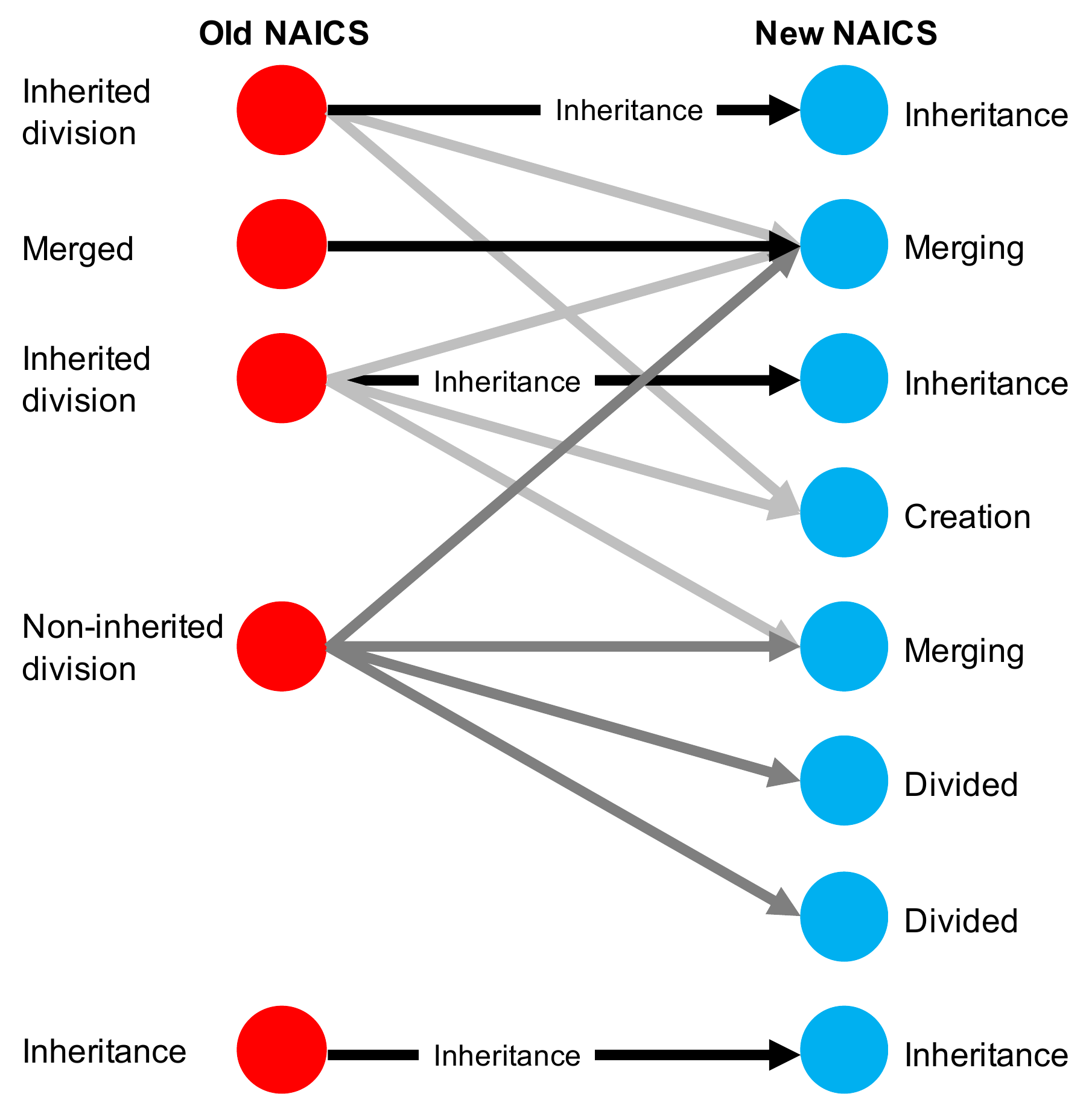}
	\caption{
	Conversion diagram for NAICS classifications. Red and Blue nodes describe old and new NAICS codes each. The arrows mean the connections between old and new codes. The colors of arrows, black, dark grey and light grey, describe whether  the number of employees $Y_{emp}$ of an old code is transferred to a new code, completely, partly and not at all, respectively.}
	\label{fig:NAICS_conversion}
\end{figure}

In our analysis, we use annual employment data from 1998 to 2013. We identify industries based on 6-digit NAICS codes, and the temporal changes of classification in 1997, 2002, 2007 and 2012 are harmonized into the classification in 2012. The intervals sharing the same NAICS classifications are 1998-2002, 2003-2007, 2008-2011, and 2012-2013 given by \href{https://www.census.gov/eos/www/naics}{U.S. Census Bureau}.
This scheme allows us to employ a single industry taxonomy for the entire time period of our study.
There are some cases of inexact matching between two consecutive NAICS revisions. For example, ``Business to Business Electronic Markets'' is a newly created sector by combining some functional parts of 68 wholesaler sectors. There are only 4 cases of creation throughout the entire period, which are ``New Housing For-Sale Builders'', ``Residential Remodelers'', ``Business to Business Electronic Markets'' and ``Wholesale Trade Agents and Brokers'' in 2003.

These temporal changes in industry classification define a bipartite network connecting old NAICS codes to new ones.
For convenience, we refer to old codes as \emph{parents} and we refer to new codes as \emph{children}. All codes are classified into 7 groups: Inheritance (I), Creation (C), Merging (MG), Divided (D), Inherited Division (ID), Non-inherited Division (ND), and Merged (MD). This scheme exclusively and entirely classifies every code change in our dataset. The classification is defined by out-degrees $k_{out}(i)$ of old codes and in-degrees $k_{in}(j)$ of new codes in the connection structure.

Figure~\ref*{fig:NAICS_conversion} shows the connection scheme for classification changes.
Inheritance is the basic connection defined for simple one-to-one conversions of NAICS code. It is also defined for a code pair when the new code has $k_{in}(j) = 1$, and the other new codes connected to the old code have $k_{in}(j) \geq 2$. Briefly, only one new code exclusively inherits its parent as the old code is the only parent of the child. 
Inherited Division and Non-inherited Division are defined for old codes with $k_{out}(i) \geq 2$ (division), and depends on whether the old code has any inherited child or not.
Creation, Merging and Divided are defined for new codes. A new code is determined as Creation when each of its parents has another inherited child. Merging is defined for the non-creation case that its parents has no other inherited child.
A new node is considered as Divided when it has $k_{in}(j) = 1$ and is the child of a Non-inherited division node.
Finally, Merged is defined for an old code that has $k_{out}(i) = 1$ and is the parent of a Merging code. 

We translate all industry classifications to the most recent classification (2012) to make a harmonized time series of number of employees. In the case of inheritance (black arrows in Fig.~\ref{fig:NAICS_conversion}), all employees in the old classification are transferred to the Inheritance node in the new classification. The division of employees colored by light grey is not transferred to the latter nodes, which means these connections are ignored. As a special case, Non-inherited division equally distribute its employment to the divided nodes (dark grey). 

\subsection{MSA revisions}

Similar to the NAICS revisions, the Metropolitan Statistical Areas (MSAs) are revised about every 5 years. 
The exact time intervals are 1993-1999, 2000-2002, 2003-2006, 2007-2011 and 2012-2013. 
By using the conversion table obtained from \href{https://www.transportation.org}{American Association of State Highway and Transportation Officials} (AASHTO), we harmonize all MSAs to a single reference. 
In 1998 and 1999, the MSAs are classified into two types: Consolidated Metropolitan Statistical Area (CMSA) and Primary Metropolitan Statistical Area (PMSA) where a CMSA contains several PMSAs. 
From 2000, MSAs have been revised to Combined Statistical Area (CBSA). 
For example, New York City is classified as `New York-Northern New Jersey-Long Island, NY-NJ-CT-PA' CBSA, and `New York-Northern New Jersey-Long Island, NY-NJ-PA' CMSA with 15 PMSAs including `Bergen-Passaic, NJ', `Bridgeport, CT', `Danbury, CT', `Dutchess County, NY', `Jersey City, NJ', `Middlesex-Somerset-Hunterdon, NJ', `Monmouth-Ocean, NJ', `Nassau-Suffolk, NY', `New Haven-Meriden, CT', `New York, NY', `Newark, NJ', `Newburgh, NY-PA', `Stamford-Norwalk, CT', `Trenton, NJ' and `Waterbury, CT'. 
For 1998 and 1999, we use CMSAs as the delineation of cities since CMSAs has a similar scale to CBSAs. 
When an MSA is divided into several MSAs or merged together with another MSA, we aggregated these MSAs to the largest one. 
Some exceptions are modified by using MSA definition provided by \href{https://www.census.gov/programs-surveys/susb.html}{U.S. Census Bureau}, and some megalopolises, such as New York, Los Angeles and Chicago, are cross-checked with the definitions in \href{https://en.wikipedia.org/wiki/List_of_Metropolitan_Statistical_Areas}{Wikipedia}.

\clearpage

\section{Revealed comparative advantage}\label{SI-sec:rca}

\begin{figure}[!th]
	\labelarial{A}
	\subfloat{\includegraphics[width=0.93\columnwidth]{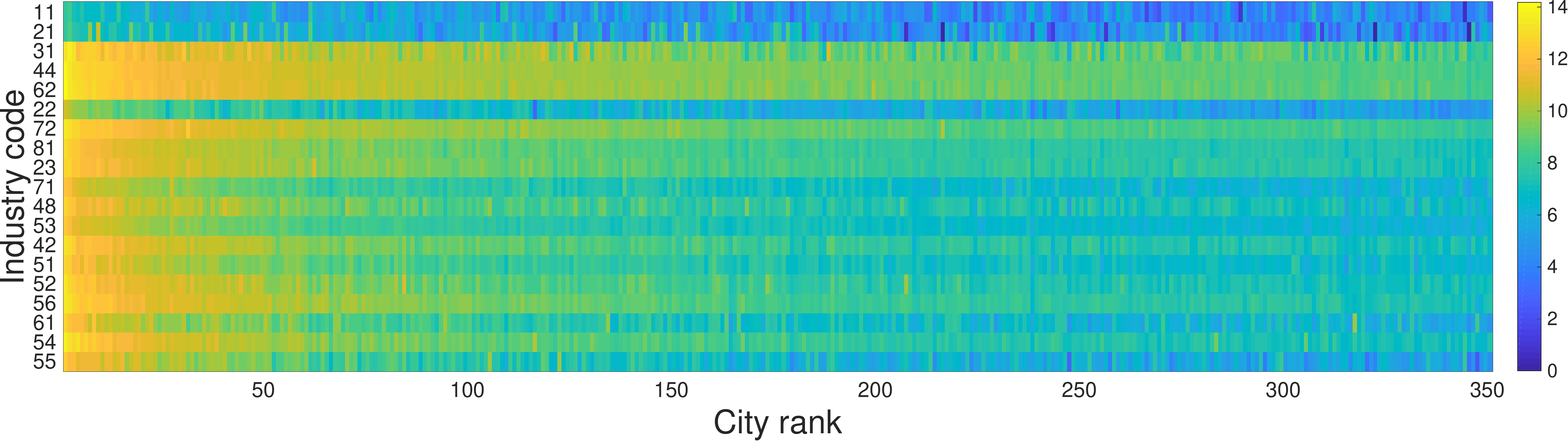}\label{fig:city-ind_size}}
	\\
	\labelarial{B}
	\subfloat{\includegraphics[width=0.93\columnwidth]{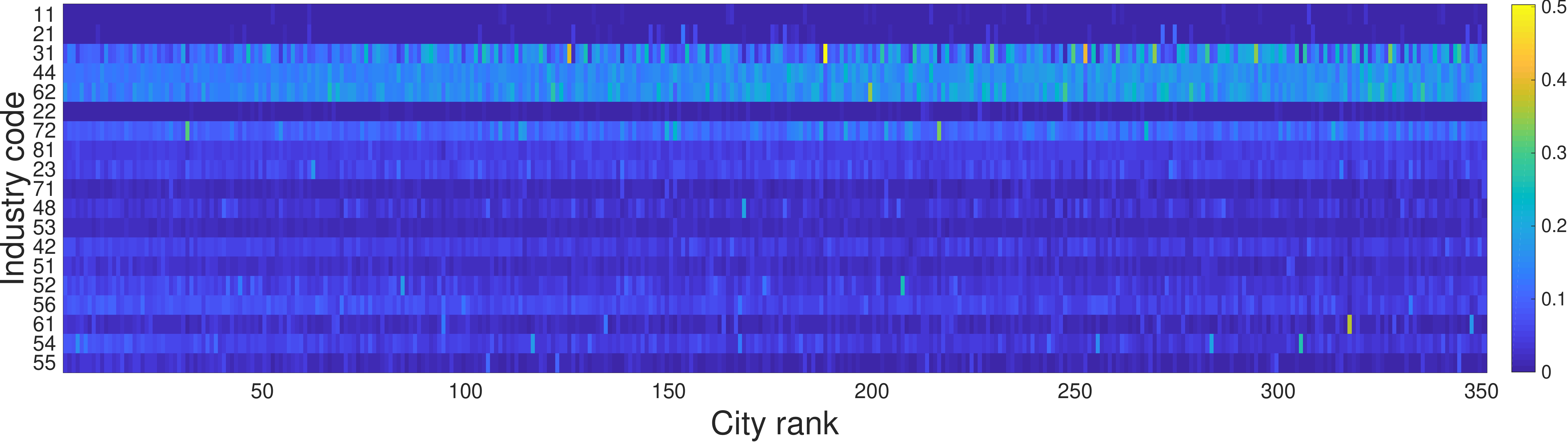}\label{fig:city-ind_share}}
	\\
	\labelarial{C}
	\subfloat{\includegraphics[width=0.93\columnwidth]{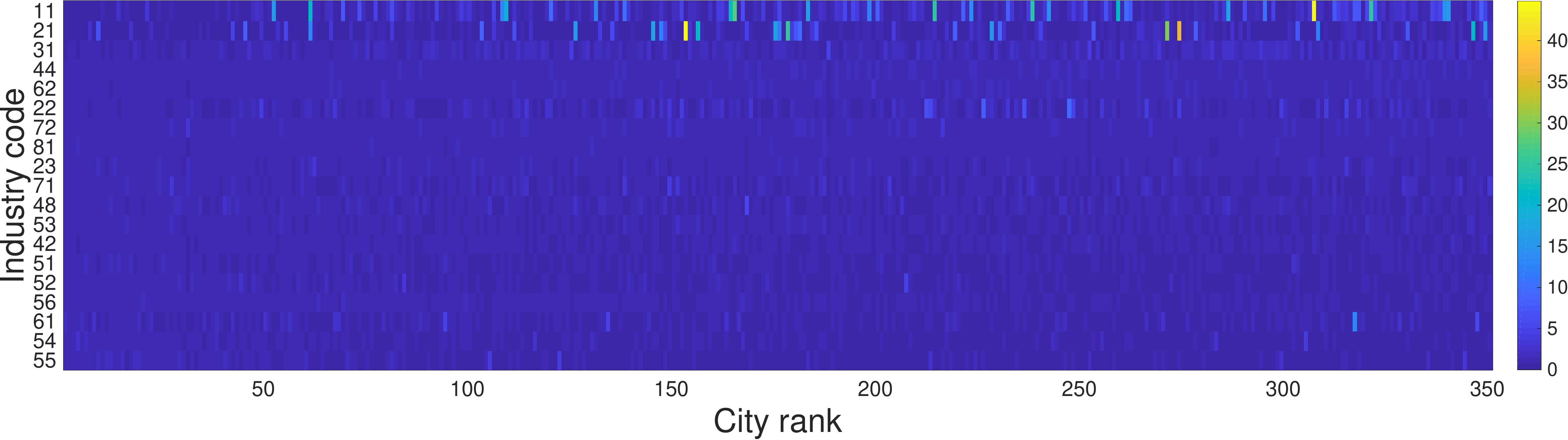}\label{fig:city-ind_rca}}
	\\
	\labelarial{D}
	\subfloat{\includegraphics[width=0.93\columnwidth]{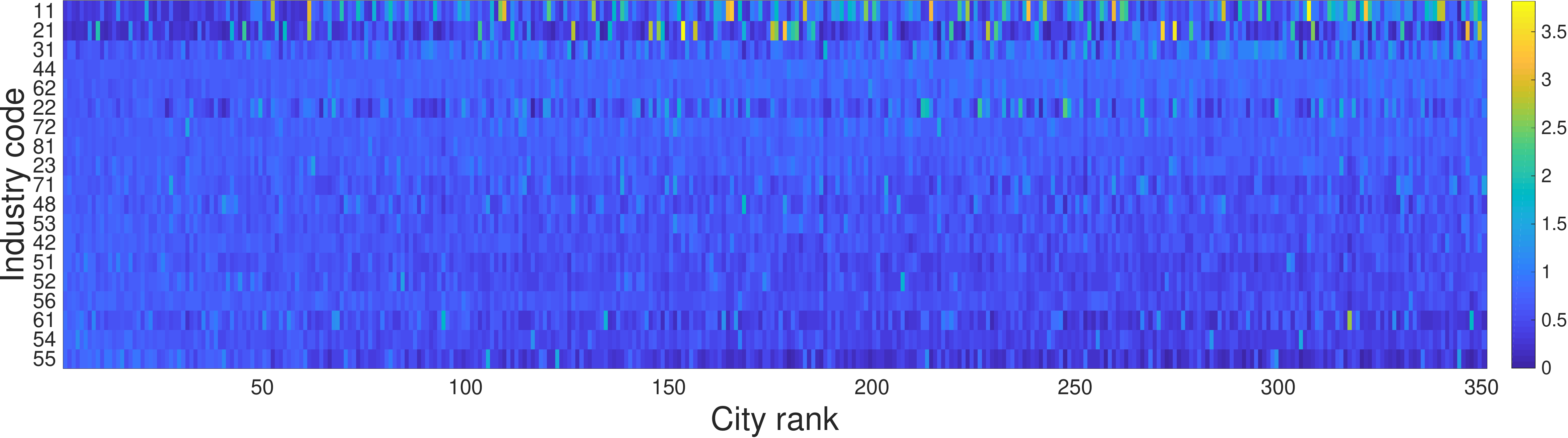}\label{fig:city-ind_lnrca}}
	\\
	\caption{
	Time-averaged size of employment of industry $i$ (y-axis) and city $c$ (x-axis) quantified by various measures from top to bottom: (A) logarithm of employee size $\log(Y(c,i)+1)$, (B) share in each city, (C) RCA and (D) logarithmic RCA ($\log(RCA+1)$). The industry code is aligned by its scaling exponent (high at bottom), and the city is aligned by its population size (low for large city).
	}
\end{figure}

\begin{figure}[!th]
	\centering
	\labelarial{A}
	\subfloat{\includegraphics[width=0.4\textwidth]{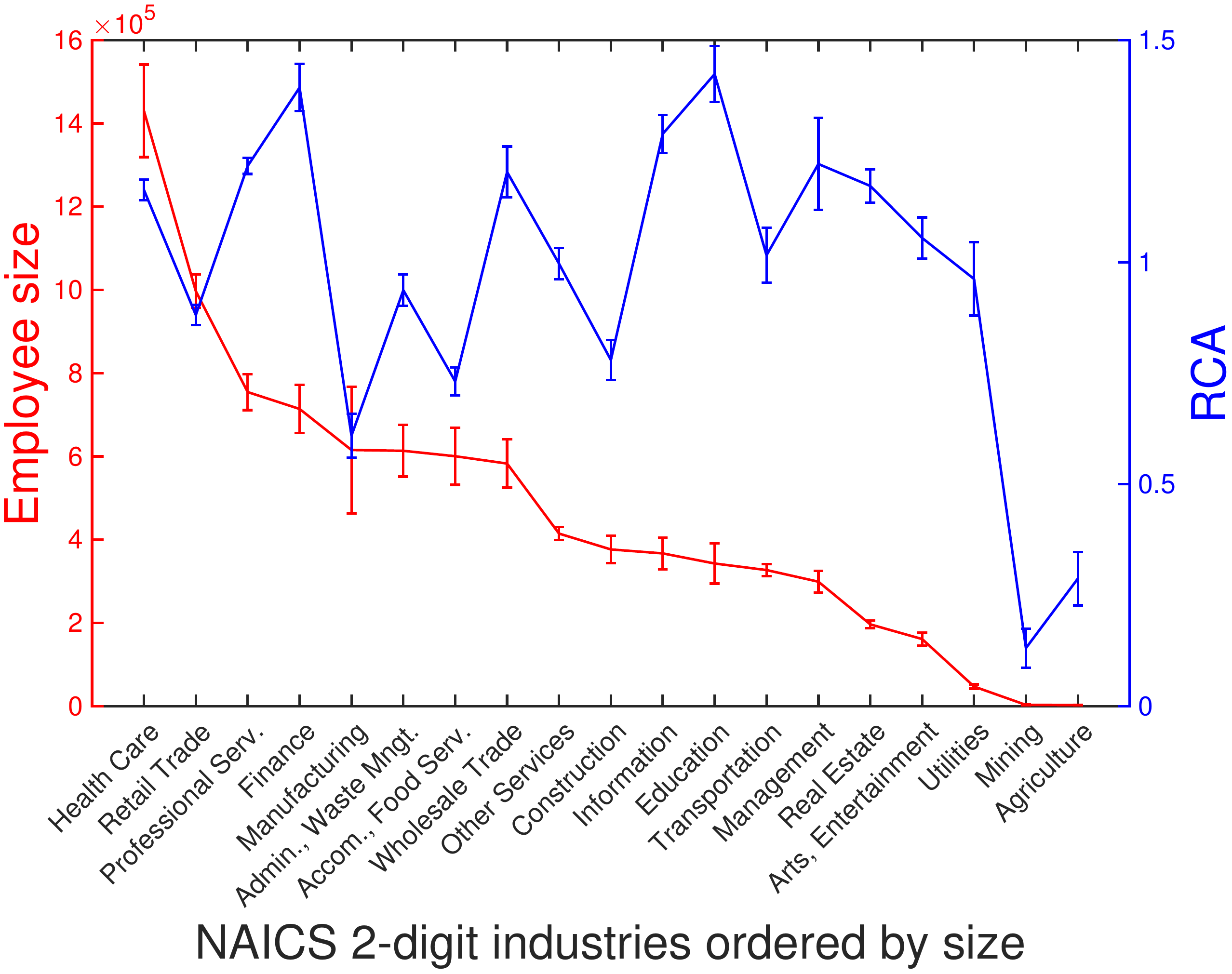}\label{fig:size_rca_ny}}\hspace{0.1cm}
	\\
	\labelarial{B}
	\subfloat{\includegraphics[width=0.4\textwidth]{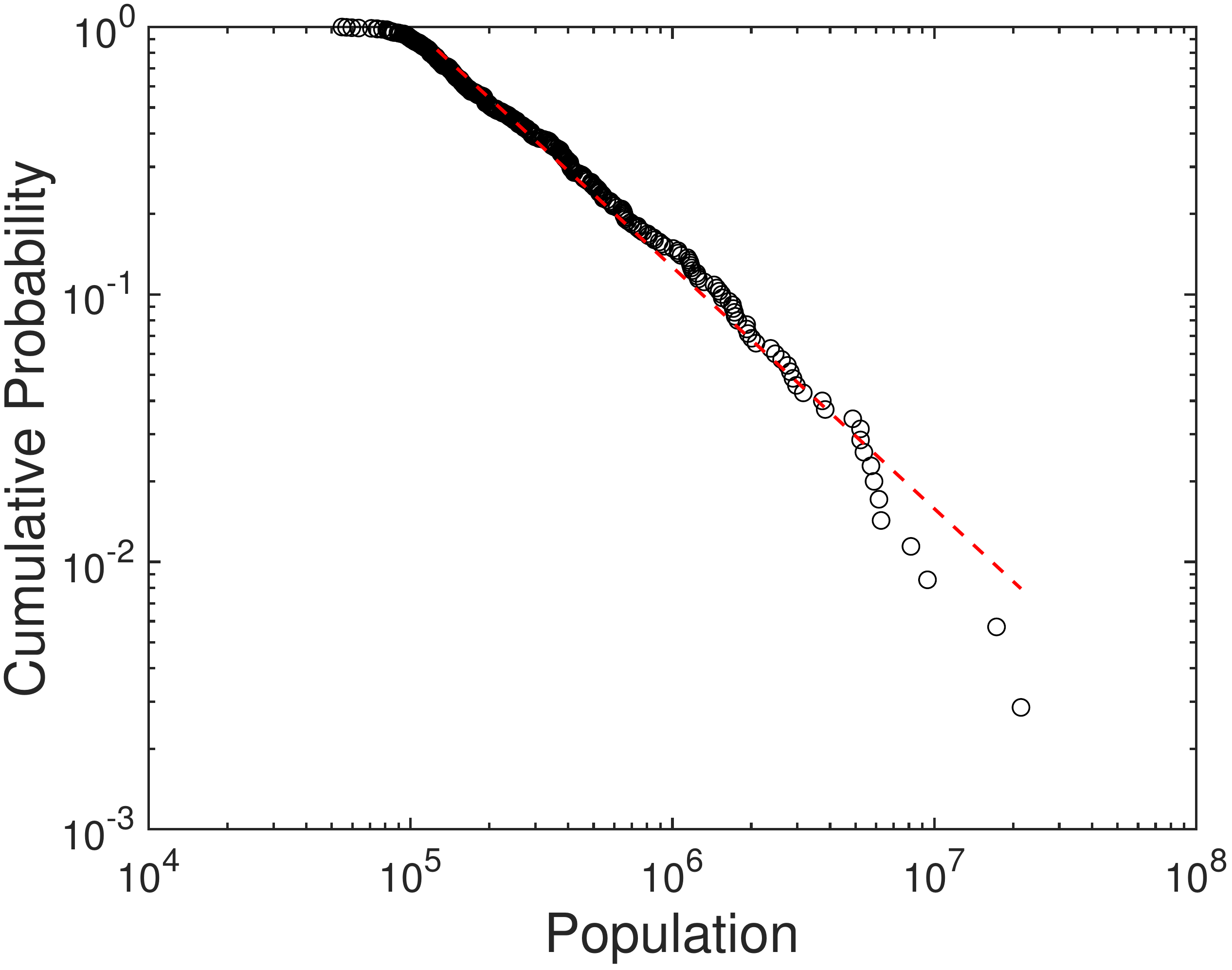}\label{fig:pop_distrib}}\hspace{0.1cm}
	\\
	\labelarial{C}
	\subfloat{\includegraphics[width=0.4\textwidth]{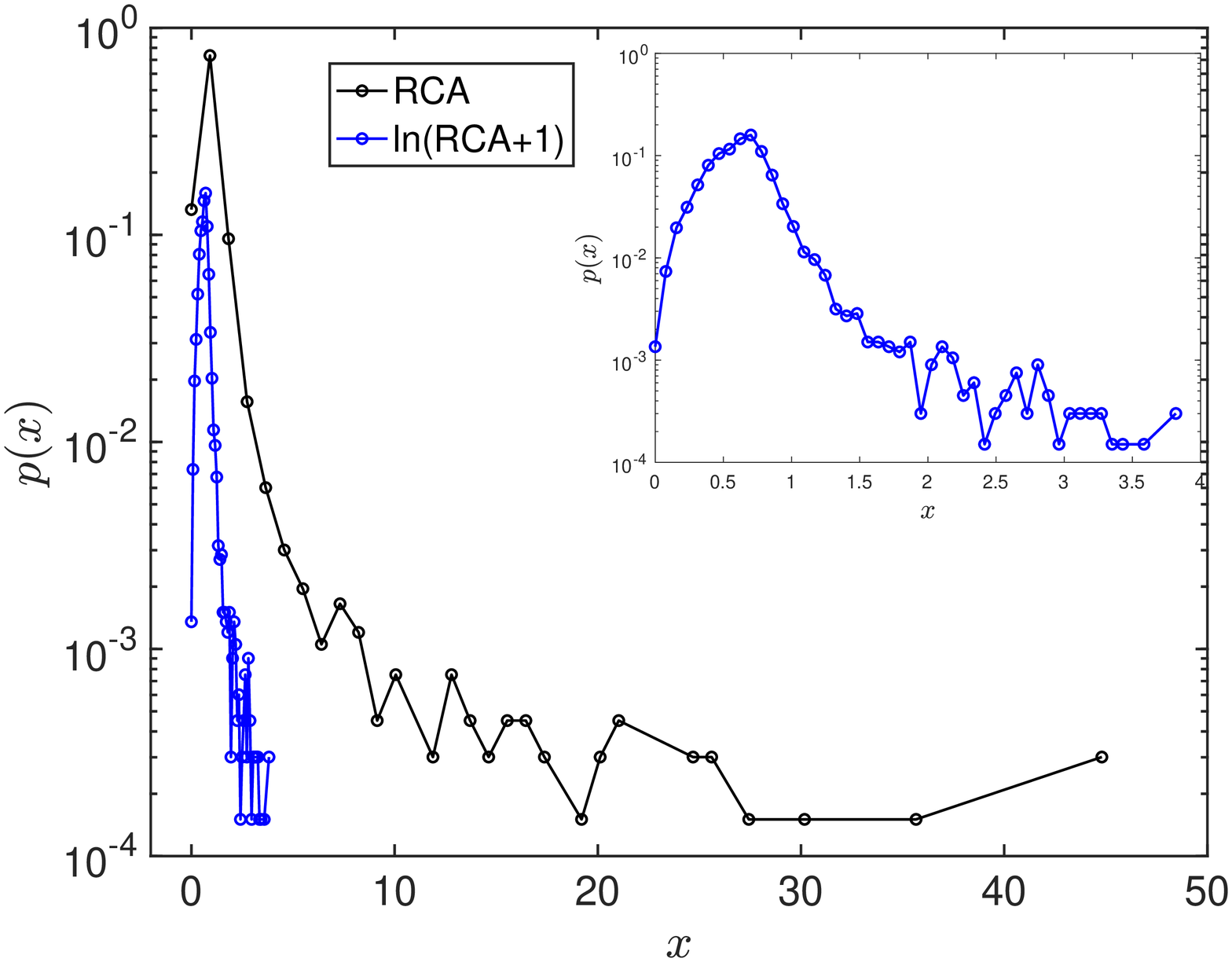}\label{fig:rca_distrib}}
	\\
	\label{fig:size_rca}
	\caption{
	(A) Size (red) and RCA (blue) of industries for 2-digit NAICS classification in New York City, NY. The values are the time-averages, and the errors are the standard deviations for time. The industries are ordered by their sizes in a descending order. 
	(B) Complementary cumulative probability distribution of time-averaged populations of all cities. The power-law exponent is measured as -0.90 corresponding to $\gamma=1.90$ for probability distribution $p(N) \sim N^{-\gamma}$.
	(C) Probability distribution of RCA (black) and logarithmic RCA (blue). The details of logarithmic RCA distribution is depicted in the inset.
	}
	\label{fig:RCA_distribution}
\end{figure}

The raw size is the most representative indicator of employment or establishments. 
However, it has two limitations in capturing the characteristics of urban industry. First, the baseline industry sizes in cities make small industries indistinguishable. For example, the share of retail trade is 100 times larger than that of agriculture on average, and most cities have similar baseline industry composition as shown as the shares of industries in Figure~\ref{fig:city-ind_share}. A small industry can easily be shaded by a large industry with this considerable size differential. 
Furthermore, huge differences in city sizes can shade the characteristics of small cities. The power-law distribution of urban populations in Figure~\ref{fig:pop_distrib} demonstrates the high heterogeneity of city sizes. Figure~\ref{fig:city-ind_size} shows that large cities are dominating almost all industries when we see the unnormalized sizes. Therefore, the city sizes need to be normalized to compare the patterns in small and large cities.

Revealed Comparative Advantage (RCA) originated from international trade analysis~\cite{balassa1965trade} and provides an ideal option for normalizing city sizes and industry sizes. 
RCA is defined as a ratio of urban share of an industry to the share of the industry across all cities. 
RCA can capture a small difference in industrial composition between cities by normalizing the urban industry size for both cities and industries.
Figure~\ref{fig:size_rca_ny} shows the industry size and its RCA for New York. The RCA normalizes tiny industries to comparable sizes.

In many studies \cite{hidalgo_product_2007,hidalgo_building_2009}, the presence of industry or production capability was considered as "revealed" if its RCA is larger than 1. Since its presence and absence have asymmetric ranges in $[0, 1]$ and $[1, \infty)$, respectively, normalized RCA measures have been developed to make the range less heterogeneous. Logarithmic RCA ($lrca$) \cite{vollrath1991theoretical,gao2017collective} is one of the normalized RCA measures. 
\begin{flalign}
    rca(c,i,t) &= \frac{Y(c,i,t)}{\sum_{i}Y(c,i,t)}/\frac{\sum_{c}Y(c,i,t)}{\sum_{c,i}Y(c,i,t)} \label{eq:RCA}\\
    lrca(c,i,t) &= \log{(rca(c,i,t)+1)} \label{eq:logRCA} \\
\end{flalign}

Compared to the logarithmic RCA, the original RCA has a highly heterogeneous distribution as in Fig.~\ref{fig:rca_distrib}. An urban industry with extraordinarily high or low RCA value makes other general industries very insignificant. 

%

\clearpage


\section{Urban scaling relation}\label{SI-sec:scaling}

\begin{figure}[ht]
	\centering
	\includegraphics[width=.45\textwidth]{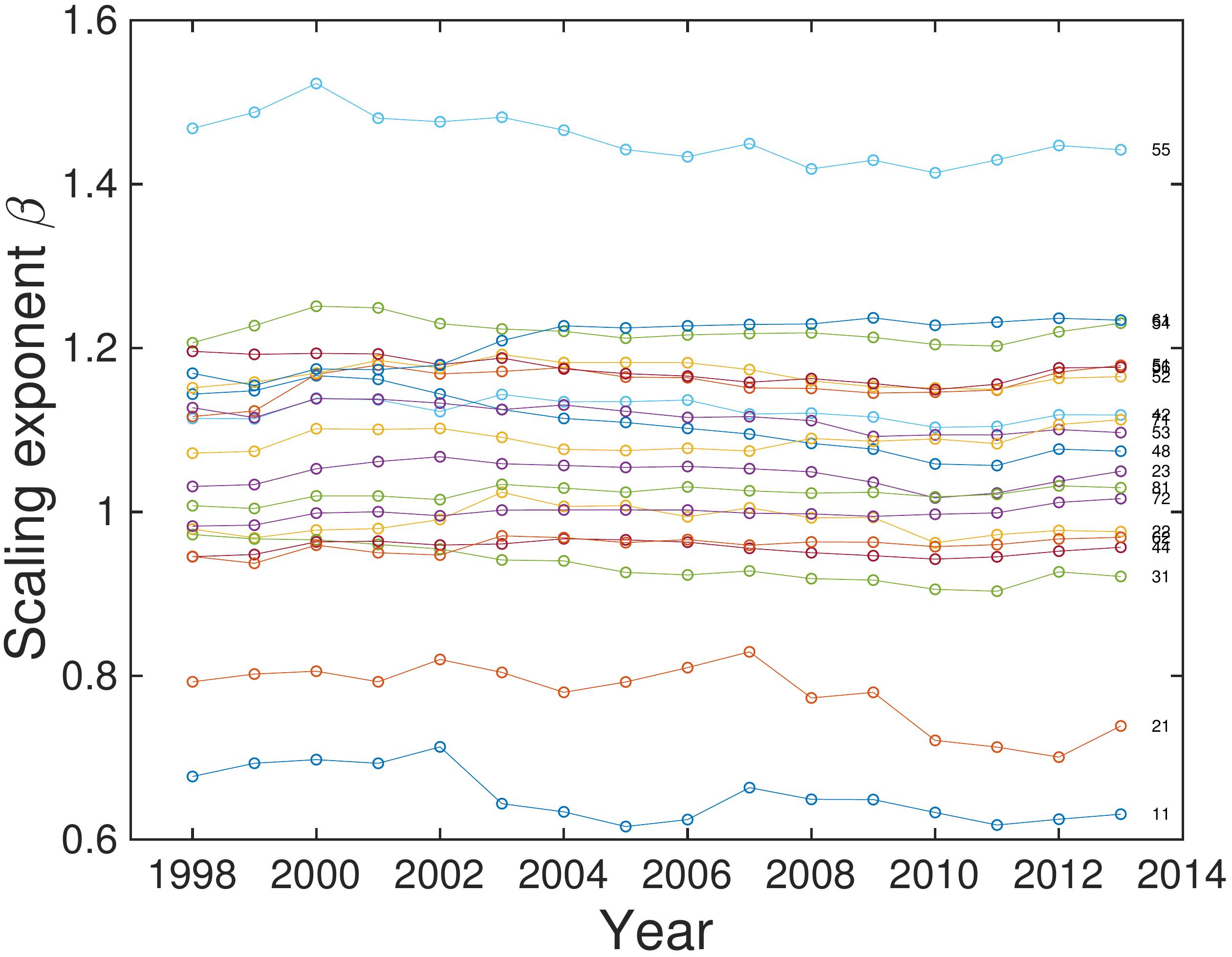}
	\caption{
	Time series of scaling exponents of industries in 2-digit NAICS classification. The numbers next to the lines denote the industry codes.}
	\label{fig:scaling_timeseries}
\end{figure}

Previous work has shown the scaling relationship of urban employment by industry~\cite{youn2016scaling}. Accordingly, we calculate the scaling exponents $\beta(i,t)$ and the pre-factors of industries $Y_{0}(i,t)$ for the size of employment $Y(c,i,t)$ and the city size $N(c,t)$ by fitting the following equation to empirical employment data: 

\begin{equation}
    Y(c,i,t) \approx Y_0(i,t)\cdot N(c,t)^{\beta(i,t)}
    \label{scaling}
\end{equation}
 
The industry size has high fluctuations in the low level of classifications. Therefore, we use 2-digit NAICS resolution. We calculate the scaling exponent by taking linear regression on the logarithms of $Y(c,i,t)$ and $N(c,i,t)$ for a fixed industry $i$ and a time $t$. The time-average of scaling exponents in 1998 to 2013 and the corresponding temporal deviation are listed in Table.~\ref{table:beta}. Scaling exponents do not show much temporal change during the time period of this study (see Fig.~\ref{fig:scaling_timeseries}).

\begin{table*}[!th]
    \centering
    \begin{tabular}{l|c|c|c}
    \hline
    Industry & NAICS code & $\beta$ & $R^2$ \\
    \hline
    Agriculture, Forestry, Fishing and Hunting  	& 11 & 0.65 $\pm$ 0.03 & 0.31 $\pm$ 0.04 \\
    Mining, Quarrying, and Oil and Gas Extraction  	& 21 & 0.78 $\pm$ 0.04 & 0.31 $\pm$ 0.04 \\
    Manufacturing  									& 31 & 0.94 $\pm$ 0.02 & 0.73 $\pm$ 0.02 \\
    Retail Trade  									& 44 & 0.96 $\pm$ 0.01 & 0.98 $\pm$ 0.00 \\
    Health Care and Social Assistance  				& 62 & 0.96 $\pm$ 0.01 & 0.94 $\pm$ 0.00 \\
    Utilities  										& 22 & 0.99 $\pm$ 0.02 & 0.66 $\pm$ 0.02 \\
    Accommodation and Food Services  				& 72 & 1.00 $\pm$ 0.01 & 0.95 $\pm$ 0.00 \\
    Other Services (except Public Administration)  	& 81 & 1.02 $\pm$ 0.01 & 0.95 $\pm$ 0.00 \\
    Construction  									& 23 & 1.05 $\pm$ 0.01 & 0.92 $\pm$ 0.01 \\
    Arts, Entertainment, and Recreation  			& 71 & 1.09 $\pm$ 0.01 & 0.86 $\pm$ 0.01 \\
    Transportation and Warehousing  				& 48 & 1.11 $\pm$ 0.04 & 0.85 $\pm$ 0.01 \\
    Real Estate and Rental and Leasing  			& 53 & 1.12 $\pm$ 0.02 & 0.93 $\pm$ 0.00 \\
    Wholesale Trade  								& 42 & 1.12 $\pm$ 0.01 & 0.91 $\pm$ 0.01 \\
    Information  									& 51 & 1.16 $\pm$ 0.02 & 0.89 $\pm$ 0.01 \\
    Finance and Insurance  							& 52 & 1.17 $\pm$ 0.01 & 0.88 $\pm$ 0.01 \\
    \makecell[l]{Administrative and Support and \\ Waste Management and Remediation Services} &	56 &	 1.17 $\pm$ 0.02  & 0.93 $\pm$ 0.01 \\
    Educational Services  							& 61 & 1.21 $\pm$ 0.03 & 0.77 $\pm$ 0.01 \\
    Professional, Scientific, and Technical Services& 54 & 1.22 $\pm$ 0.01 & 0.92 $\pm$ 0.01 \\
    Management of Companies and Enterprises  		& 55 & 1.46 $\pm$ 0.03 & 0.76 $\pm$ 0.02 \\
    \hline
    \end{tabular}
    \caption{
    Scaling exponent of industries in 2-digit NAICS classification. The scaling exponent is the time average of scaling exponents in 1998-2013, and the $R^2$ is also the time average.}
    \label{table:beta}
\end{table*}

\clearpage

\section{Lead-follow matrix}\label{SI-sec:lead-follow}

\subsection{Definition of lead-follow matrix}

We normalize the size of urban employment using RCA in Eq.~\ref{eq:logRCA}. By using the RCA values, we characterize the industries in a city as industry vector $\vec{I}(c,t)$ of city $c$ in year $t$. The spatial vector of industry $i$ can be defined in a similar way.
\begin{flalign}
	\vec{I}(c,t) &= (..., lrca(c,i,t),...)_{i \in I} \\
	\vec{L}(i,t) &= (..., lrca(c,i,t),...)_{c \in C}	
\end{flalign}
where $lrca = \log{(rca + 1)}$, $I$ and $C$ denote the logarithmic RCA, the set of industries, and the set of cities, respectively.
The length of the industry vector is equal to the number of industries (e.g. 19 for 2-digit NAICS industries). The industrial similarity between two cities $c$ and $c'$ with a time lag $\tau$ is now measured by a Pearson correlation between two vectors $\vec{I}(c, t)$ and $\vec{I}(c', t+\tau)$.
\begin{equation}
	\phi(c,t;c',t+\tau) = \rho(\vec{I}(c, t), \vec{I}(c', t+\tau))
	\label{eq:ind_similarity}
\end{equation}
where $\rho$ measures the Pearson correlation between two vectors. The inter-industry locational similarity $\psi(i,t;i',t+\tau)$ can also be obtained for the location vector $\vec{L}(i,t)$. 

As the dynamics of each city fluctuates, we group the cities by their populations, and use the average population as the reference.
\begin{equation}
	\vec{G}(g,t)=(\dots,\langle \log{(rca(c,i,t)+1)} \rangle_{c\in g},\dots)_{i\in I}
\end{equation}
where $\langle\cdot\rangle_{c\in g}$ denotes the average for the cities $c$ in group $g$. We usually group the cities into 20 groups of equal sizes. The industrial similarity between groups can also be defined using their correlations as in Eq.~\ref{eq:ind_similarity}.
\begin{equation}
	\phi(g,g',t,\tau) = \rho(\vec{G}(g, t), \vec{G}(g', t+\tau)) 
\end{equation}

\begin{figure*}[thp!]
	\centering
	\labelarial{A}
	\subfloat{\includegraphics[width=.28\textwidth]{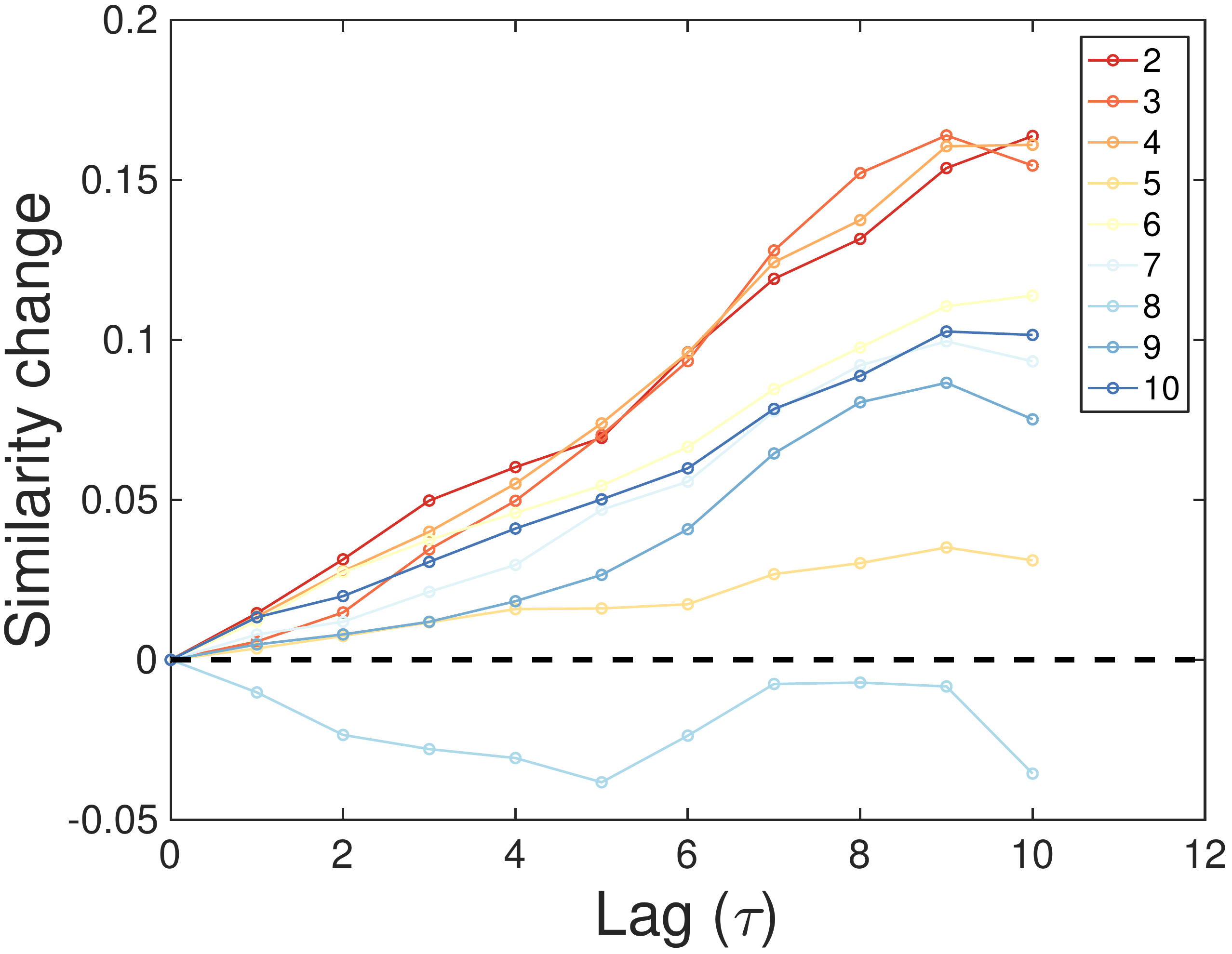}\label{fig:lf_ref1}}
	\labelarial{B}
	\subfloat{\includegraphics[width=.28\textwidth]{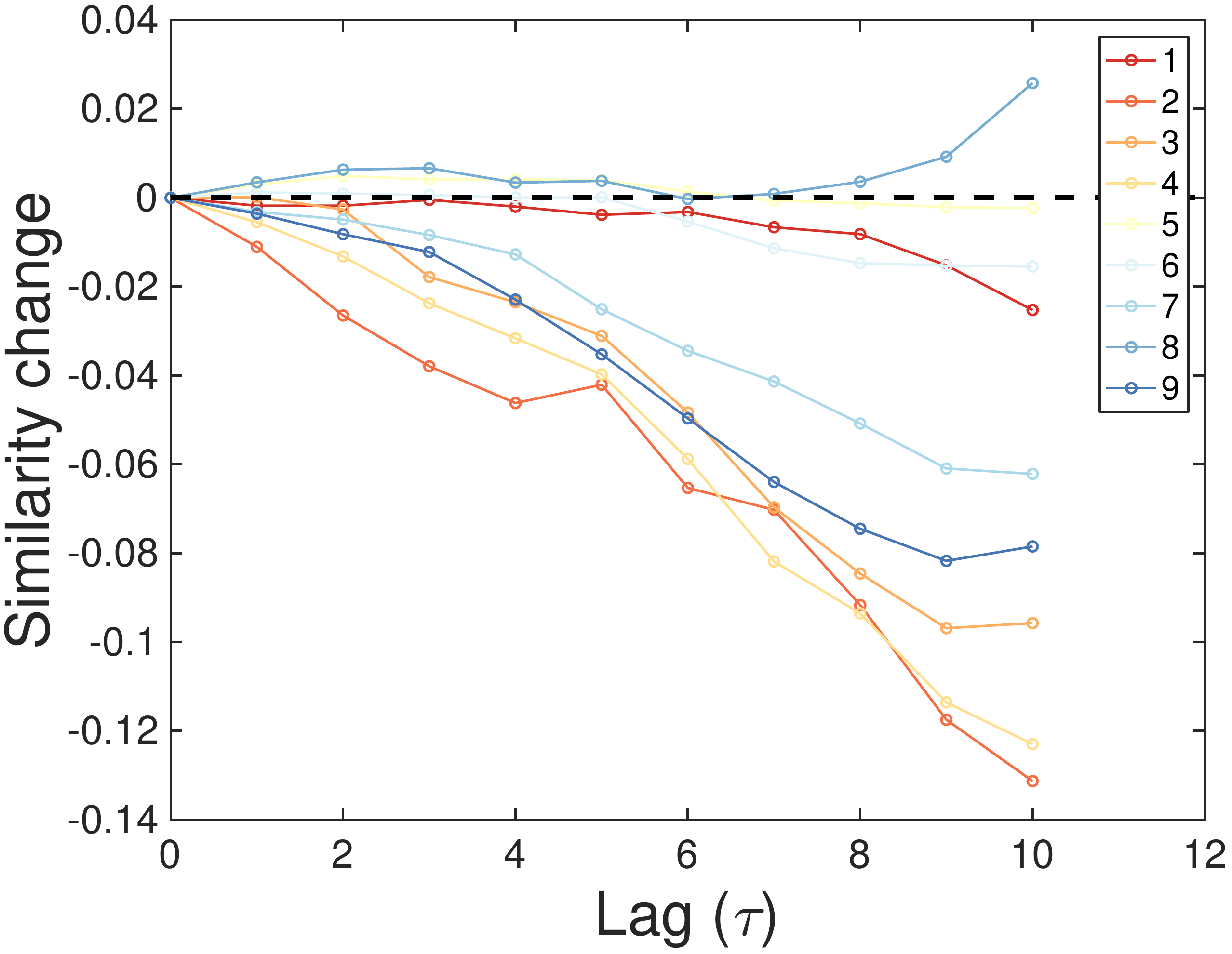}\label{fig:lf_ref10}}
	\labelarial{C}
	\subfloat{\includegraphics[width=.28\textwidth]{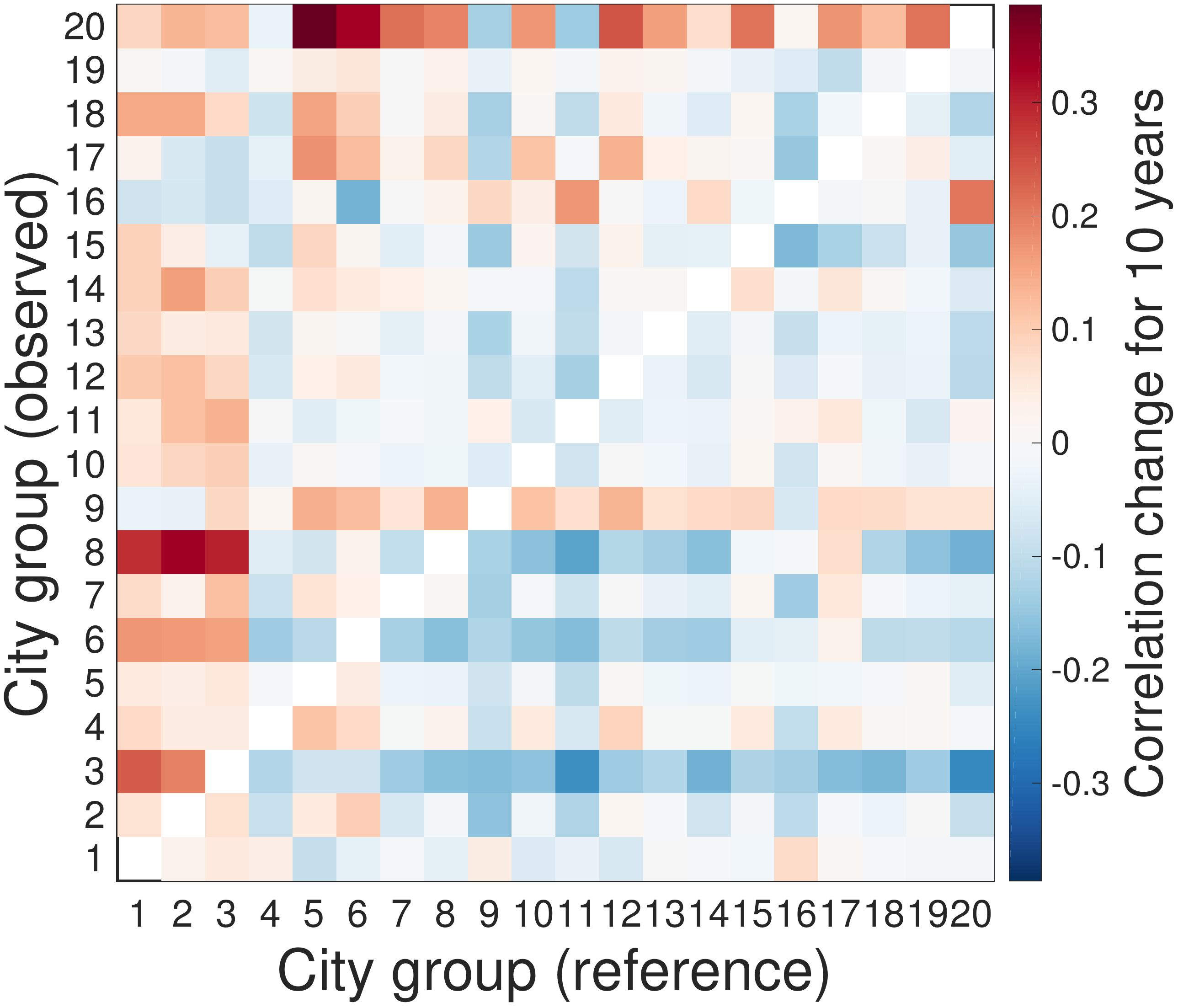}\label{fig:lead-follow}}
	\caption{
	(A-B) Similarity change within time lag $\tau$ between a reference group fixed in time and the rest city groups evolving in time. The reference groups are the largest group (A) and the smallest group (B) for 2-digit NAICS. Cities are grouped into 10 groups, and colored by population size. 
	(C) The Lead-Follow Matrix shows increase (red) or decrease (blue) in industrial similarity between cities grouped, and ranked, by size for employment in 2-digit NAICS classification. Each cell represents the similarity increase (or decrease) of a group of cities on $y$-axis (target) in one, two, and five years, when a reference group on $x$-axis is fixed in time. The positive upper triangle (mostly colored red) means that smaller cities in the future resemble larger cities at the present.} 
\end{figure*}

Then, we fix one group $g$ as a reference group, let another group $g'$ evolve in time ($\tau$), and measure their industrial similarity $\phi(g,g',t,\tau)$. By following the change of industry similarity $\phi$ as time lag $\tau$ changes, we can observe if group $g'$ is getting similar or dissimilar to group $g$ in time. To focus on the effect by time lag $\tau$, we average the change over the reference times $t$.
\begin{flalign}
	\Delta \phi(g,g',t,\tau) &= \phi(g,g',t,\tau) - \phi(g,g',t,0) \\
	\Delta \phi(g,g',\tau) &= \langle\Delta \phi(g,g',t,\tau)\rangle_{t}
\end{flalign}
where $\Delta\phi(g,g',\tau)$ is the average change of similarity between reference group $g$ and observed group $g'$ for time difference $\tau$. Figure~\ref{fig:lf_ref1} and Figure~\ref{fig:lf_ref10} show the trends of $\Delta\phi(g,g',\tau)$ versus on lag $\tau$ for a fixed reference group $g$ and various observed group $g'$ denoted by colored lines. In Fig.~\ref{fig:lf_ref1}, the similarity generally increases for the largest reference city, which means that small cities get similar to the largest cities in time. On the other hand, large cities get dissimilar to the smallest group in Fig.~\ref{fig:lf_ref10}. These trends show the directional evolution of urban economy led by large cities and followed by small cities. We summarize these trend in the form of matrix.

Finally, we measure the average similarity change between groups for $T$ years, which is called ``Lead-follow matrix'' denoted as $LF$. We set $T$ as 10 years considering the time span of our data. The average rate is obtained by the least squares method.
\begin{flalign}
	LF(g,g') &= \frac{\displaystyle\sum_{\tau=1}^{10}10\tau\cdot\Delta \phi(g;g,\tau)}{\displaystyle\sum_{\tau=1}^{10} \tau^{2}} \\
	&= \frac{\displaystyle\sum_{\tau=1}^{10}10\tau\cdot\langle\phi(g,g',t,\tau)-\phi(g,g',t,0) \rangle_t}{\displaystyle\sum_{\tau=1}^{10}\tau^2}.
	\label{eq:lead-follow}	
\end{flalign}

Figure~\ref{fig:lead-follow} summarizes the time-lag similarities across cities as a ``Lead-Follow Matrix''. The general pattern observed in the figures is that small cities get similar to larger cities whereas large cities get dissimilar from smaller cities.
As the city groups are ordered by their population sizes, the upper triangle represents the temporal behavior of small cities on the larger cities on x-axis, whereas the lower triangle describes the behavior of larger cities. 
The positive values in the upper triangle mean that small cities become more and more similar to the past of large cities, recapitulating, in terms of their industrial characteristics, while the negative values in the lower triangle represent that large cities grow away from the past of smaller cites as time goes. 
The asymmetric shape indicates that the evolution of urban industry has a clear direction that large cities lead the change and small cities follow the path from behind. In other words, the largest 3 city groups (top 60 cities) are considered as main leaders from their significant patterns in the lead-follow matrix. 

\subsection{Robustness check}
We check the robustness of the lead-follow matrix for various conditions: length of time lags, NAICS level, and number of city groups. We test the time lags for 3 and 5 years, NAICS for 3- and 4-digits, and city groups for 50 groups and 70 groups based on the the reference parameters for 10-years time lags, 2-digit NAICS, and 20 city groups. 
As a result, the lead-follow matrix is robust for the variants of time lags, NAICS digits, and city group sizes.



\begin{figure*}[th!]
	\centering
	\labelarial{A}
	\subfloat{\includegraphics[width=.30\textwidth]{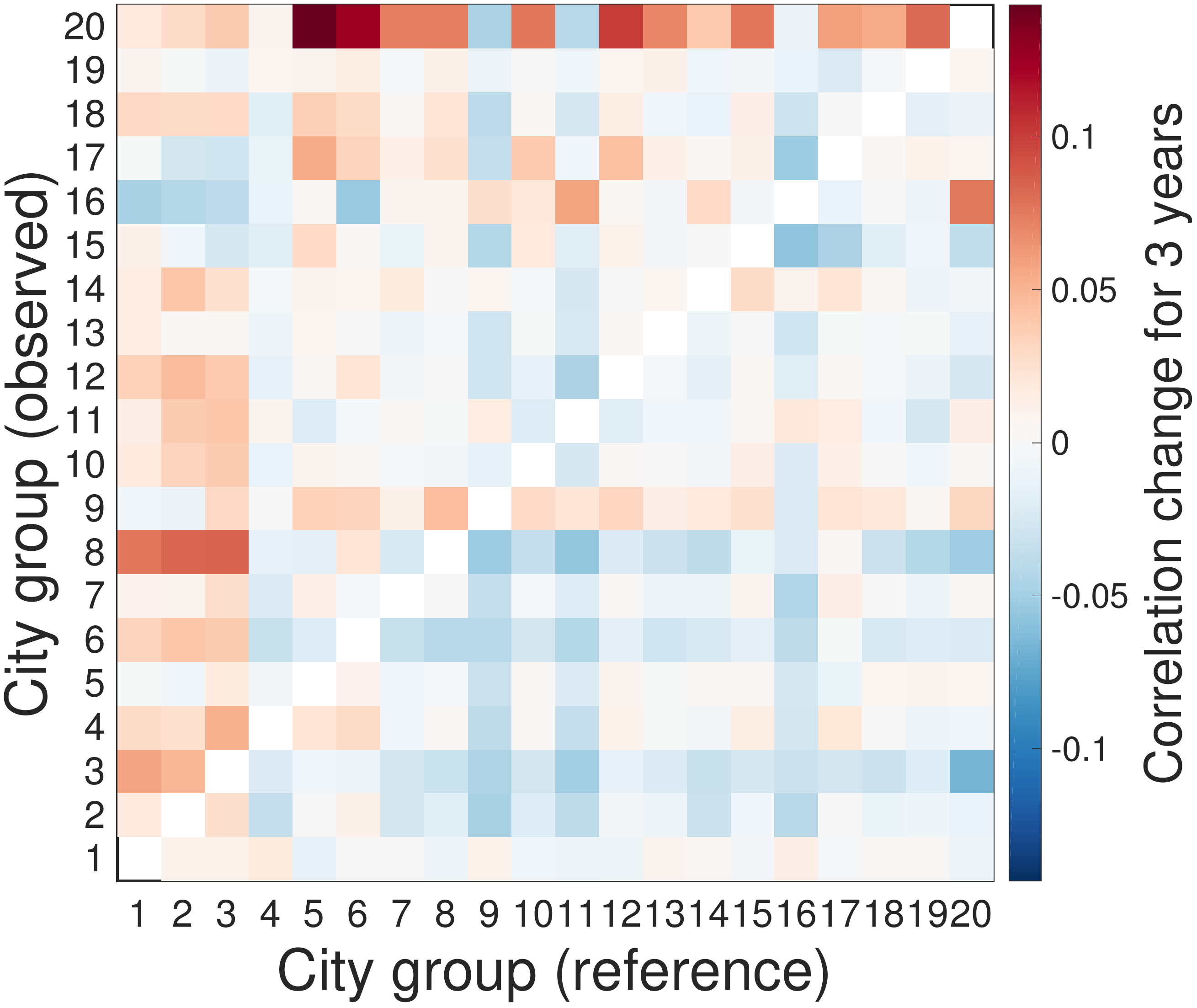}}
	\labelarial{B}
	\subfloat{\includegraphics[width=.30\textwidth]{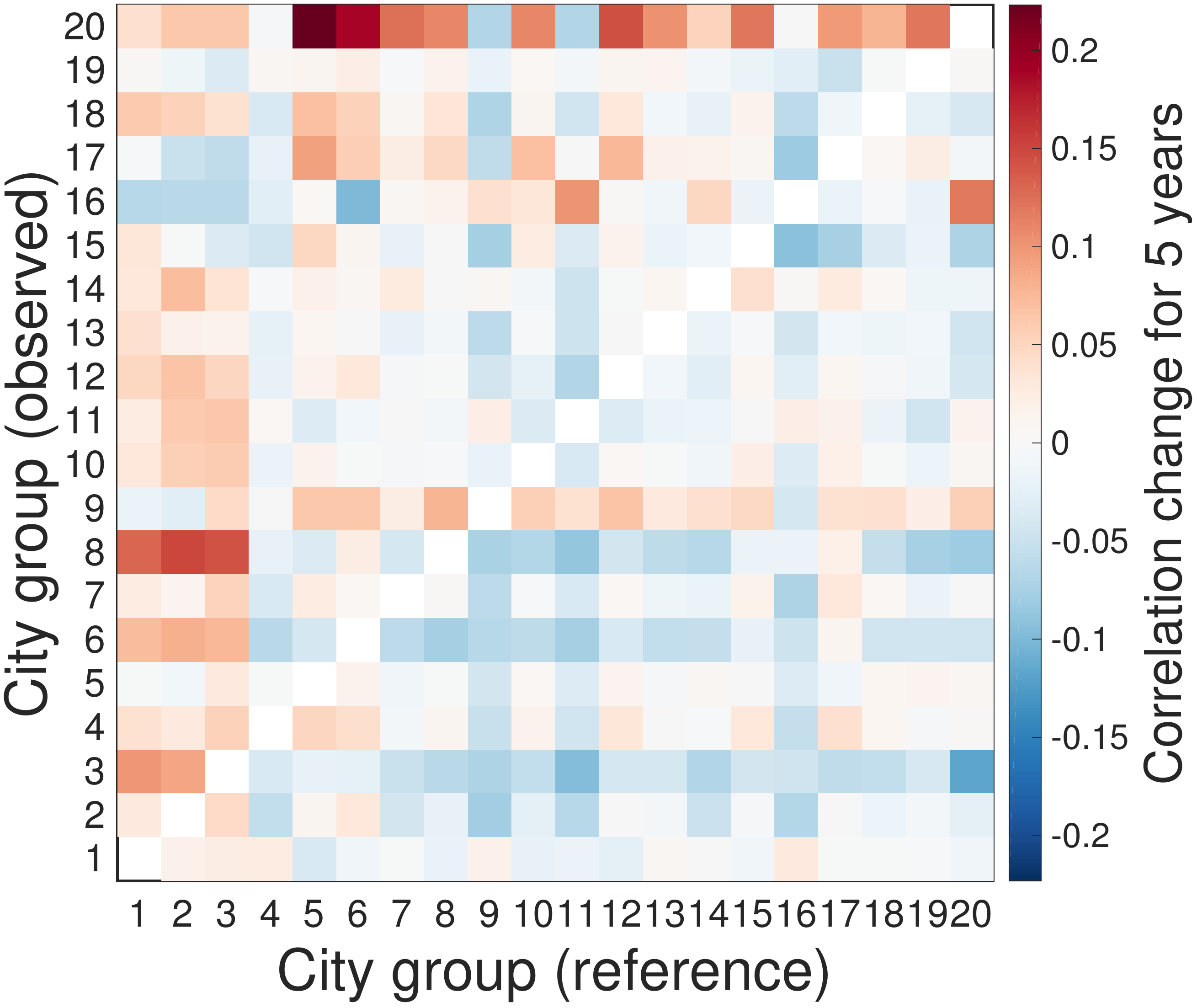}}
	\labelarial{C}
	\subfloat{\includegraphics[width=.30\textwidth]{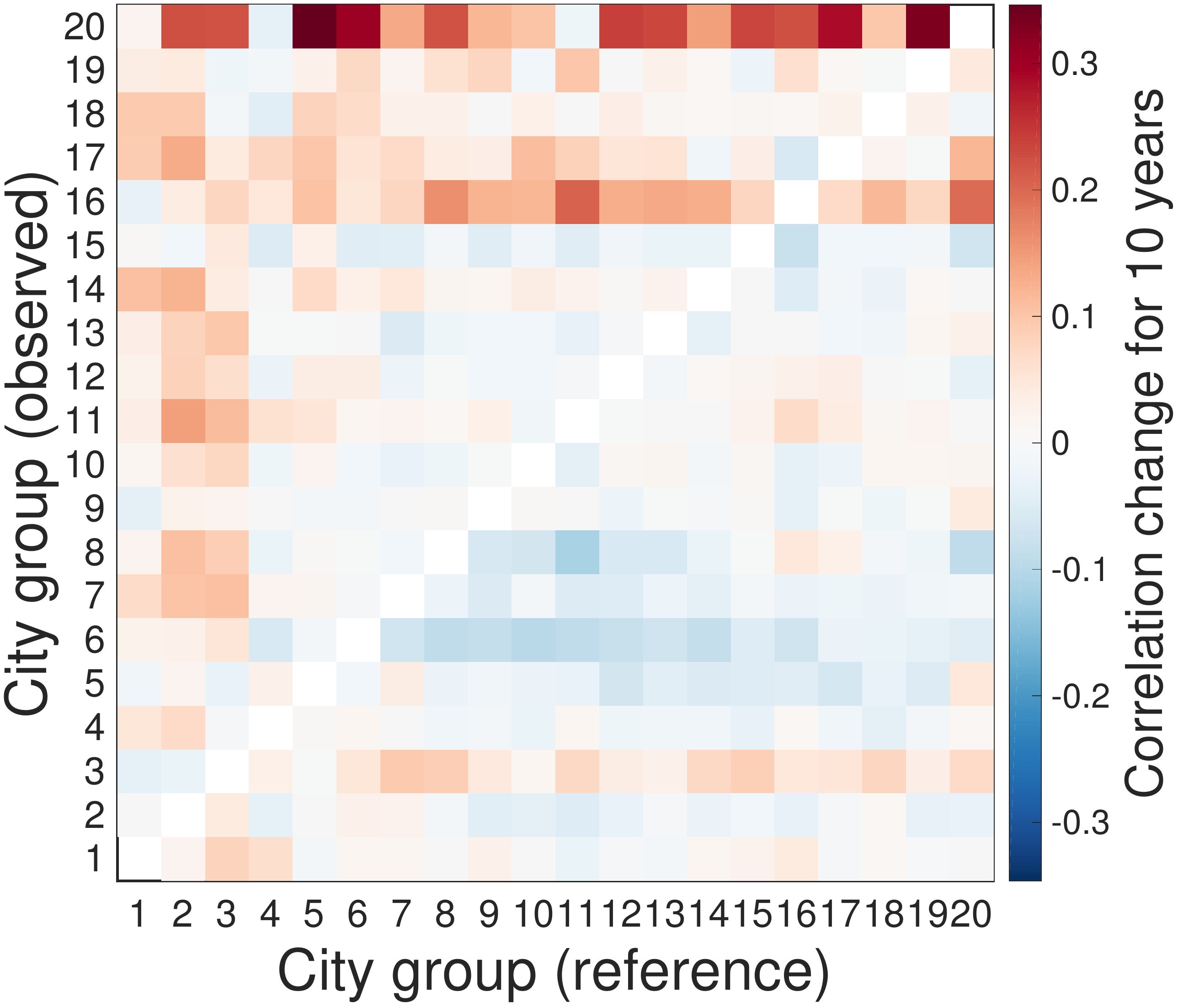}}
	\\
	\labelarial{D}
	\subfloat{\includegraphics[width=.30\textwidth]{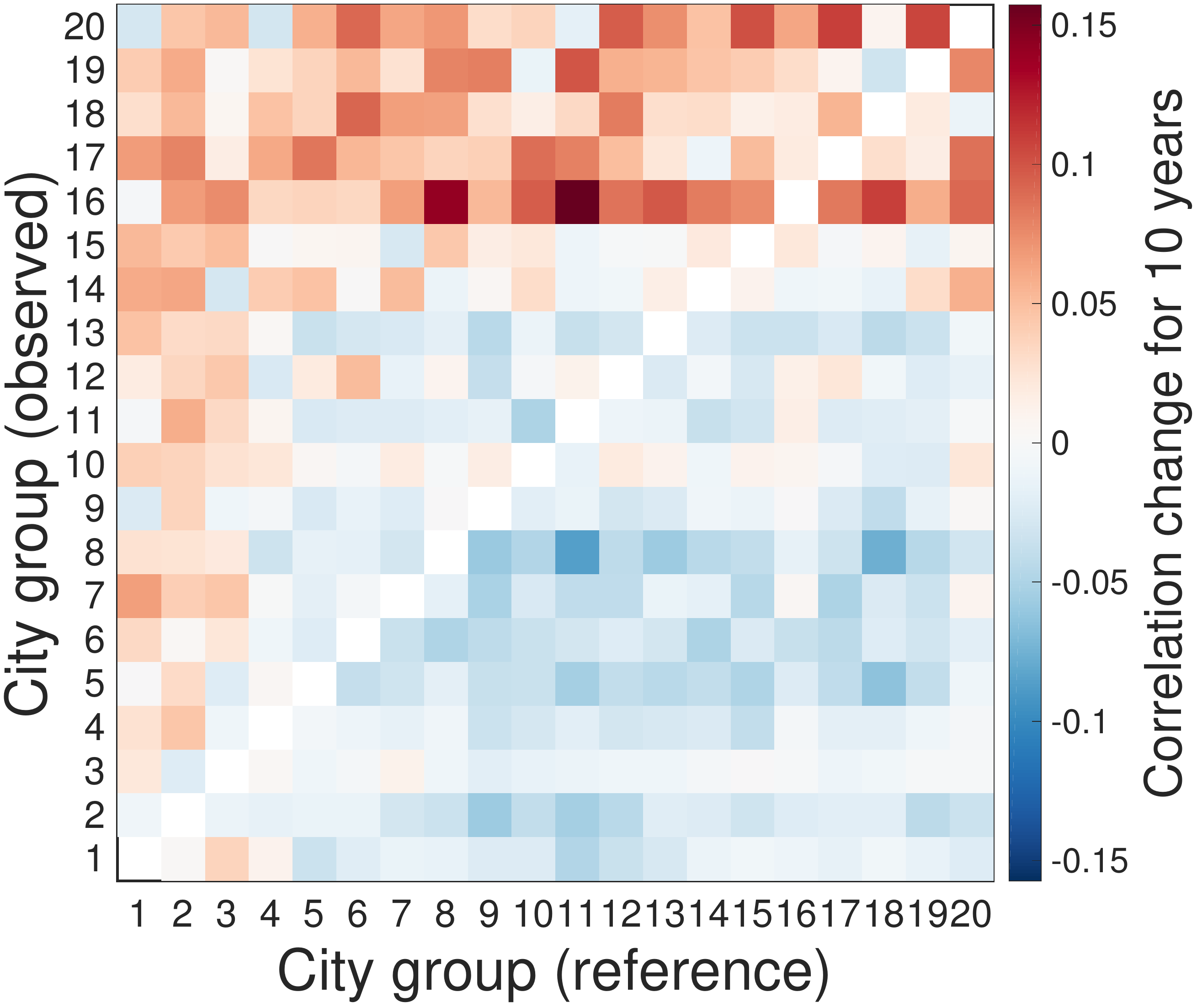}}
	\labelarial{E}
	\subfloat{\includegraphics[width=.30\textwidth]{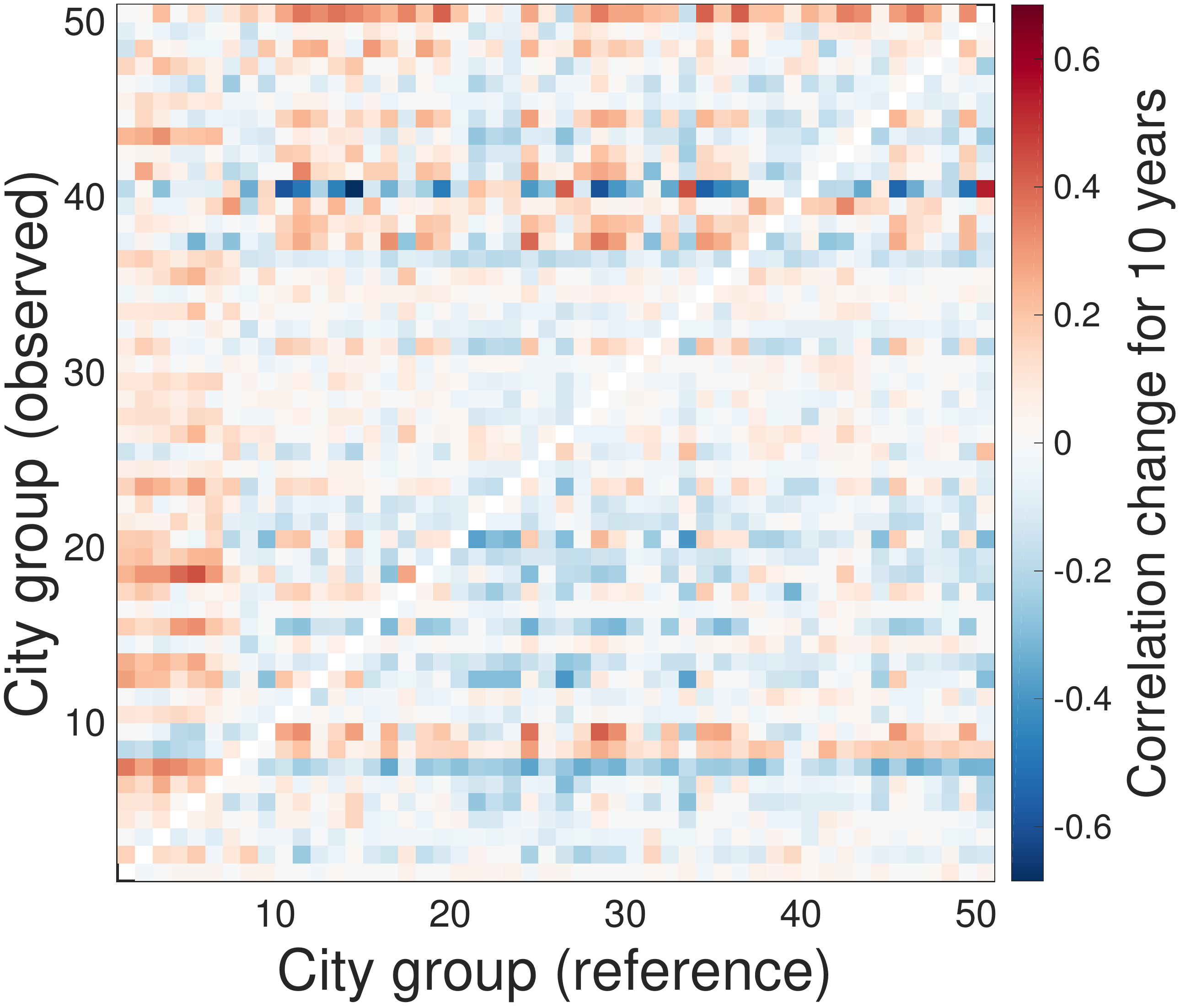}}
	\labelarial{F}
	\subfloat{\includegraphics[width=.30\textwidth]{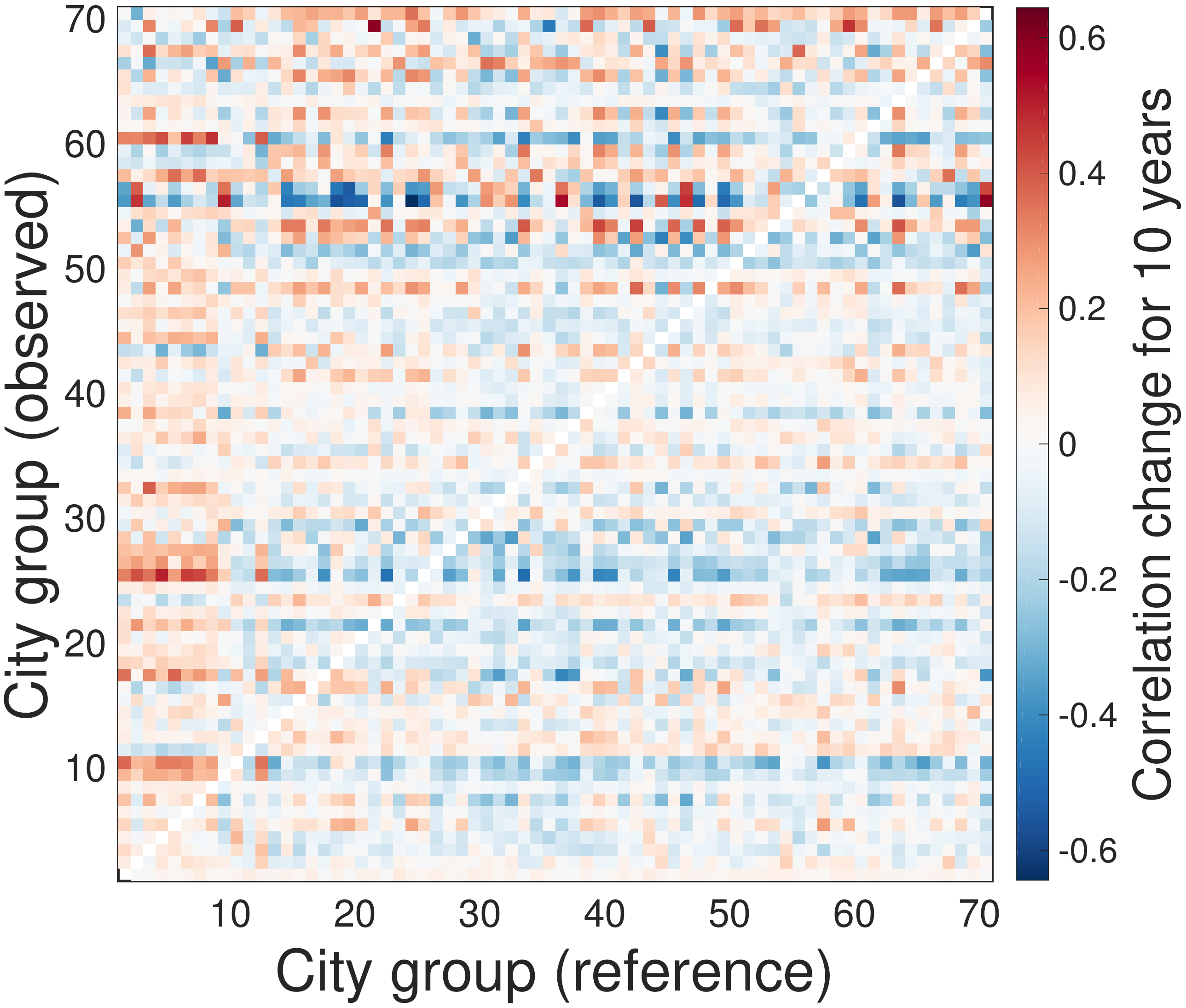}}
	\caption{
	Robustness check on the Lead-Follow Matrices for various parameters. We vary one condition in each plot from the reference conditions, 10-years time lags, 2-digit NAICS, and 20 city groups. 
	 (A) 3-years time lag. 
	 (B) 5-years time lag. 
	 (C) 3-digits NAICS. 
	 (D) 4-digits NAICS. 
	 (E) 50 city groups. 
	 (F) 70 city groups.
	}
\end{figure*}

\clearpage

\section{Integrated framework}\label{SI-sec:integrated_framework}

\subsection{Time-independent integrated framework of RCA and scaling relation}

\begin{figure*}[h!t]
	\centering
	\includegraphics[width=0.9\textwidth]{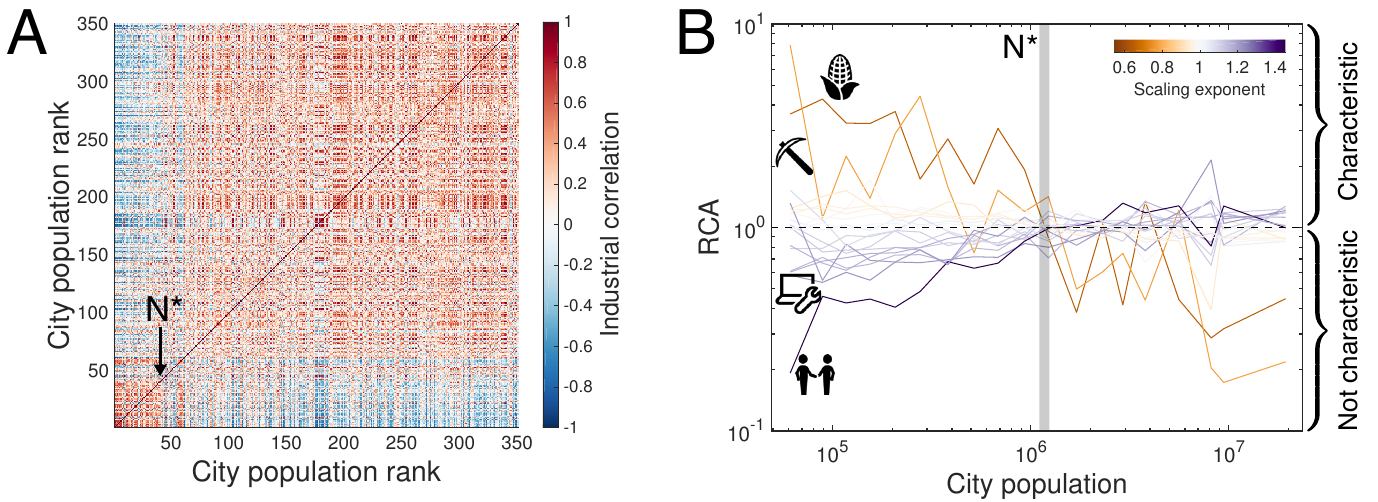}
	\caption{
    How small cities transition into an innovative large-city economy.
	(A) Industrial correlation between cities. We measure the Pearson correlations between cities characterized as $\log{(rca+1)}$ in 2-digit NAICS classifications, and average them for the entire time span. The cities are ordered by their population (large to small).
	(B) RCA of each industry averaged for log-binned population groups. The trends are reversed by their scaling exponents, and they meet at the characteristic city size predicted by the fixed point analysis, where $N^*\approx 1.2$ (grey bar).
	}
	\label{fig:clustering}
\end{figure*}

By the scaling relation, the size $Y(c,i)$ of industry $i$ in a city $c$ with population size $N(c)$ is given as follows:
\begin{equation}
	Y(c,i) = Y_{0}(c,i)N(c)^{\beta(i)}. 
\end{equation}
The pre-factor $Y_{0}$ and the scaling exponent $\beta$ are obtained by taking a regression on $\log{Y(c,i)}$ and $\log{N(c)}$ for each industry $i$. By definition, the revealed comparative advantage (RCA) can be expressed as a function of population and scaling exponent.
\begin{flalign}
	rca(c,i) &= \frac{Y(c,i)}{\sum_{i}Y(c,i)}/\frac{\sum_{c}Y(c,i)}{\sum_{c,i}Y(c,i)} \nonumber\\
	&\sim \frac{Y_{0}(i)N(c)^{\beta(i)}}{\sum_{i}Y(c,i)\sum_{c}Y_{0}(i)N(c)^{\beta(i)}} \nonumber\\
	&= \frac{N(c)^{\beta(i)}}{\sum_{i}Y(c,i)\sum_{c}N(c)^{\beta(i)}}
\end{flalign}
where $\sum_{c,i}Y(c,i)$ is a constant.
Since $\sum_{i}Y(c,i)$ is the total size of industries in a city, it can be approximated to be proportional with $N(m)$. The total industry size for all cities, $\sum_{c}N(c)^{\beta(i)}$, is estimated by continuous population approximation, whose distribution follows $P(N)\sim N^{-\gamma}$. For $\beta \neq \gamma-1$, the summation becomes,
\begin{flalign}
	\sum_{c}N(c)^{\beta} &\simeq \int_{N_{min}}^{N_{max}} P(N)dN N^{\beta} \nonumber\\
	& \sim \int_{N_{min}}^{N_{max}} dN N^{\beta-\gamma} \nonumber\\
	& \sim \frac{1}{\beta-\gamma+1}(N_{max}^{\beta-\gamma+1} - N_{min}^{\beta-\gamma+1}).
\end{flalign}

For $\beta=\gamma-1$, the summation can be determined by either taking limitation $\beta \to \gamma-1$ or changing the integration to $\int_{N_{min}}^{N_{max}} dN N^{-1}$.

\begin{flalign}
	\lim_{\beta \to \gamma-1}\sum_{c}N(c)^{\beta} 
	& \sim \lim_{\beta \to \gamma-1}\frac{1}{\beta-\gamma+1}(N_{max}^{\beta-\gamma+1} - N_{min}^{\beta-\gamma+1}) \nonumber \\
	&= \lim_{\delta\to0} \frac{1}{\delta}((N_{max}^{\delta}-1) - (N_{min}^{\delta}-1)) \nonumber \\
	&= \log N_{max} - \log N_{min} 
\end{flalign}
where $\delta=\beta-\gamma+1$ and $\log x=\lim_{\delta\to0}(\frac{x^{\delta}-1}{\delta})$ by definition.
The same result can be achieved by changing the integration as,
\begin{flalign}
	\sum_{c}N(c)^{\beta} & \sim \int_{N_{min}}^{N_{max}} dN N^{-1} \nonumber\\
	& = \log N_{max} - \log N_{min}.
\end{flalign}

As a result, the RCA is approximated as,
\[
rca(\beta,N)\sim 
\begin{cases}
    N^{\beta-1}[\frac{(\beta-\gamma+1)}{N_{max}^{\beta-\gamma+1}-N_{min}^{\beta-\gamma+1}}]& \text{for } \beta\neq\gamma-1,\\
    \frac{N^{\beta-1}}{\log N_{max} - \log N_{min}} & \text{for } \beta=\gamma-1.
\end{cases}
\]

The scaling exponent of population distribution $\gamma$ generally equals to 2 for cities following the Zipf's law. For $\gamma=2$, the RCA is approximated as, 
\begin{equation}
	rca(\beta,N)\sim (\beta-1)N^{\beta-1}[\frac{1}{N_{max}^{\beta-1}-N_{min}^{\beta-1}}]. \\
	\label{eq:rca_approx}
\end{equation}

The derived RCA has different trends by population sizes. Approximately, RCA is proportional to $\beta$ for large $N$ while it is inversely proportional to $\beta$ for small $N$. In that case, if we correlate the RCA vectors of small and large cities, the correlation becomes negative while the correlations inside small cities and large cities remain positive. This property generates two clusters of small and large cities. \\

\subsection{Fixed point of city clustering}

Now, we find the characteristic city size that distinguishes the small and large cities. The characteristic size is given as a saddle point of Eq.~\ref{eq:rca_approx}. The saddle point should satisfy both $\partial rca / \partial \beta = 0$ and $\partial rca / \partial N = 0$. Since $rca$ is continuous, we only consider the case for $\beta\neq\gamma-1$ in the calculation of the saddle point.

\begin{flalign}
	\frac{\partial rca(\beta,N)}{\partial N} &= (\beta-\gamma+1)(\beta-1)N^{\beta-2}[\frac{1}{N_{max}^{\beta-\gamma+1}-N_{min}^{\beta-\gamma+1}}] = 0 \nonumber
\end{flalign}
Therefore, $\beta^{*}=1$.
The above conditions mean that the saddle point exists only if the probability of scaling exponent around $\beta=1$ is not zero. For example, if every scaling exponent is larger than 1 or smaller than 1, the saddle point needed for clustering of cities does not exist. 

\begin{flalign}
	\frac{\partial rca(\beta,N)}{\partial \beta} 
	&= (\beta-\gamma+1)(\beta-1)N^{\beta-1}[\frac{1}{n-m}]^2 \nonumber \\ 
	&\times [\frac{n-m}{\beta-\gamma+1}+(n-m)\log{N}-(n\log{N_{max}}-m\log{N_{min}})] \nonumber \\ 
	&= 0 \nonumber
\end{flalign}
where $n=N_{max}^{\beta-\gamma+1}$ and $m=N_{min}^{\beta-\gamma+1}$. Therefore, the saddle point population is given at, 
\begin{flalign}
	N^{*} &= N_{norm}(\beta^{*}=1, \gamma)e^{-1/(\beta-\gamma+1)} \\ 
	&= N_{norm}(1, \gamma)e^{1/(\gamma-2)}
\end{flalign}
where $N_{norm} = \exp(\frac{n \log{N_{max}} - m \log{N_{min}}}{n - m})$. 
The saddle-point population $N^{*}$ is calculated as 1.2 millions for $\gamma=1.9$, $N_{min}=10^5$ and $N_{max}=10^7$, which is similar with the observation in data.

As a result of connection between RCA and scaling relations, we can expect the clustering of cities by their industrial compositions. Fig.~\ref{fig:clustering}A is the result obtained by the Pearson correlations between industry vectors of cities characterized by RCA with 2-digit industry classifications. The clustering of cities is clearly observed with a fixed point near 50th largest city.

We can also figure out the fixed point using the trends in Fig.~\ref{fig:clustering}B. RCA as a function of $N$ and $\beta$ in Eq.~\ref{eq:rca_approx} can be simplified by taking logarithm on both sides.

\begin{equation}
	\log{rca(\beta,N)} \sim (\beta-1)\log{N} + \log{(\beta-1)} - \log{(N_{max}^{\beta-1}-N_{min}^{\beta-1})}
	\label{eq:rca_approx_log}
\end{equation}.

Logarithm of RCA becomes a linear function of $N$ with a slope $\beta-1$. It makes the trend inverted when $\beta$ changes from $\beta < 1$ to $\beta > 1$, and generates two clusters by their populations.

%
%

\subsection{Time-dependent integrated framework of RCA and scaling relation}

To understand the time-dependent behavior of RCA following the scaling relation, we derive the time-dependent equation of RCA under several assumptions. In addition to two assumptions (i.e., (1) power-law distributed population, (2) total employee size in a city) used in the time-independent derivation, we include two additional assumptions: (3) total national employee size proportional to the summation of the total populations, and (4) time-independent $Y_{0}$, $\beta$, $\gamma$, $N_{max}$ and $N_{min}$. The third assumption is given from the second assumption by definition.

\begin{flalign}
    \sum_{c,i}Y(c,i) &= \sum_{c}\sum_{i}Y(c,i) \nonumber\\
    &\simeq \sum_{c}C(t)N(c)
\end{flalign}
where the second assumption is given as $\sum_{i}Y(c,i,t) \simeq C(t)N(c)$ for time-dependent constant $C(t)$. For the fourth assumption, the scaling exponent $\beta$ and the pre-factor $Y_{0}$ do not change much in time as in Fig.~\ref{fig:scaling_timeseries}, and the scaling exponent of the population distribution $\gamma$ is also not expected to change much in time. For the maximum and minimum populations, the average urban population change between the first year (1998) and the last year (2013) is about 23\%, so we can marginally assume its time-independency when we observe the dynamics in a short period. Four assumptions are listed as follows:

\begin{enumerate}
\item $\begin{aligned}[t]
    P(N) \simeq \frac{1-\gamma}{N_{max}^{1-\gamma}-N_{min}^{1-\gamma}}N^{-\gamma} \\
\end{aligned}$
\item $\begin{aligned}[t]
    \sum_{i}Y(c,i,t) \simeq C(t)N(c,t) \\
\end{aligned}$
\item $\begin{aligned}[t]
    \sum_{c,i}Y(c,i,t) \simeq C(t)\sum_{c}N(c,t) \\
\end{aligned}$
\item 
    Time-independent $Y_{0}$, $\beta$, $\gamma$, $N_{max}$ and $N_{min}$.
\end{enumerate}

We can derive the time-dependent RCA by substituting the time-dependent industry size $Y(c,i,t)$ in the definition.

\begin{flalign}
    rca(c,i,t) &= \frac{Y(c,i,t)}{\sum_{c}Y(c,i,t)}/\frac{\sum_{i}Y(c,i,t)}{\sum_{c,i}Y(c,i,t)} \\
    rca(\beta, N, t) &= \frac{Y_{0}N(c,t)^\beta}{\sum_{c}Y_{0}N(c,t)^\beta}/\frac{\sum_{i}Y(c,i,t)}{\sum_{c,i}Y(c,i,t)} \\
    &= \frac{N(c,t)^{\beta}}{\sum_{c}N(c,t)^\beta}/\frac{C(t)N(c,t)}{C(t)\sum_{c}N(c,t)} \nonumber\\
    &= N(c,t)^{\beta-1}\frac{\sum_{c}N(c,t)}{\sum_{c}N(c,t)^{\beta}}
\end{flalign}

\begin{figure}[t!h]
	\centering
	\includegraphics[width=0.45\textwidth]{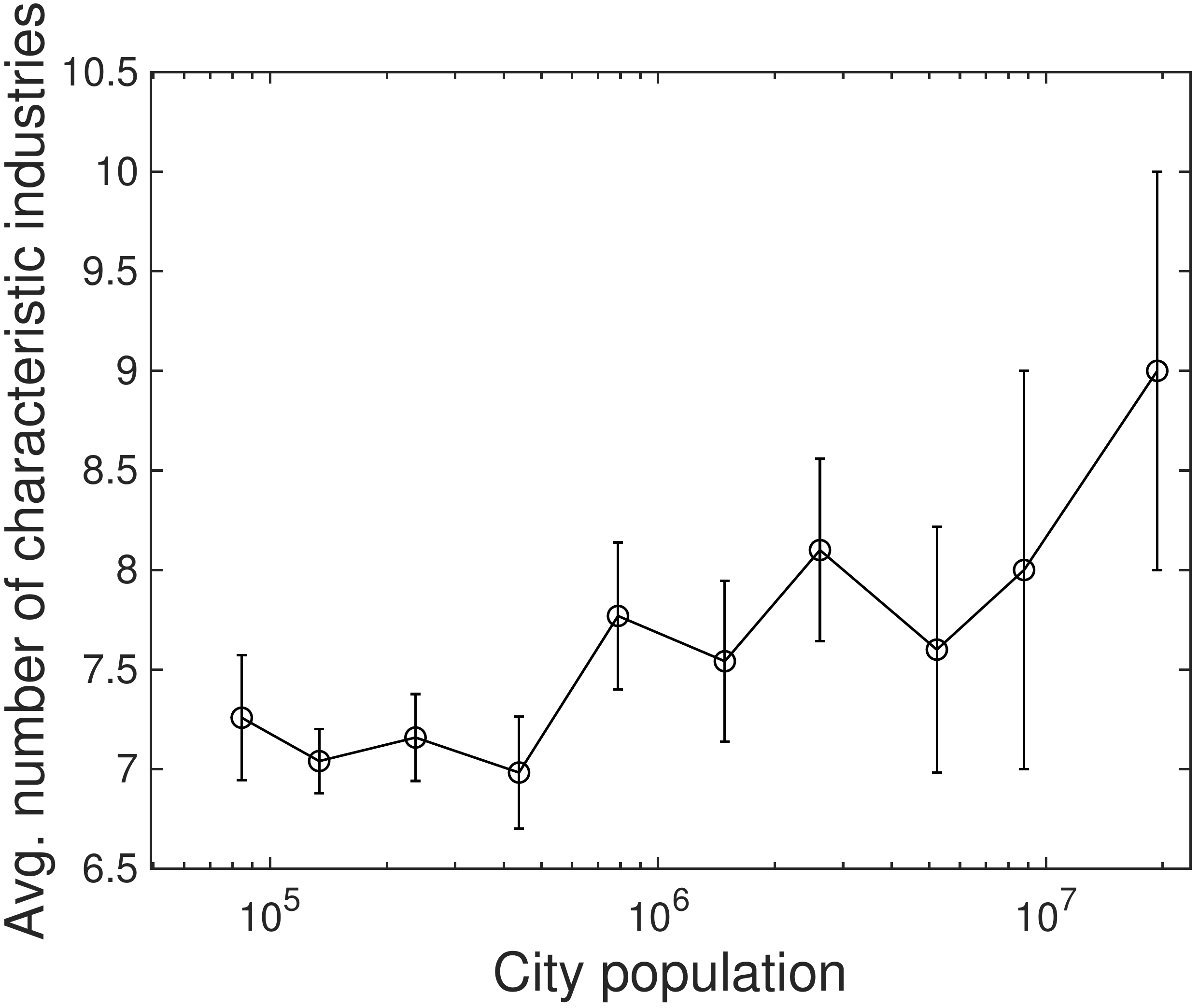}
	\caption{
	Average number of characteristic industries as a function of populations for 2-digit NAICS industries. We count the number of characteristic industries of every city, and take the averages for 10 city groups determined by their log-binned populations. The error bar denotes the standard error of each group.
	}
	\label{fig:n_prominent}
\end{figure}

The summations can be obtained using the approximation of power-law distributed continuous populations (assumption 1).

\begin{flalign}
    \sum_{c}N(c, t) &\simeq \int_{N_{min}}^{N_{max}}dN \frac{1-\gamma}{N_{max}^{1-\gamma}-N_{min}^{1-\gamma}}N^{1-\gamma} \nonumber\\
    &= \frac{1-\gamma}{2-\gamma}\frac{N_{max}^{2-\gamma}-N_{min}^{2-\gamma}}{N_{max}^{1-\gamma}-N_{min}^{1-\gamma}}
\end{flalign}

\begin{flalign}
    \sum_{c}N(c, t)^{\beta} &\simeq \int_{N_{min}}^{N_{max}}dN \frac{1-\gamma}{N_{max}^{1-\gamma}-N_{min}^{\beta-\gamma}}N^{1-\gamma} \nonumber\\
    &= \frac{1-\gamma}{\beta-\gamma+1}\frac{N_{max}^{\beta-\gamma+1}-N_{min}^{\beta-\gamma+1}}{N_{max}^{1-\gamma}-N_{min}^{1-\gamma}}
\end{flalign}

Finally, we can obtain the closed-form time-dependent RCA as a function of $\beta$, $N$, and $t$.

\begin{flalign}
    rca(\beta, N, t) &= N(t)^{\beta-1}\frac{\beta-\gamma+1}{2-\gamma}\frac{N_{max}^{2-\gamma}-N_{min}^{2-\gamma}}{N_{max}^{\beta-\gamma+1}-N_{min}^{\beta-\gamma+1}} \\
    &= CN(t)^{\beta-1}
\end{flalign}
where $C$ is a time-invariant constant for simplifying the equation. Since the only time-variant term is $N(t)^{\beta-1}$, the time-derivative of RCA depends only on the time-derivative of population. For easy calculation, we take logarithm on both sides and take the derivatives. 

\begin{flalign}
    \log{rca(\beta, N, t)} &= (\beta-1)\log{N} + \log{C} \label{eq:logrca-N}\\
    \frac{d}{dt}(\log{rca(\beta, N, t)}) &= (\beta-1)\frac{d}{dt}(\log{N})
\end{flalign}

One interesting thing of these equations is the analogy with the scaling relation of size. Following the scaling relation $Y = Y_{0}N^{\beta}$, the log size and its time-derivative have similar forms with RCA except the difference in the proportional constant, 1.

\begin{flalign}
    \log{Y(\beta, N, t)} &= \beta\log{N} + \log{Y_{0}} \\
    \frac{d}{dt}(\log{Y(\beta, N, t)}) &= \beta\frac{d}{dt}(\log{N})
\end{flalign}

In that sense, RCA is similar to a coordinate transformation that makes a projection of scaling relation onto $x$-axis, because a scaling relation usually have a slope close to 1. One remarkable advantage of using RCA is its sensitivity to super- or sub-linearity. For the situation that the urban scaling relation is valid, the size of every industry always grows if its urban population grows, however, the RCA depends on whether each industry has superlinear or sublinear scaling.

\begin{enumerate}
\item For $dN/dt > 0$ and $\beta > 1$, $drca/dt > 0$.
\item For $dN/dt > 0$ and $\beta < 1$, $drca/dt < 0$.
\item For $dN/dt < 0$ and $\beta > 1$, $drca/dt < 0$.
\item For $dN/dt < 0$ and $\beta < 1$, $drca/dt > 0$.
\end{enumerate}

The dependency of RCA on populations can be a more intuitive measure of superlinearity or sublinearity of the industry, i.e., a superlinear scaling for positive correlation and a sublinear scaling for negative correlation. 

\subsection{Diversity by the number of characteristic industries}

Fig.~\ref{fig:clustering}B and Eq.~\ref{eq:logrca-N} show that industries with $\beta(i)<1$ is characteristic in small cities, and industries with $\beta(i)>1$ is characteristic in large cities. $N^{*}$ becomes the reference of small cities and large cities. We can expect that the distribution of $\beta$ will determine the number of characteristic industries by city size. There are fewer industries with $\beta(i)<1$ than industries with $\beta(i)>1$, therefore, the number of characteristic industries is expected to increase with city size.

Fig.~\ref{fig:n_prominent} shows that this expectation is true in our urban employment data. Small cities with $N < N^{*}$ have fewer characteristic industries, about 7 industries out of 19 industries, while large cities have more characteristic industries, about 8-9 industries. It shows the increase of industrial diversity as city size increases. 

\clearpage

\section{Urban recapitulation}\label{SI-sec:recapitulation}

\subsection{Trajectory of cities for each industry}

The growth of population and industry size of cities can be represented as trajectories in the population-industry plane like the scaling relation. We present the trajectory of each city as arrows for 2-digit NAICS industries in Fig.~\ref{fig:scaling_trajectory}. In some industries, most of cities grow following the fitting line of scaling relation, but they drift or irregularly move in some industries. The scaling relation for the cross-sectional view gives an insight on explaining the dynamics. We linearize and differentiate the time-dependent scaling relation to mathematically formulate the dynamics. Since the scaling exponent $\beta$ does not change much in our time scope (Fig.~\ref{fig:scaling_timeseries}), we can consider the scaling exponent as constant in time (i.e., denoted as $\beta(i)$). 
\begin{flalign}
	Y(c,i,t) &= Y_{0}(i,t) + N(c,t)^{\beta(i)} \\ 
	\log{Y(c,i,t)} &= \log{Y_{0}}(i,t) + \beta(i) \log{N(c,t)} \\
	\Delta \log{Y(c,i)} &= \Delta \log{Y_{0}}(i) + \beta(i) \Delta \log{N(c)} \label{eq:slope}
\end{flalign}
where $\Delta$ means the difference of a value of $X$ between the start year (1998) and the end year (2013) of our time scope, i.e., $\Delta X = X(2013) - X(1998)$.
Eq.~\ref{eq:slope} decomposes the growth into time-dependent global growth $\Delta\log{Y_{0}}(i)$ and population-dependent growth $\beta(i)\Delta\log{N(c)}$. The trajectories in Fig.~\ref{fig:scaling_trajectory} directly present these decomposition of growth that the global shift is based on $\Delta\log{Y_{0}}(i)$ and the city-specific local movement is based on $\beta(i)\Delta\log{N(c)}$. 

Based on Eq.~\ref{eq:slope}, we measure the change of industry size $\Delta\log{Y(c,i)}$ versus on the change of population $\Delta\log{N(c)}$ of each city. Fig.~\ref{fig:scaling-trend} shows common positive correlations between them. 
By Eq.~\ref{eq:slope}, the slope of each industry, longitudinal size-dependency, $\hat{\beta}(i)$ should be the same with the cross-sectional scaling exponent $\beta(i)$. 
We obtain the longitudinal size-dependency by taking a linear regression on the scatter plot in Fig.~\ref{fig:scaling-trend}.
The comparison of the slope and the scaling exponent is depicted in \Cref{fig:growth-scaling_2dig}. If the dynamics perfectly follows the theoretical expectation by scaling relations, all data points should be on the prediction line (red dot). Although they are not perfectly matched, the scaling relation predicts its general trend, that is, the direction of industry size change is coherent with the direction of population change.

\subsection{Recapitulation scores}

The degree of recapitulation can be measured from how much each city follows the fitting line. Let us consider the case of perfect recapitulation. In that case, the trajectory of every city will lie exactly on the fitting line. When a city population grows, its industry size should accordingly grow following the fitting line. In that case, the measured growth terms $\Delta \log{\hat{Y}_{0}(i)}$, and $\hat{\beta}(i)$ using Eq.~\ref{eq:slope} and Fig.~\ref{fig:scaling-trend} quantify how much the real data follows the dynamics predicted by theory. 
Since the amount of drift $\Delta \log{Y_{0}(i)}$ only shifts the fitting line, we can measure the ``degree of recapitulation'' by comparing longitudinal size-dependency $\hat{\beta}(i)$ and cross-sectional scaling exponent $\beta(i)$. The difference between them becomes a direct measure of recapitulation of each industry. We define the recapitulation score $S(i)$ of each industry $i$ using the relative error of $\hat{\beta}(i)$ to average $\beta(i)$ in our time scope. 
\begin{equation}
	S(i) = 1 - \Big\lvert\frac{\hat{\beta}(i)-\beta(i)}{\beta(i)}\Big\rvert
	\label{eq:recap_score}
\end{equation}

The score becomes maximum ($S(i) = 1$) when $\hat{\beta}(i)=\beta(i)$, and it becomes zero when $\hat{\beta}(i)$ has no tendency ($\hat{\beta}(i) = 0$. When $\hat{\beta}(i)$ has the opposite sign of $\beta(i)$, the score becomes negative. The recapitulation scores for 2-digit and 3-digit industries are shown in Fig.~\ref{fig:recap_score_N2} and Fig.~\ref{fig:recap_score_N3}. The average recapitulation score of industries with significant scaling relation ($R^2 \geq 0.5$) is 0.70 which validates the existence of recapitulation in cities. The significantly low scores of `Agriculture' and `Mining' industries seem to originate from their strong dependency on regional characteristics. 

Now, how can we measure the degree of recapitulation in each city? For that, we need to define a measure of industrial growth of individual city in the scaling framework. Using the measured global growth $\Delta \log{\hat{Y}_{0}(i)}$ in Eq.~\ref{eq:slope}, we can estimate the coefficient of growth $\hat{B}(c,i)$ of a city $c$ for a specific industry $i$. As the equation was defined for an industry, it can easily be generalized using an industry index $i$.
\begin{flalign}
	\Delta\log{Y}(c,i) &= \Delta \log{\hat{Y}_{0}(i)} + \hat{B}(c,i)\Delta\log{N}(c)\\
	\hat{B}(c,i) &= \frac{\Delta\log{Y}(c,i) - \Delta\log{\hat{Y}_{0}(i)}}{\Delta\log{N}(c)}	
\end{flalign}

The coefficient $\hat{B}(c,i)$ denotes the ratio of industry growth to the population growth in city $c$. Since many cities have a small relative population change $\Delta\log{N}$ as depicted in Fig.~\ref{fig:scaling-trend}, $\hat{B}(c,i)$ in a city scale is very fluctuating. To observe a general trend, we group the cities into 20 groups and measure the averaged slope $\hat{B}(g,i)$ of $\Delta\log{Y}(c,i) - \hat{\Delta\log{Y_{0}(i)}}$ versus on ${\Delta\log{N}(c)}$ using least square errors.
\begin{equation}
	\hat{B}(g,i) = \frac{\sum_{c \in g}(\Delta\log{Y}(c,i) - \Delta\log{\hat{Y}_{0}(i)})\cdot\Delta\log{N}(c)}{\sum_{c \in g}(\Delta\log{N}(c))^{2}}	
\end{equation}

The measured $\hat{B}(g,i)$ is the coefficient of population-dependent growth of city group $g$ for industry $i$. It should be the same with $\beta(i)$ when every city in $g$ perfectly recapitulates. The city group recapitulation score of group $g$ for industry $i$ can be measured in the same way of Eq.~\ref{eq:recap_score}. 
\begin{equation}
	S(g, i) = 1 - \Big\lvert\frac{\hat{B}(g,i)-\beta(i)}{\beta(i)}\Big\rvert	
\end{equation}

Finally, the city group recapitulation score for every urban industry can be defined as the average of recapitulation scores for each industry $i$.
\begin{flalign}
	S(g) &= \frac{1}{|I|}\sum_{i \in I}(S(g,i)) \\
	&= 1 - \frac{1}{|I|}\sum_{i\in I}(\Big\lvert\frac{\hat{B}(g,i)-\beta(i)}{\beta(i)}\Big\rvert)	
\end{flalign}
where $|I|$ is the number of industries. 

Fig.~\ref{fig:city_recap_score_N2} and Fig.~\ref{fig:city_recap_score_N3} show the recapitulation score of each city. We can see that most of cities have positive recapitulation scores. Generally, recapitulation is strong except for very small cities. It validates the general existence of recapitulation in most of cities. 

\subsection{Detailed explanation on the drifts}

\begin{figure*}[thp!]
	\centering
	\subfloat{\includegraphics[width=.23\textwidth]{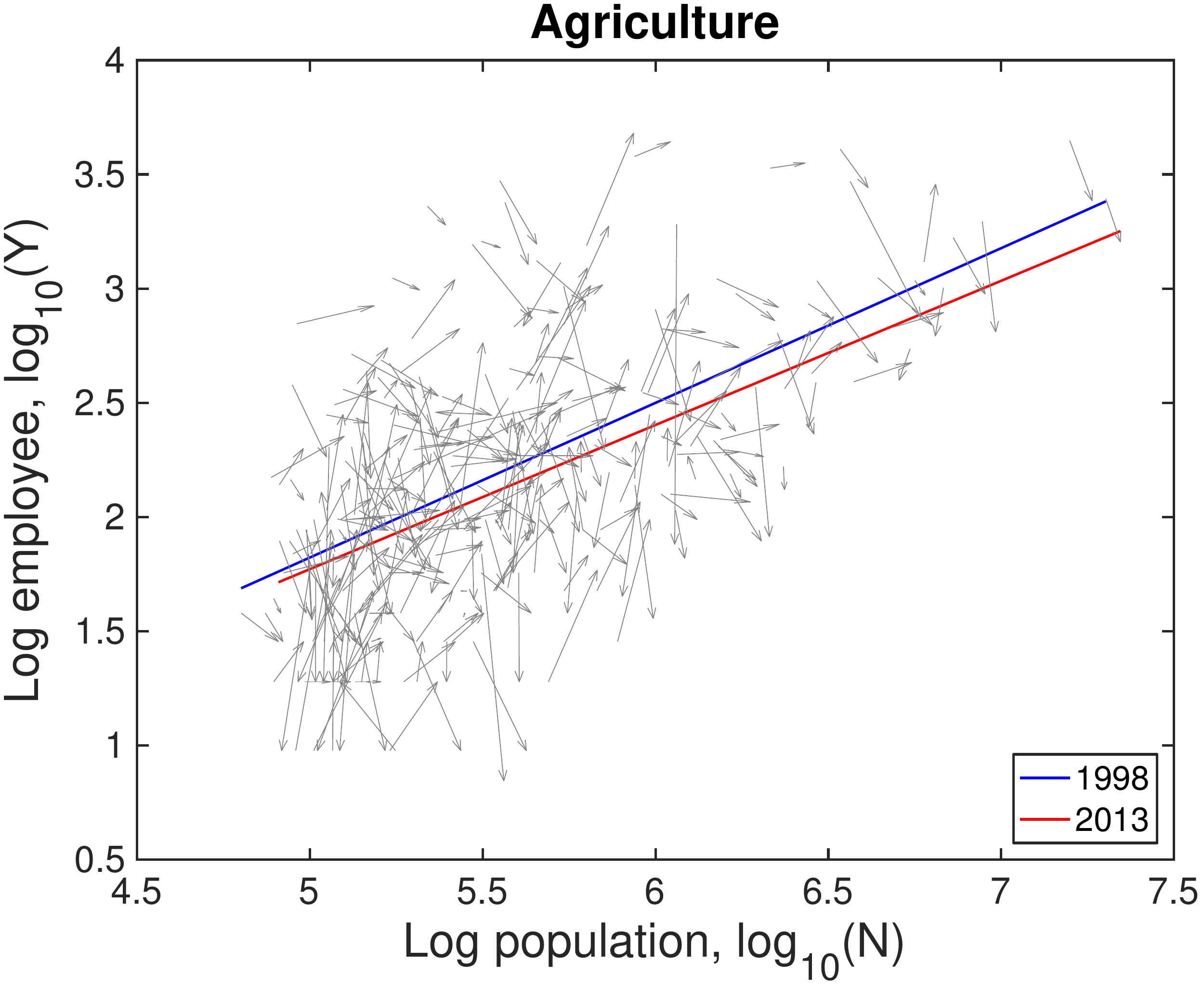}}\label{fig:tra_agri}\hspace{0.1cm}
	\subfloat{\includegraphics[width=.23\textwidth]{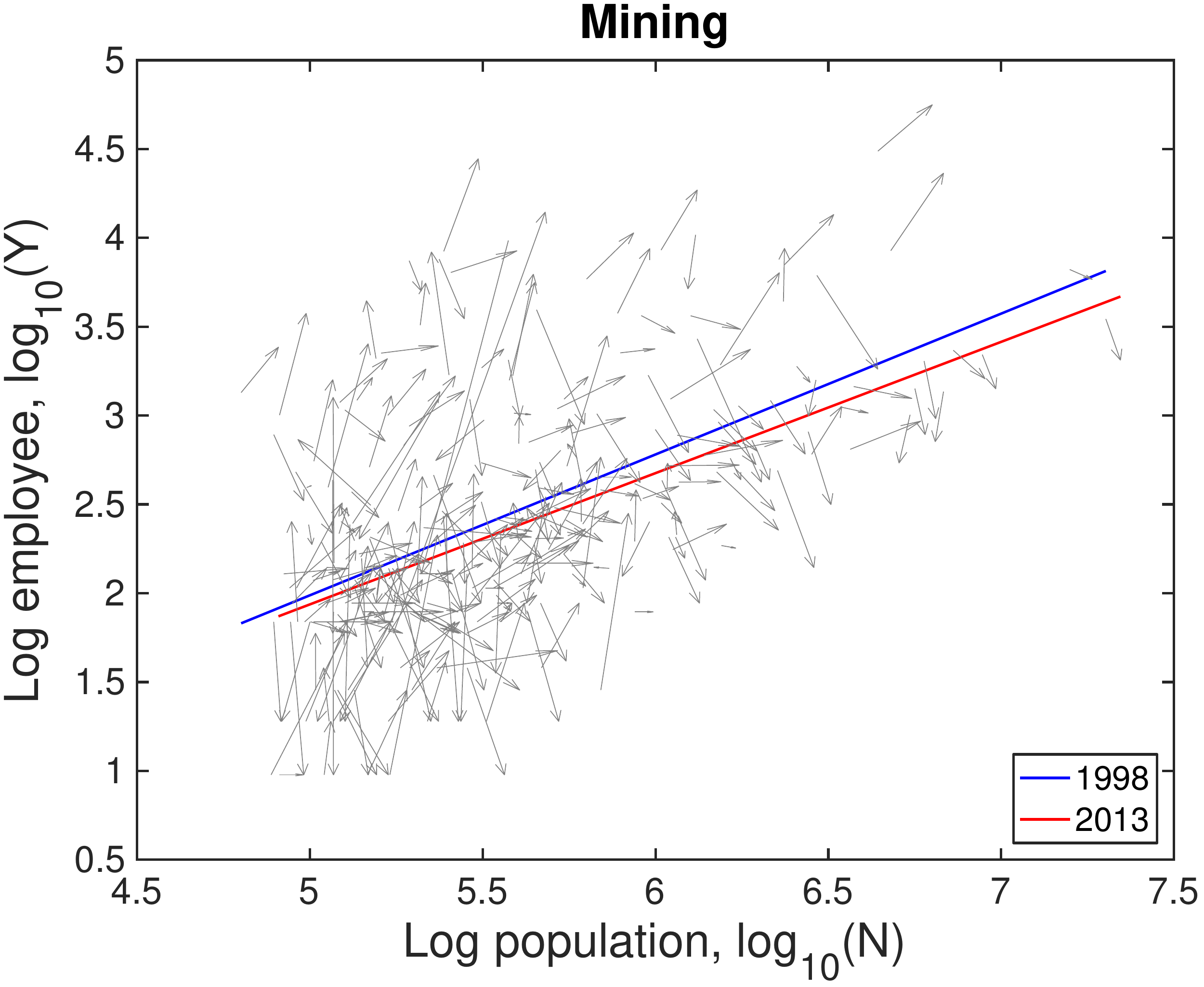}}\label{fig:tra_mining}\hspace{0.1cm}
	\subfloat{\includegraphics[width=.23\textwidth]{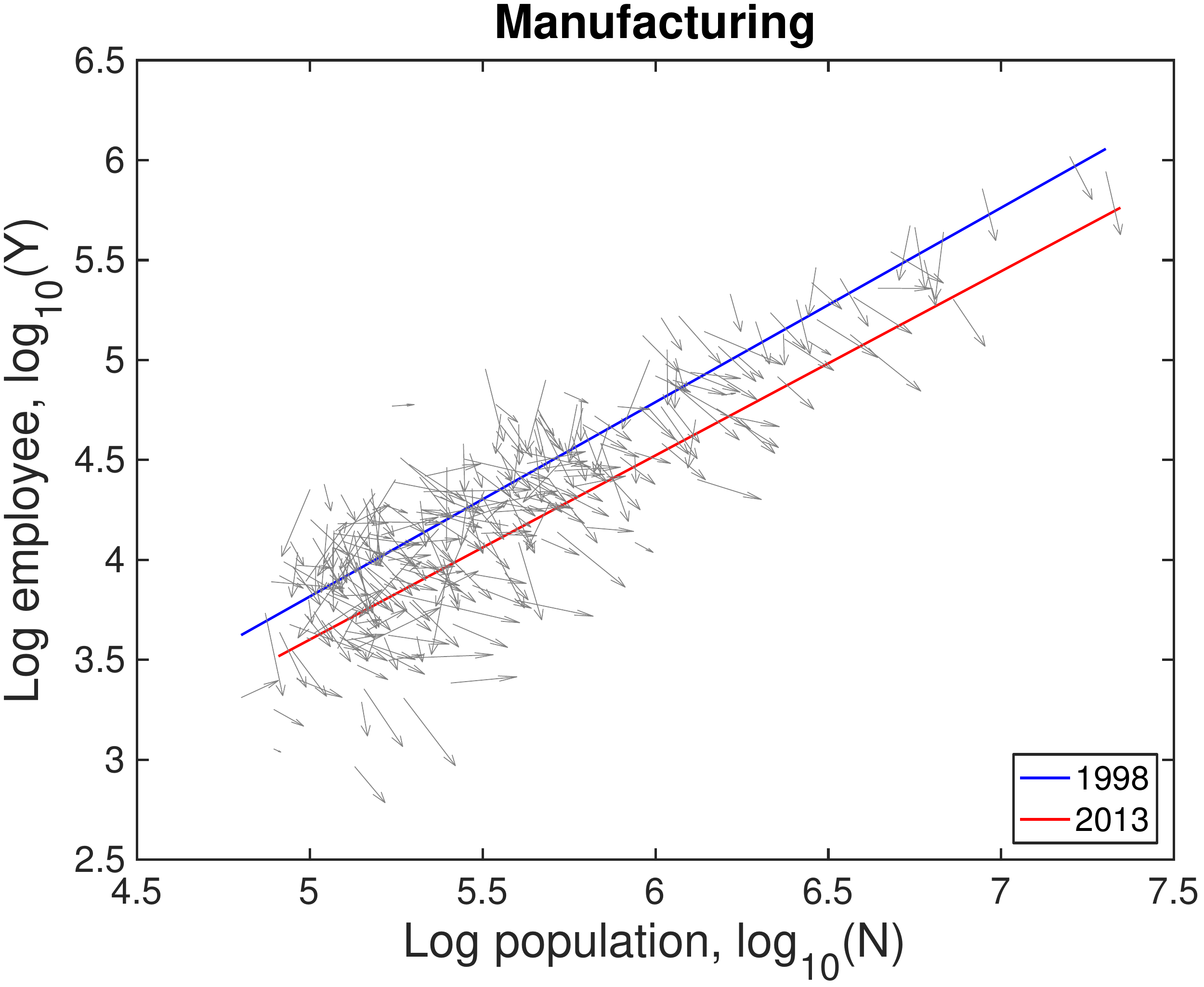}}\label{fig:tra_manufac}\hspace{0.1cm}
	\subfloat{\includegraphics[width=.23\textwidth]{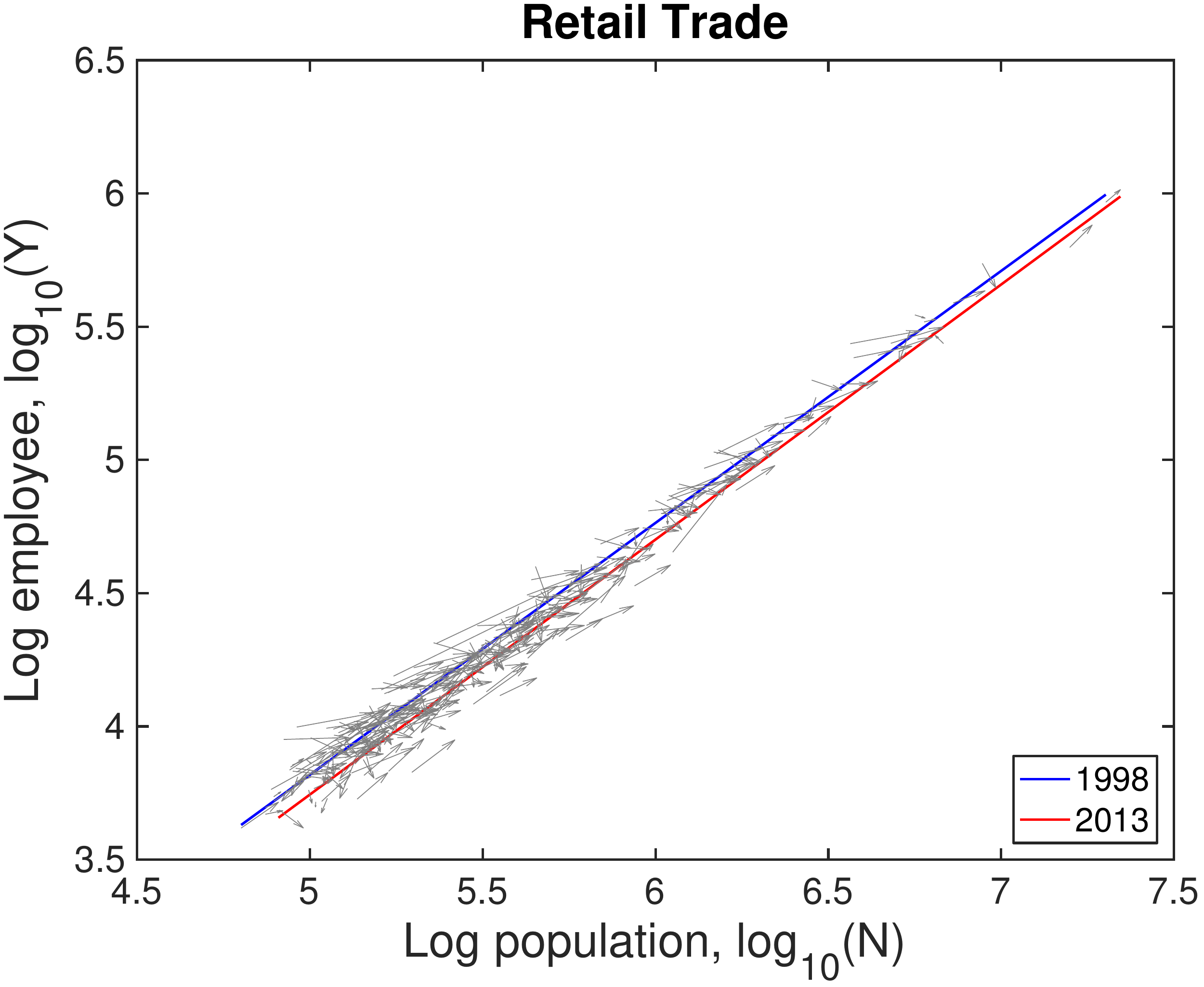}}\label{fig:tra_retail}
	\\
	\subfloat{\includegraphics[width=.23\textwidth]{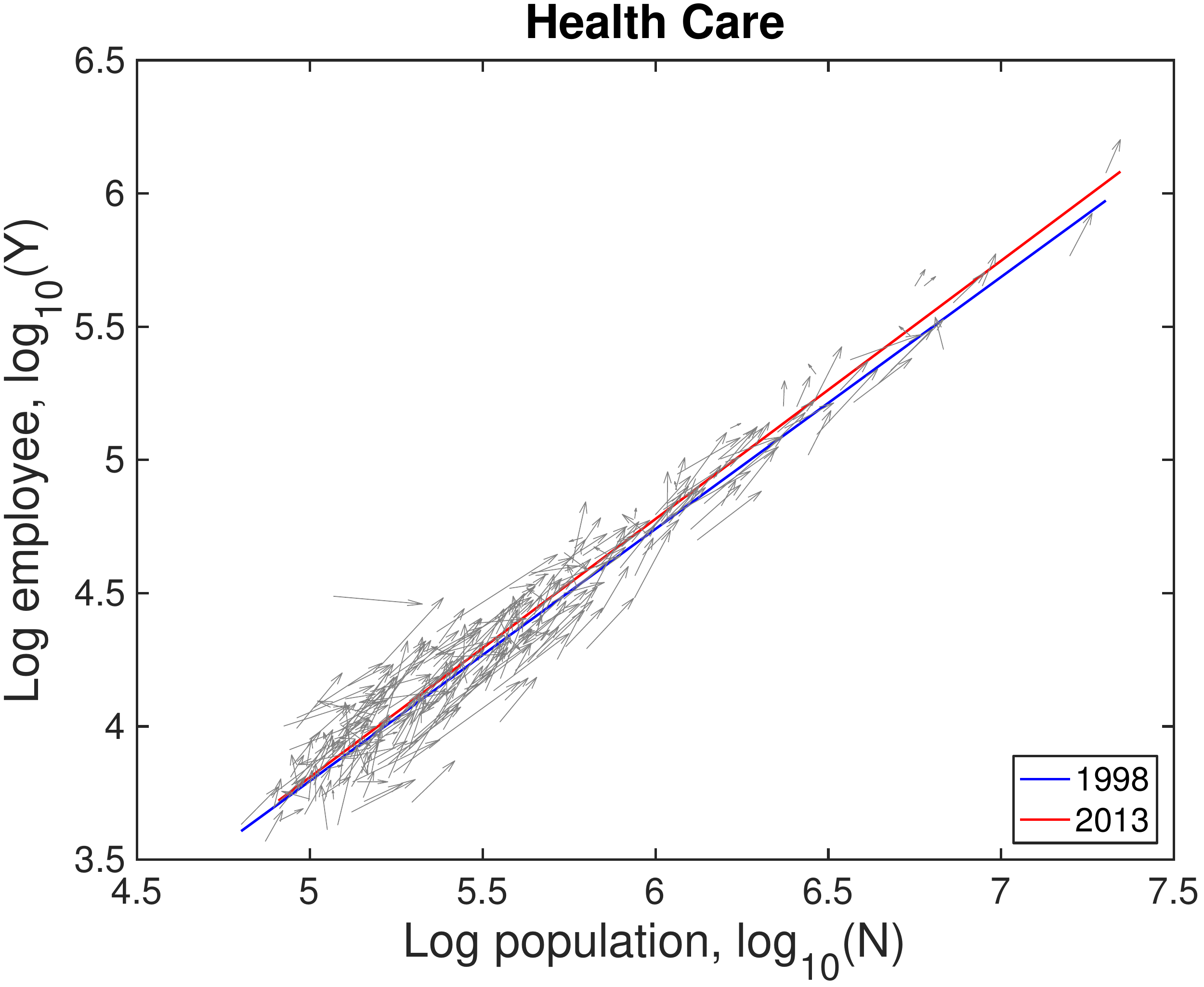}}\label{fig:tra_health}\hspace{0.1cm}
	\subfloat{\includegraphics[width=.23\textwidth]{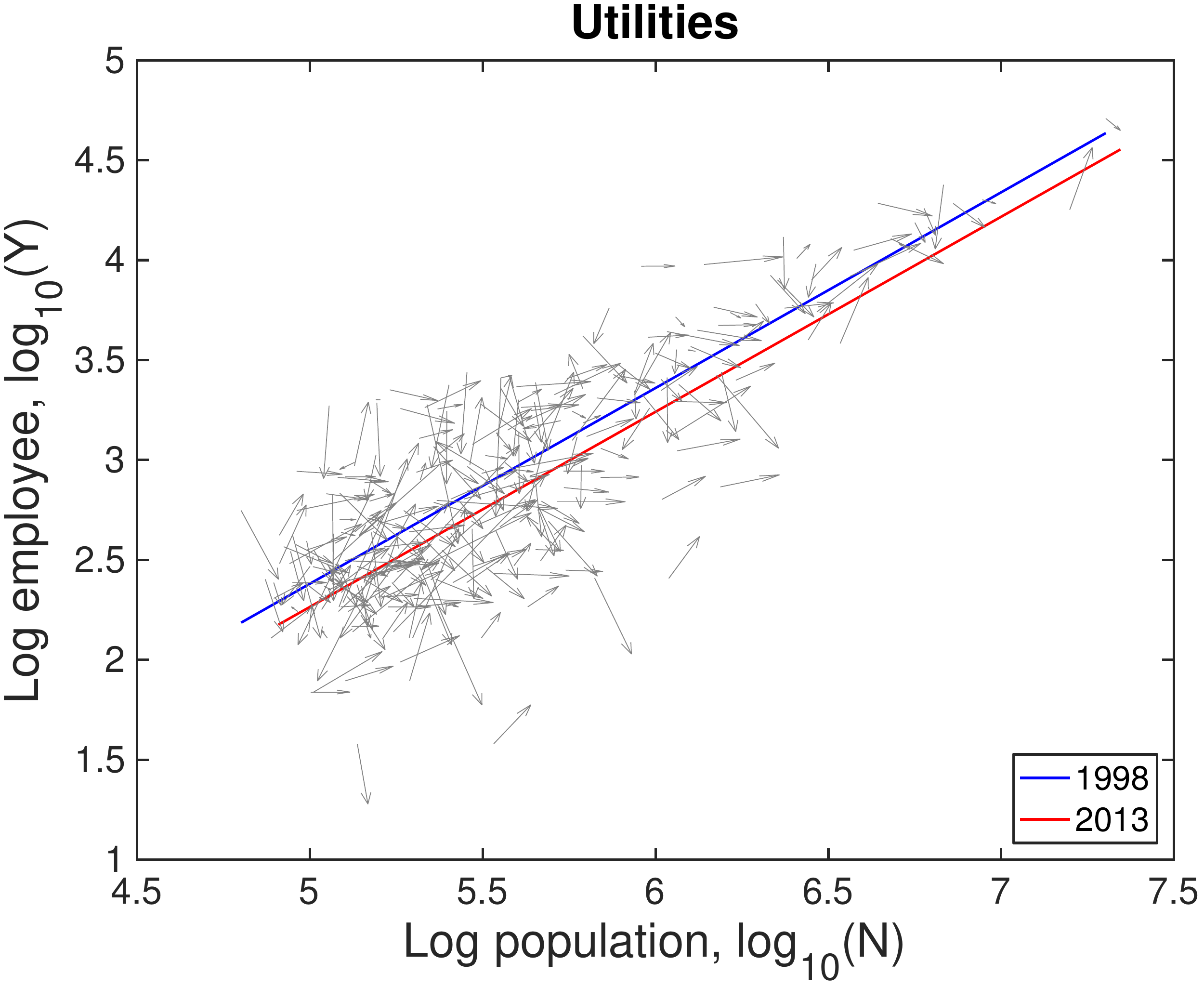}}\label{fig:tra_utilities}\hspace{0.1cm}		
	\subfloat{\includegraphics[width=.23\textwidth]{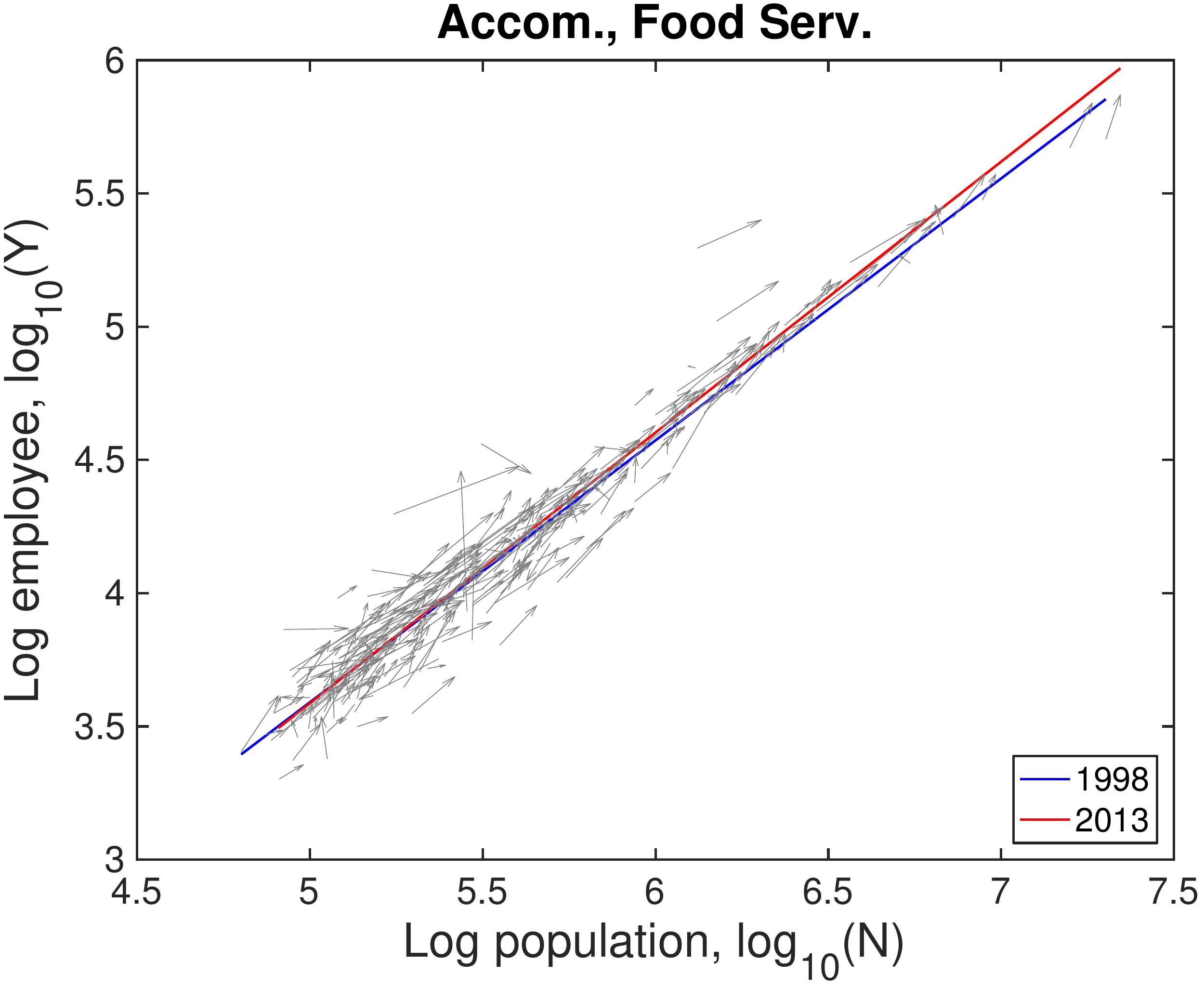}}\label{fig:tra_accomo}\hspace{0.1cm}
	\subfloat{\includegraphics[width=.23\textwidth]{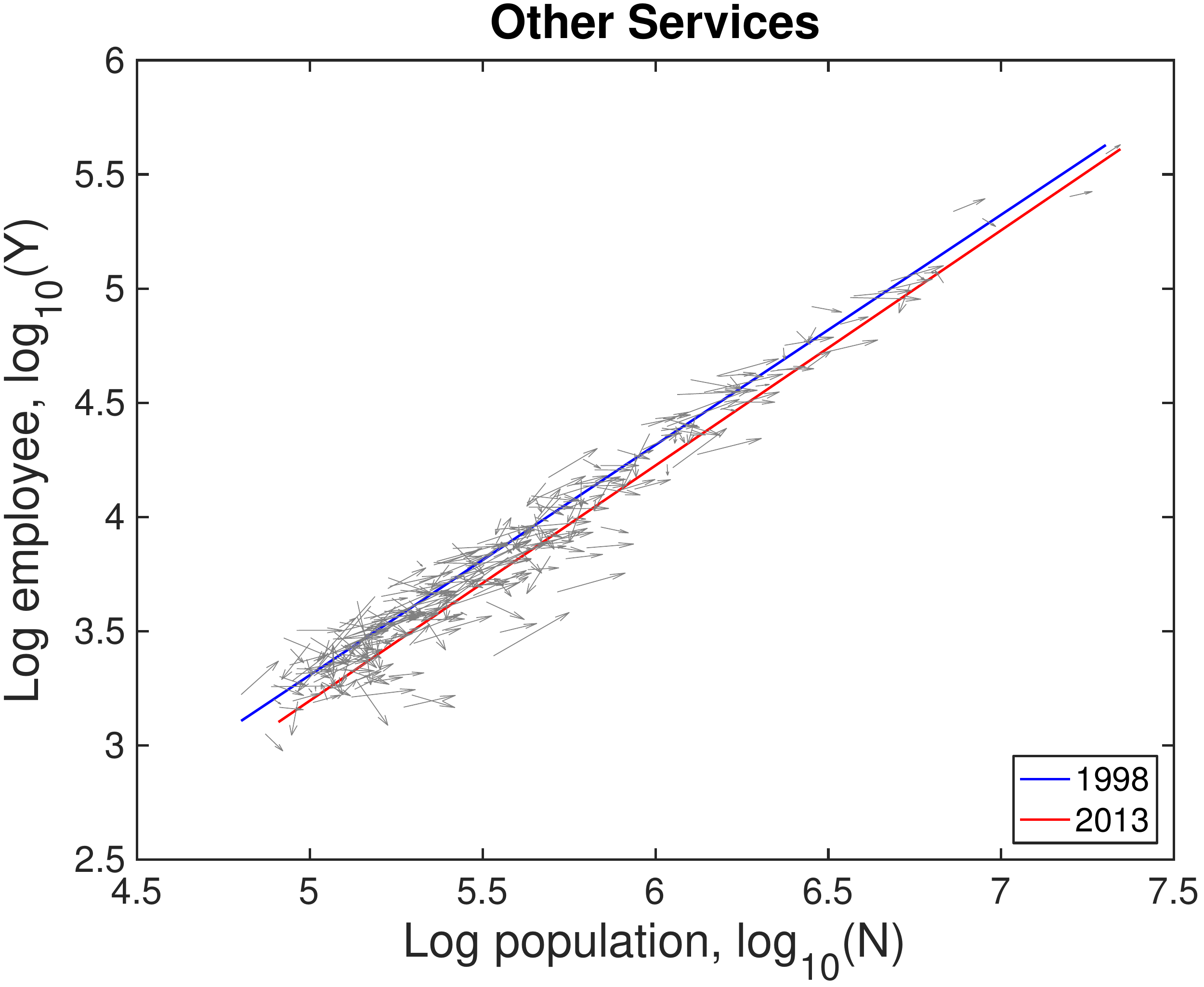}}\label{fig:tra_other}
	\\
	\subfloat{\includegraphics[width=.23\textwidth]{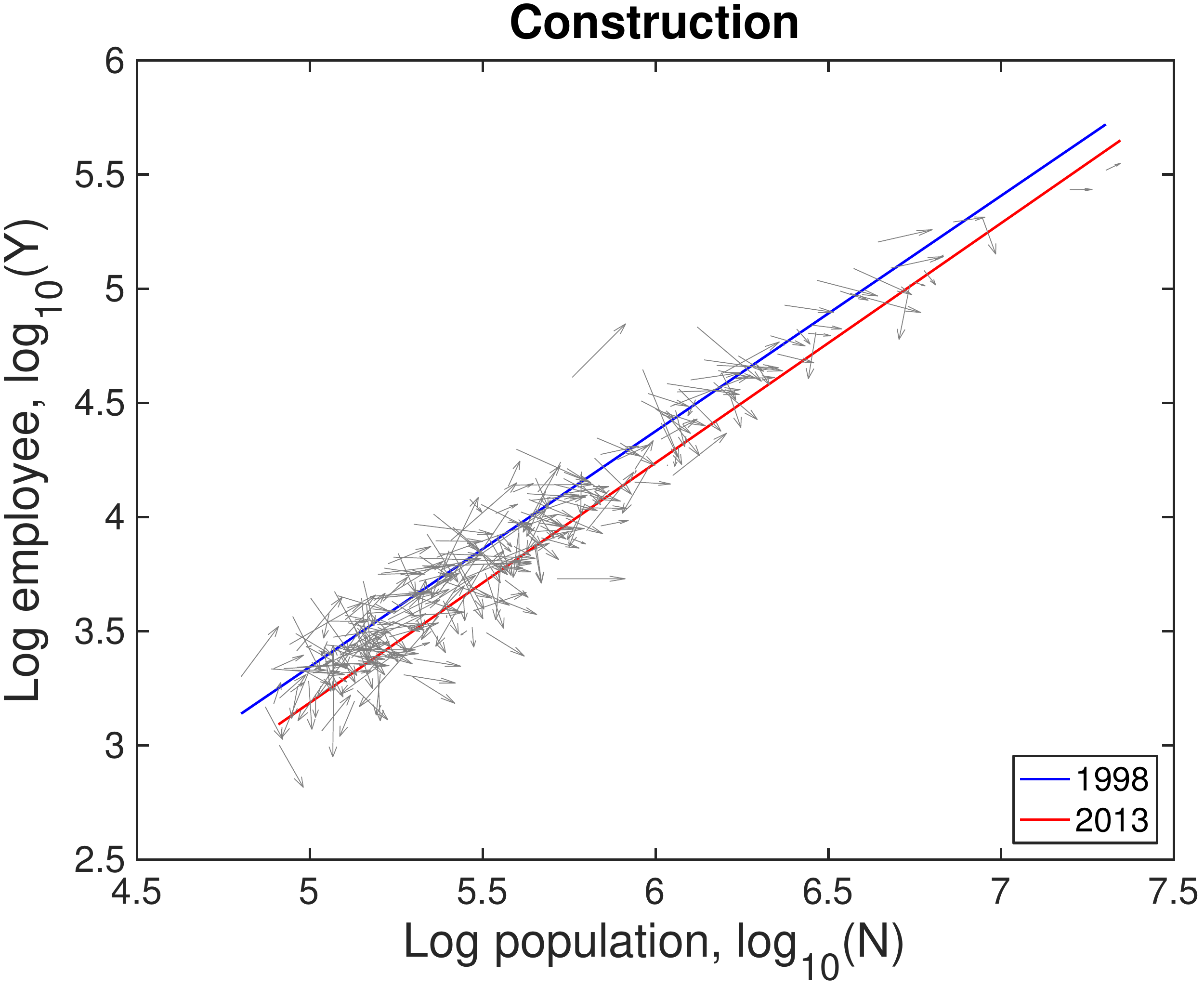}}\label{fig:tra_construc}\hspace{0.1cm}
	\subfloat{\includegraphics[width=.23\textwidth]{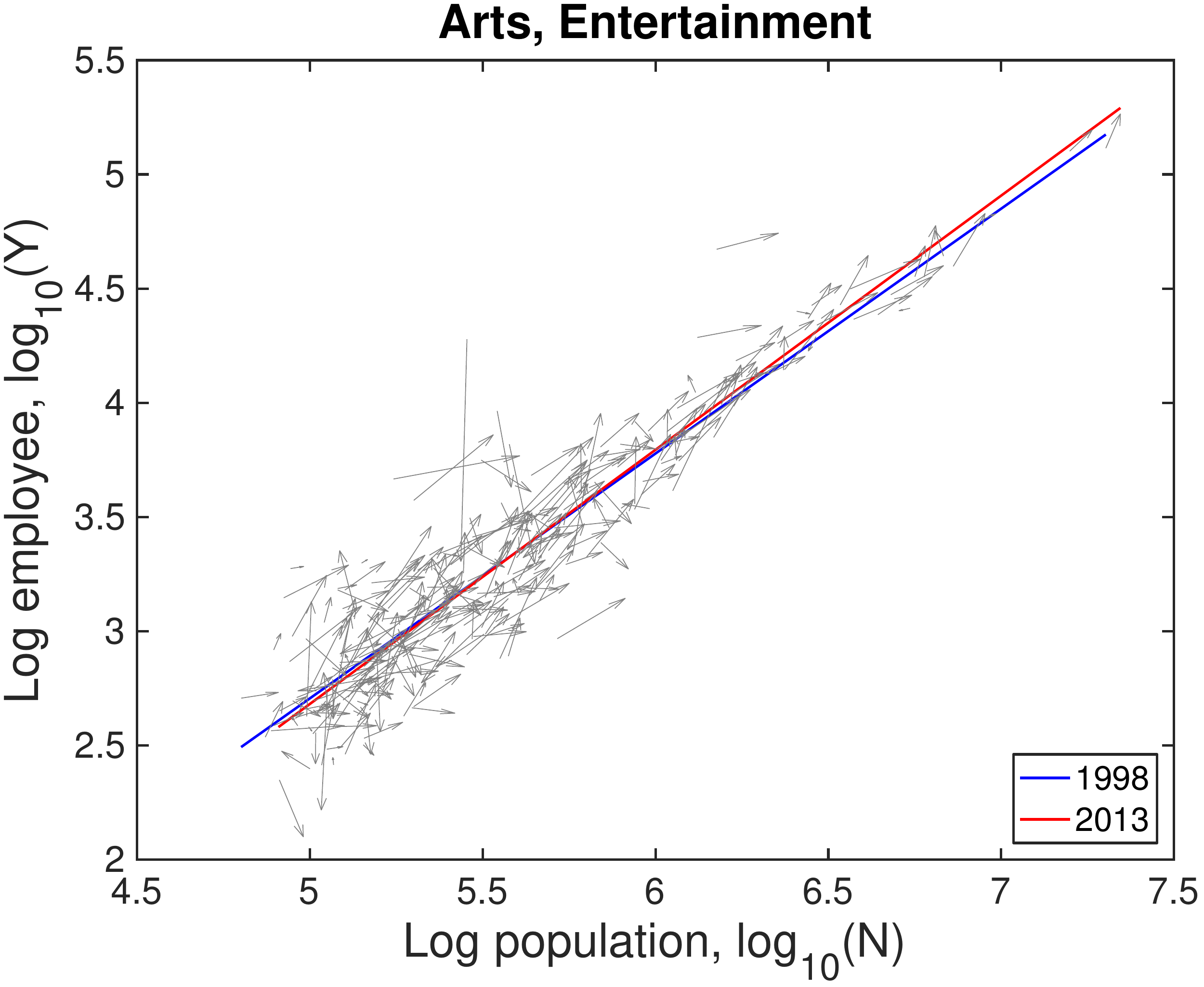}}\label{fig:tra_art}\hspace{0.1cm}
	\subfloat{\includegraphics[width=.23\textwidth]{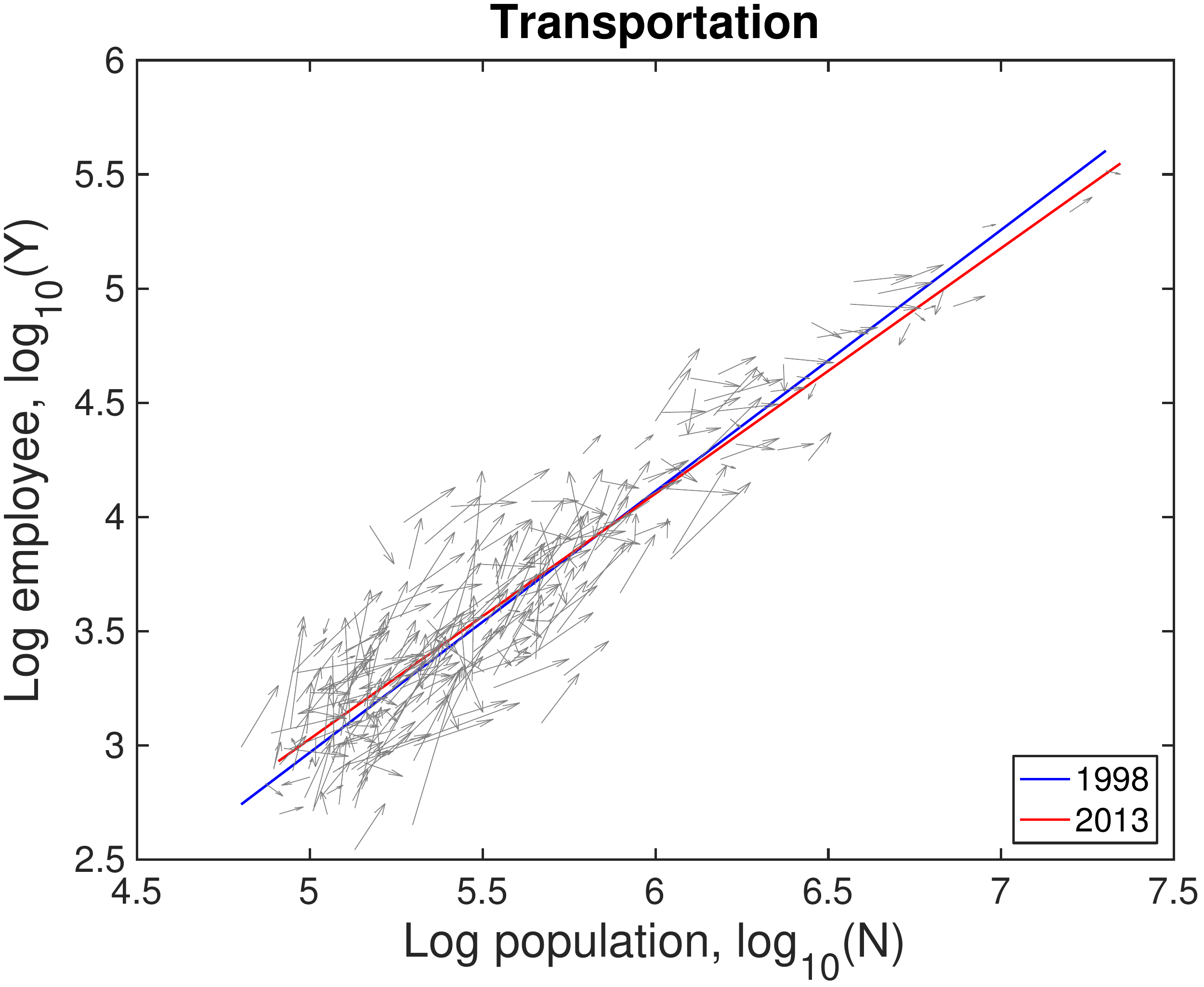}}\label{fig:tra_transp}\hspace{0.1cm}
	\subfloat{\includegraphics[width=.23\textwidth]{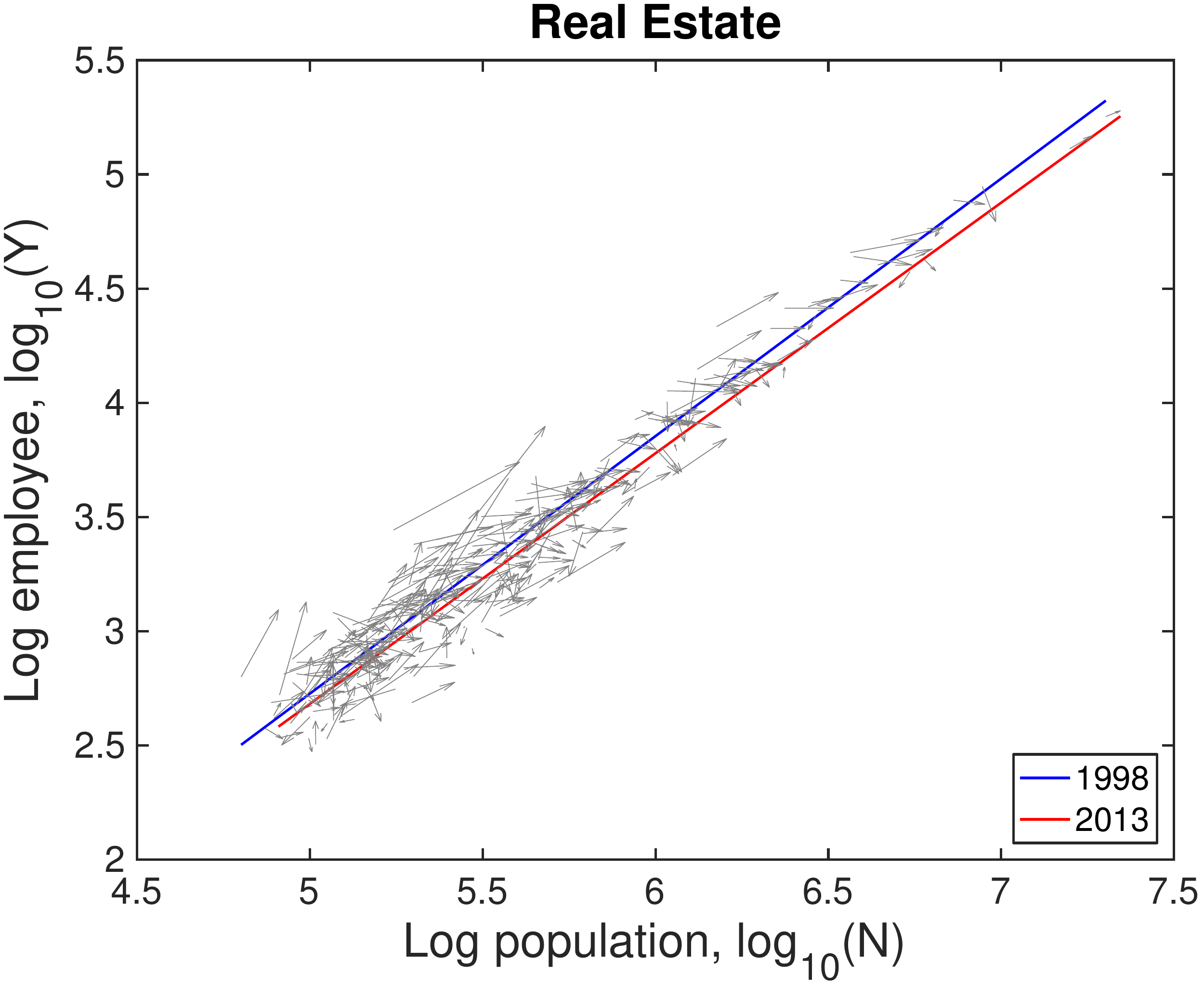}}\label{fig:tra_real}
	\\
	\subfloat{\includegraphics[width=.23\textwidth]{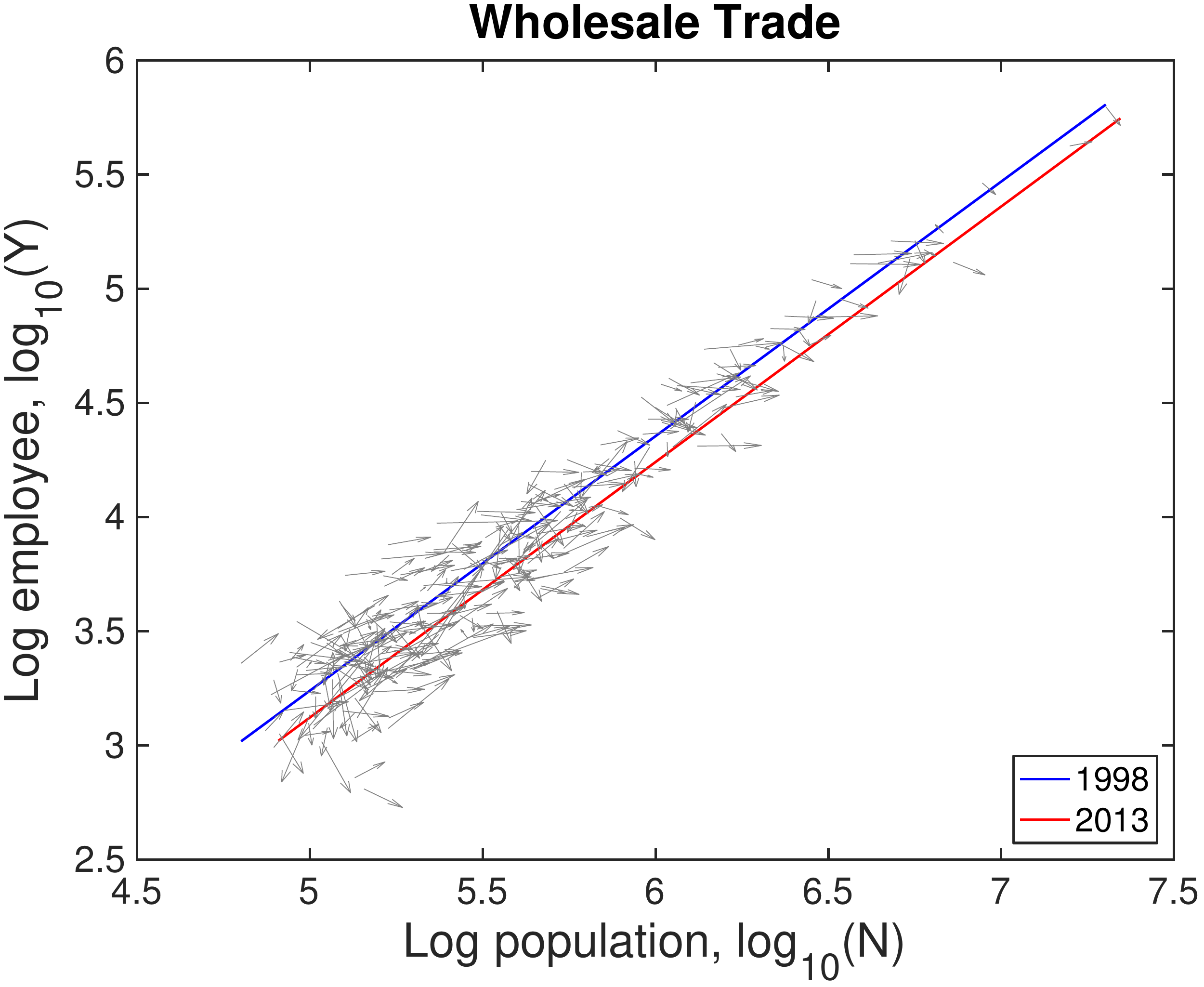}}\label{fig:tra_wholesale}\hspace{0.1cm}
	\subfloat{\includegraphics[width=.23\textwidth]{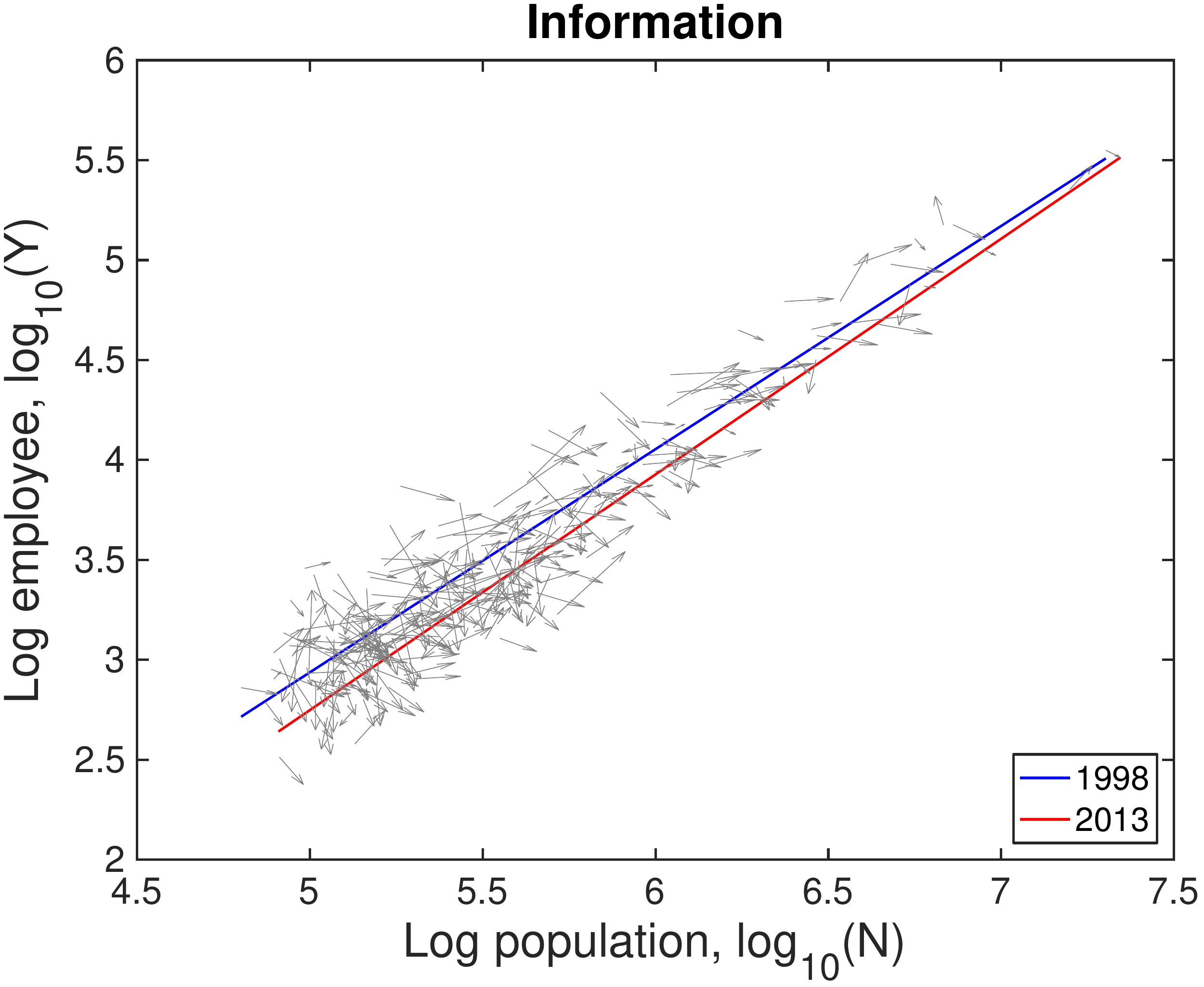}}\label{fig:tra_infor}\hspace{0.1cm}
	\subfloat{\includegraphics[width=.23\textwidth]{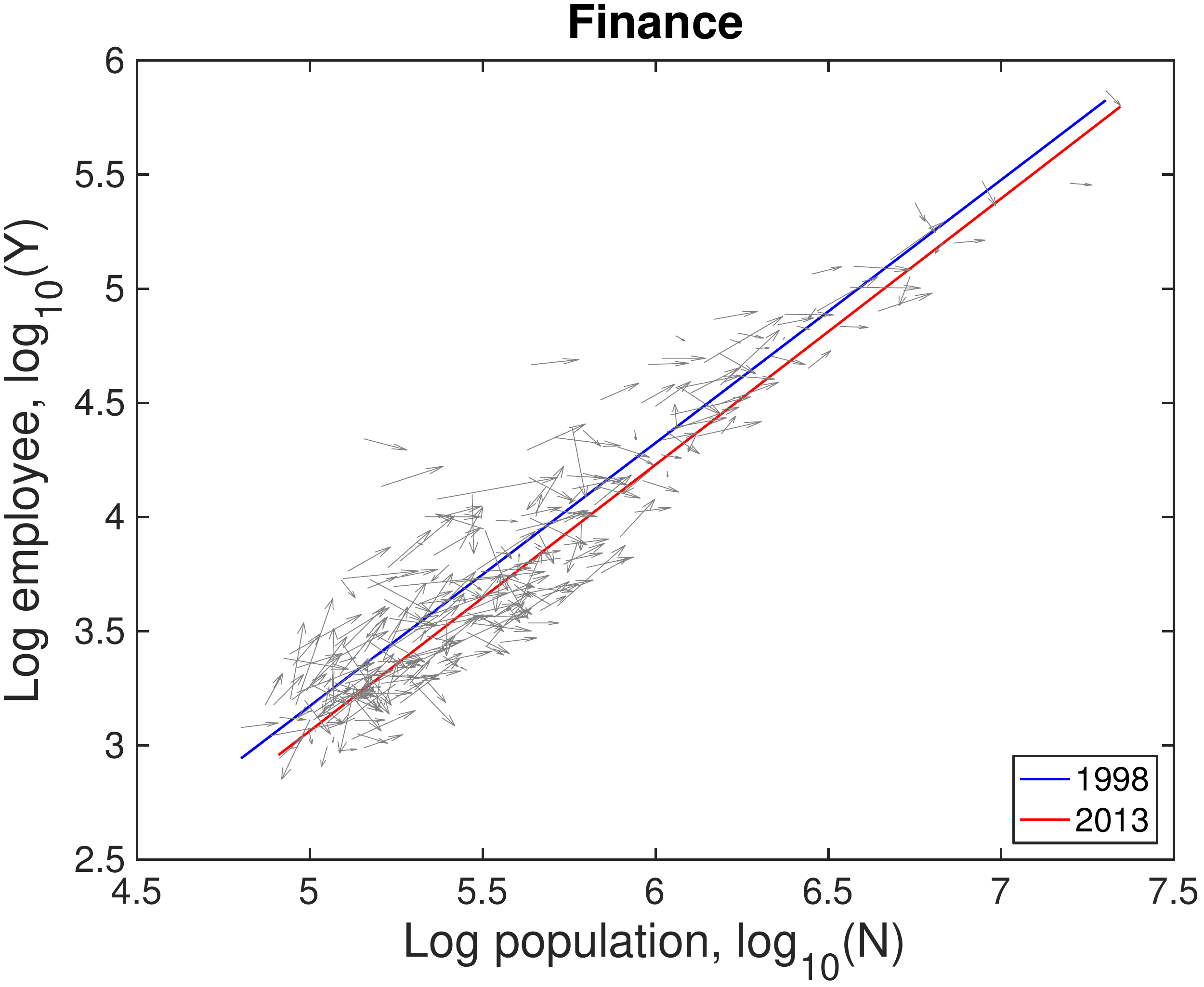}}\label{fig:tra_finan}\hspace{0.1cm}
	\subfloat{\includegraphics[width=.23\textwidth]{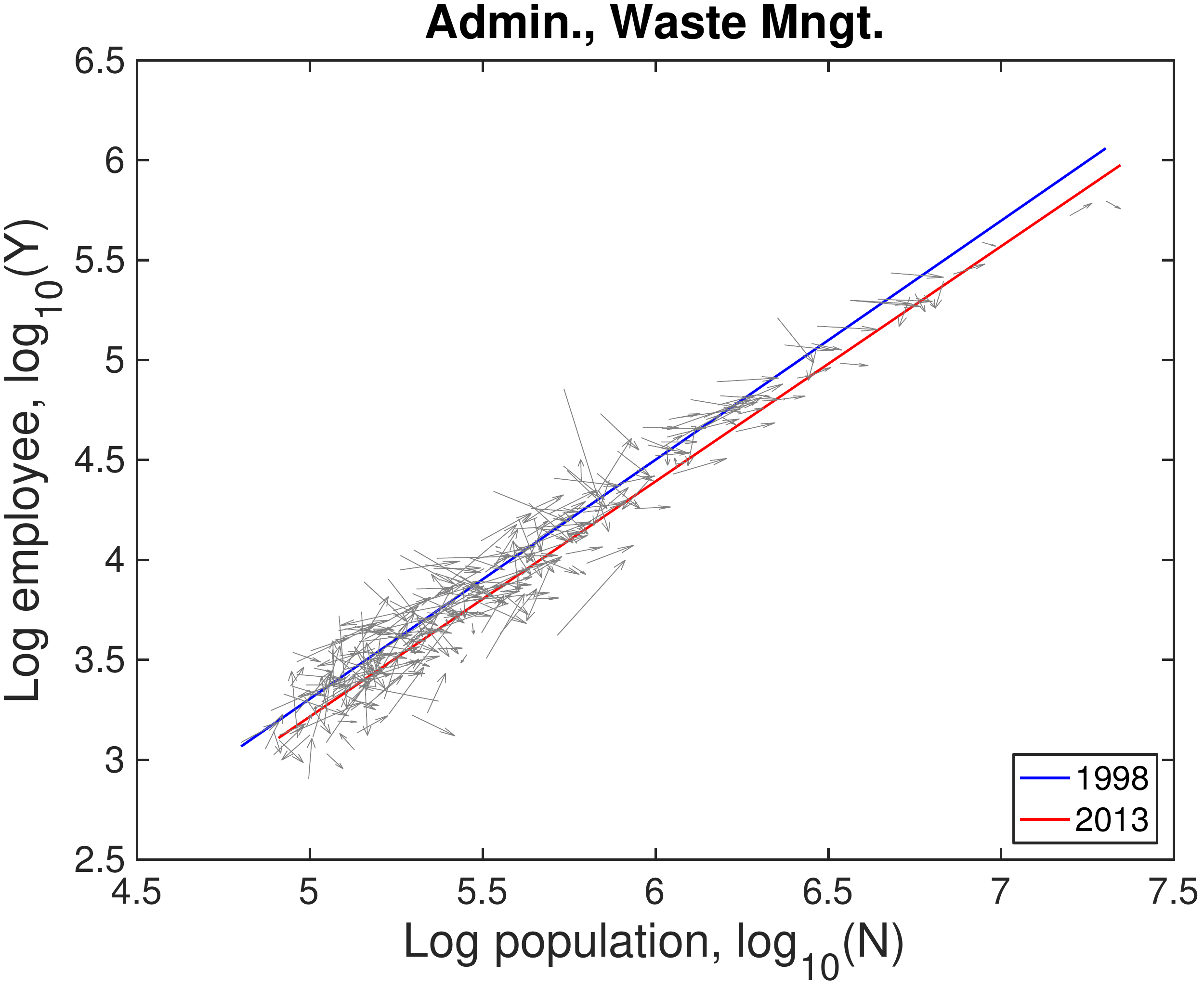}}\label{fig:tra_admin}
	\\
	\subfloat{\includegraphics[width=.23\textwidth]{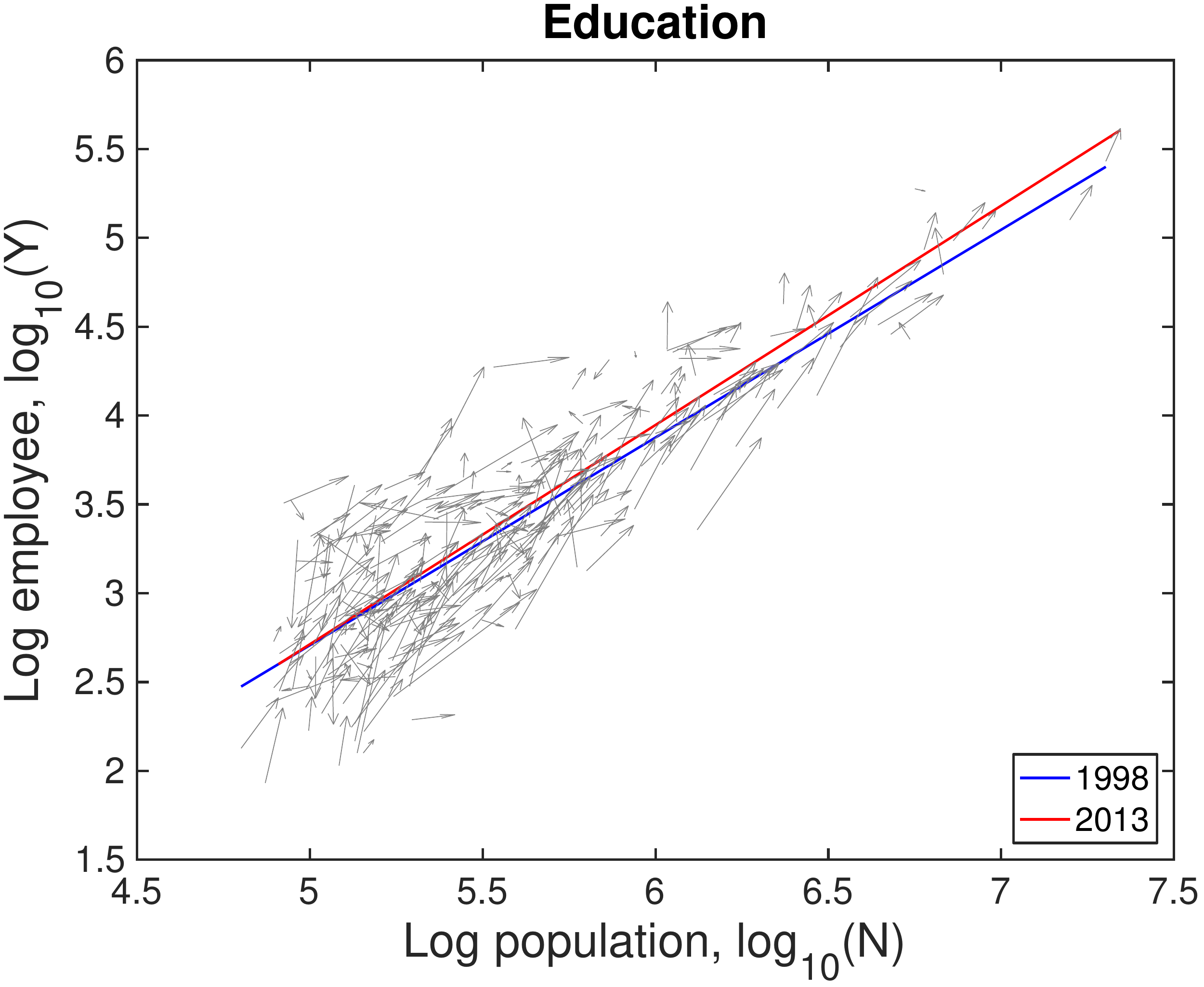}}\label{fig:tra_edu}\hspace{0.1cm}
	\subfloat{\includegraphics[width=.23\textwidth]{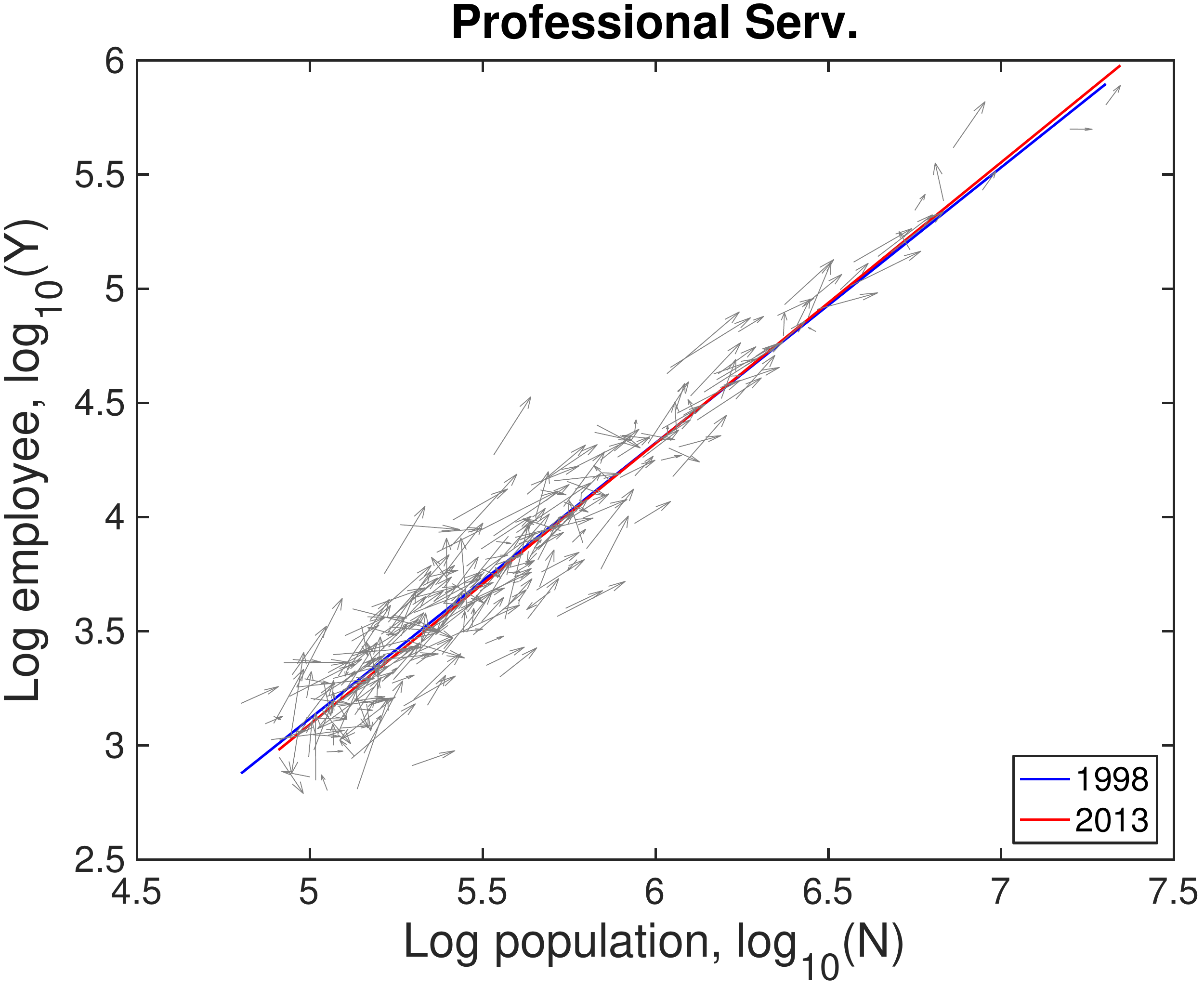}}\label{fig:tra_prof}\hspace{0.1cm}
	\subfloat{\includegraphics[width=.23\textwidth]{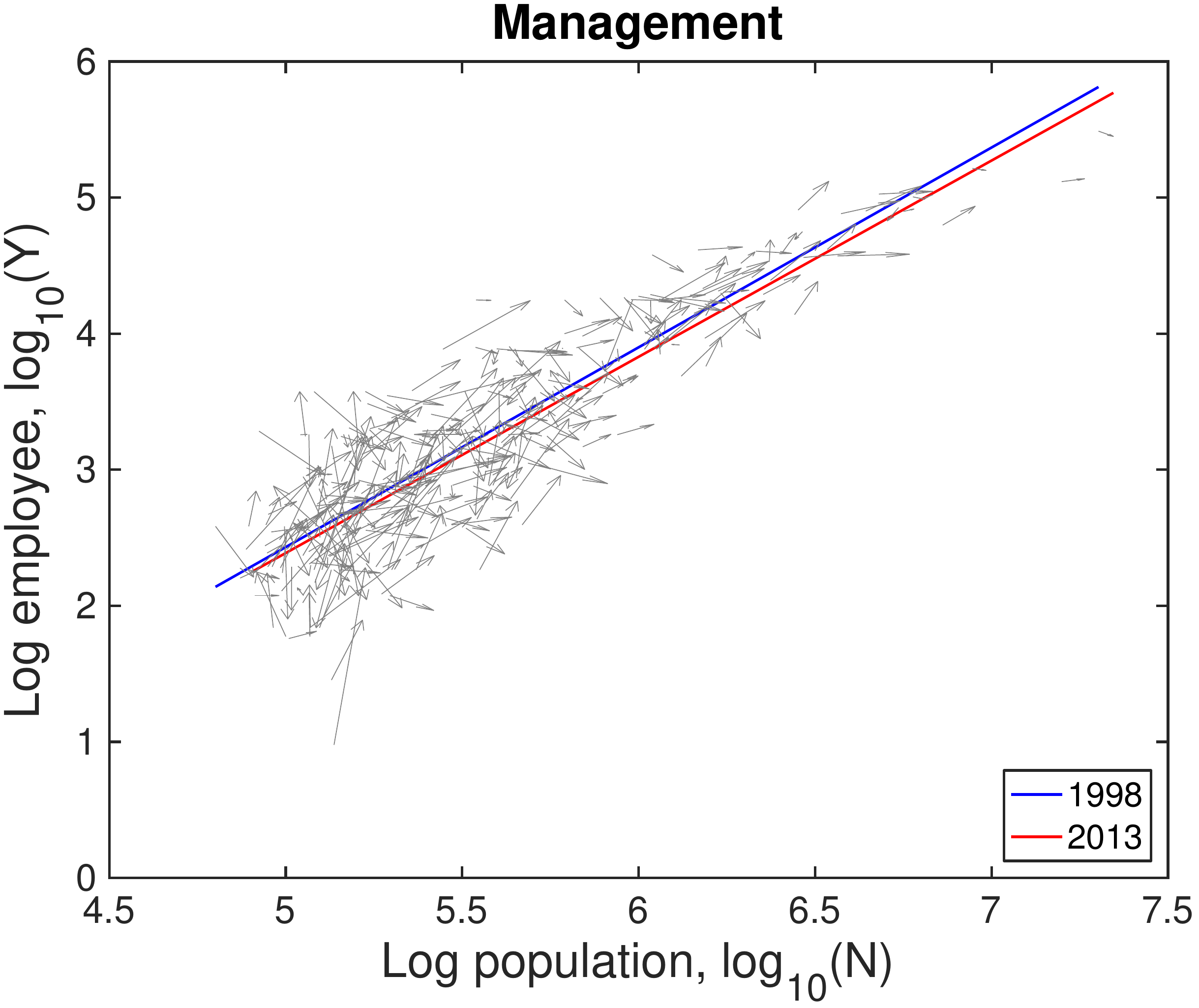}}\label{fig:tra_manage}\hspace{0.1cm}
	\hbox to .23\textwidth{}
	\\
	\caption{
	Trajectory of each city on the population-industry plane in the order of increasing scaling exponent. The arrows depict the change of population and industry size of each city from 1998 to 2013. The fitting lines of scaling relation are colored as blue for 1998 and red for 2013.} 
	\label{fig:scaling_trajectory}
\end{figure*}

In Fig.~\ref{fig:scaling_trajectory}, each city moves on the population-industry size space as time proceeds, and their trend can be described by the scaling relation. In many cases such as ``Manufacturing'', the fitting line in 2013 is at the bottom-right side of the fitting line in 1998 with little change of slope. Does it mean decrease in industry size contradicting to the overall increase in Fig.~\ref{fig:data_timeseries}? It depends on the ratio of relative industry growth to relative population growth, and it does not contradict to the overall increase. Let us see the details with mathematical formulation.

As a fitting line is obtained for the scaling relation in the log-log scale, this shift of the fitting line can be caused by a uniform relative change of populations or industry sizes. Let these relative changes of populations and industry sizes be $\Delta\log{N}(c)$ and $\Delta\log{Y}(c,i)$ for city $c$ and industry $i$. For convention, let us consider only the change between the initial year, 1998, and the last year, 2013, in our scope.
\begin{flalign}
	\Delta\log{N}(c) &\simeq \frac{\Delta N(c)}{N(c)} = \frac{N(c, 2013)-N(c, 1998)}{N(c, 1998)}\\
	\Delta\log{Y}(c,i) &\simeq \frac{\Delta Y(c,i)}{Y(c,i)} = \frac{Y(c,i,2013)-Y(c,i,2013)}{Y(c,i,1998)}
\end{flalign} 

Since the difference of logarithmic values is equivalent to the relative change, $d(\log{N})=dN/N \simeq \Delta\log{N}(c)$, the fitting line of scaling relation may shift in the horizontal direction when every city undergoes a uniform relative population growth, $\Delta\log{N}$. In the same way, the fitting line moves in the vertical direction when every city undergoes a uniform relative industry growth, $\Delta\log{Y}(i)$. The fitting line moves to a lower side when the population growth is not followed by a sufficient industry growth. To stay in the current position, the fitting line should move vertically in $\Delta\log{Y}(i)=\beta(i)\Delta\log{N}$ when it moves horizontally in $\Delta\log{N}$. We can also derive it from the time derivative of scaling relation in Eq.~\ref{eq:slope}. 	
\begin{flalign}
	\beta(i) = \frac{d(\log{Y}(i))}{d(\log{N})} = \frac{\Delta\log{Y}(i)}{\Delta\log{N}} \\
	\Delta\log{Y}(i) = \beta(i)\Delta\log{N}	
\end{flalign}

Therefore, the red lines under the blue lines in Fig.~\ref{fig:scaling_trajectory} do not mean the actual decrease in the industry size. It is caused by a relatively lower industry growth than the population growth, i.e., $\Delta\log{Y(i)} < \beta(i)\Delta\log{N}$. Similarly, the fitting line moving upward as in `Health Care' industry is given by a relatively higher industry growth with $\Delta\log{Y(i)} > \beta(i)\Delta\log{N}$. 

\subsection{Time-dependent scaling relation from industry size and population changes}

\begin{figure*}[th!]
	\centering
	\labelarial{A}
	\subfloat{\includegraphics[width=0.45\textwidth]{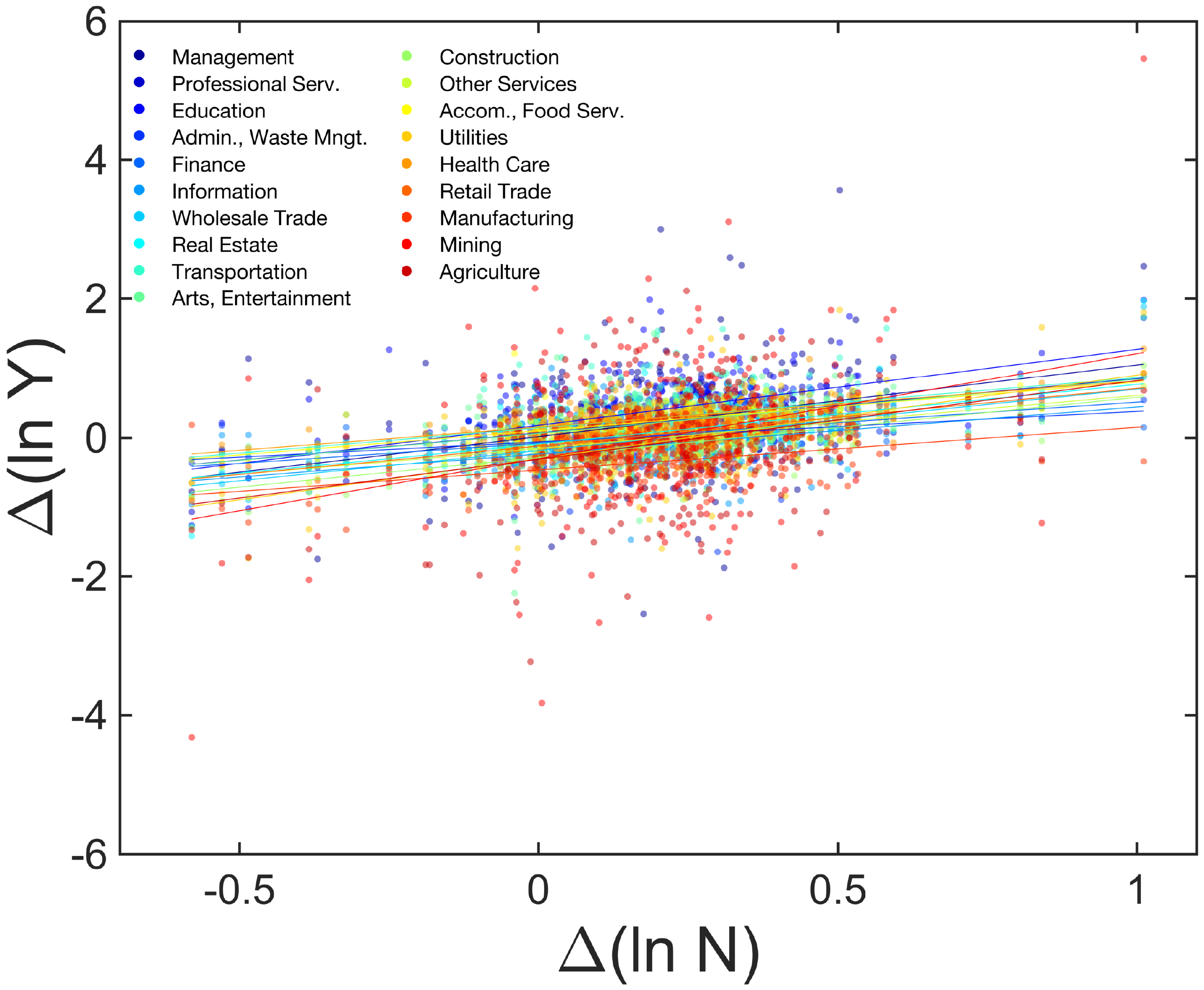}\label{fig:scaling-trend}}\hspace{0.3cm}
	\labelarial{B}
	\subfloat{\includegraphics[width=0.45\textwidth]{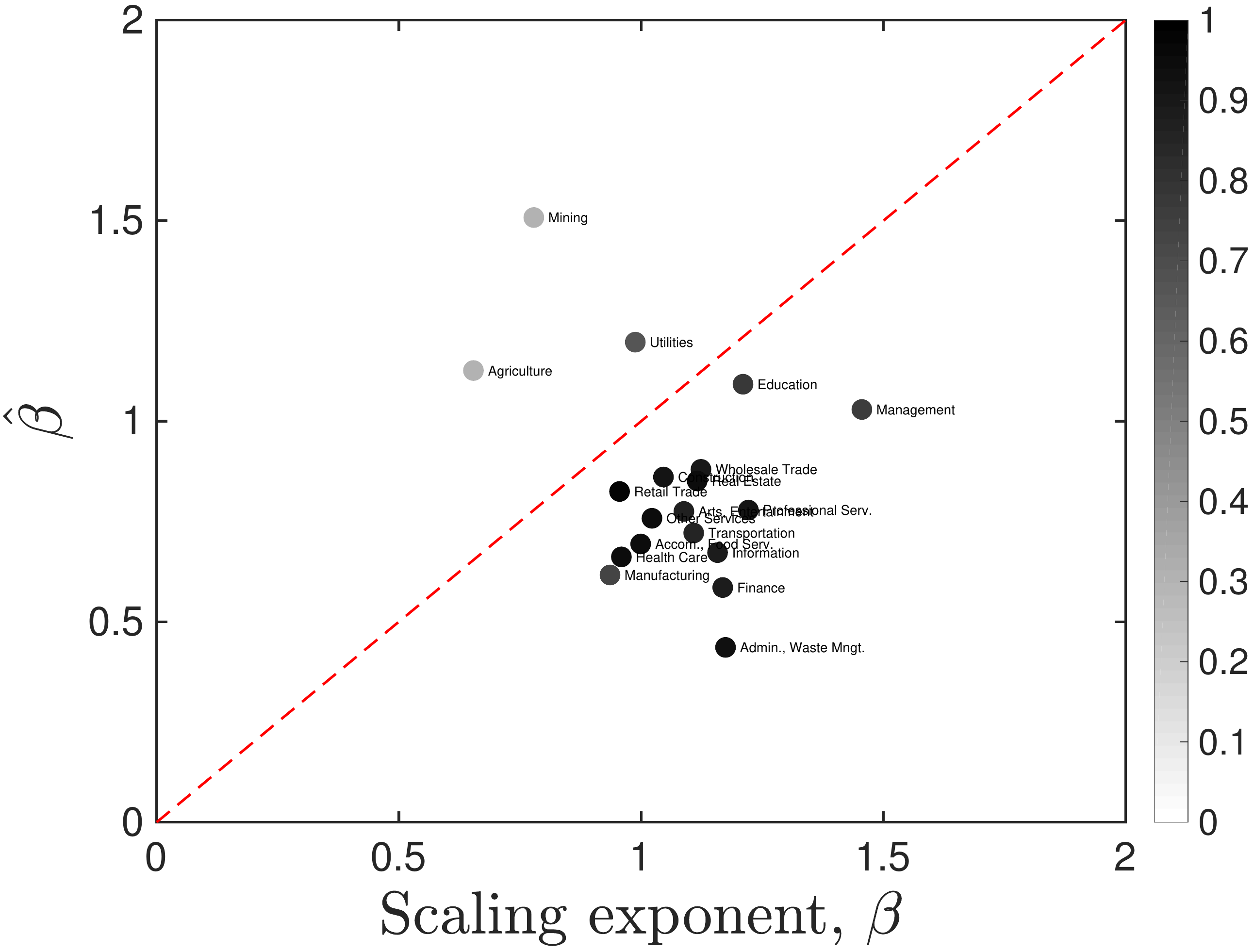}\label{fig:growth-scaling_2dig}}
	\\
	\labelarial{C}
	\subfloat{\includegraphics[width=0.45\textwidth]{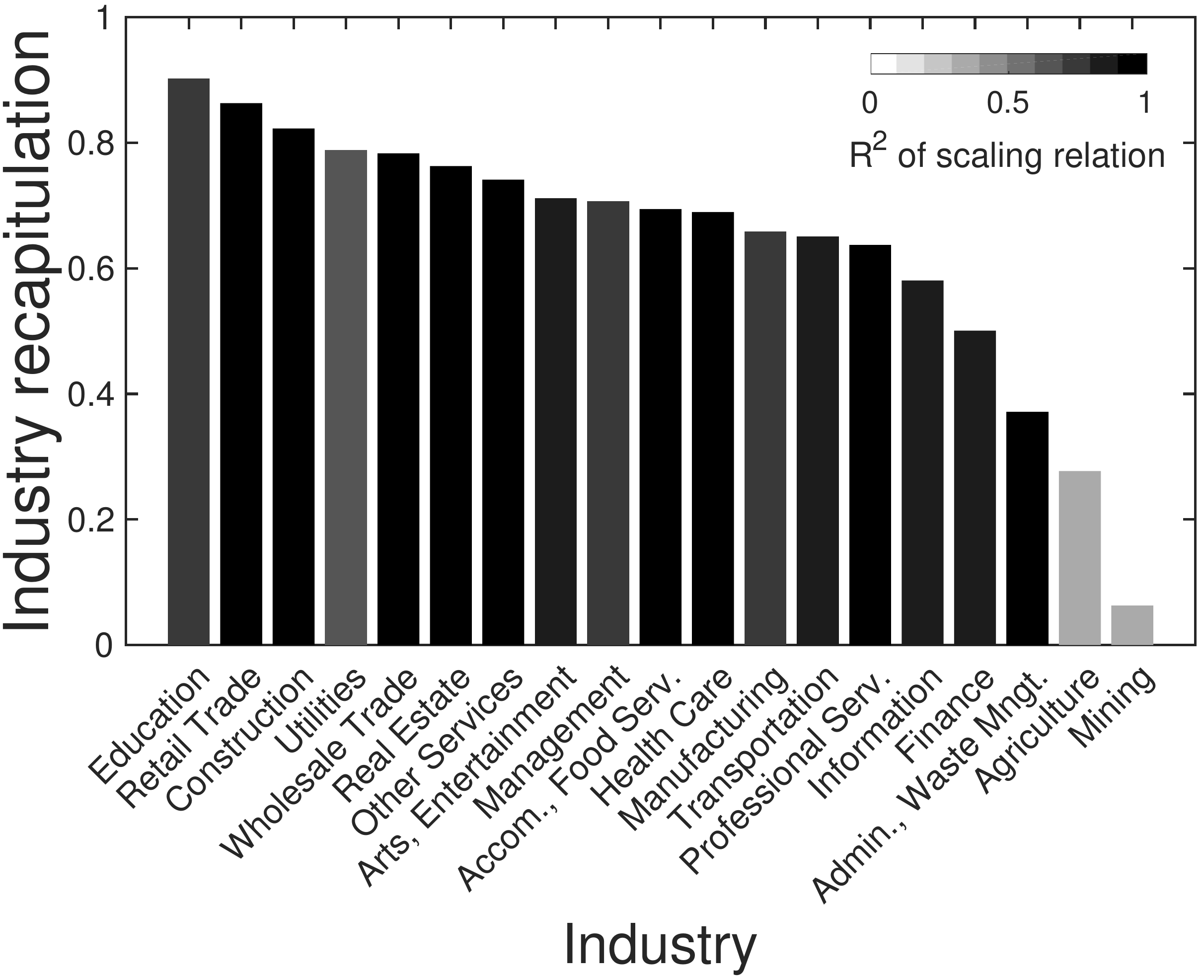}\label{fig:recap_score_N2}}\hspace{0.3cm}
	\labelarial{D}
	\subfloat{\includegraphics[width=0.45\textwidth]{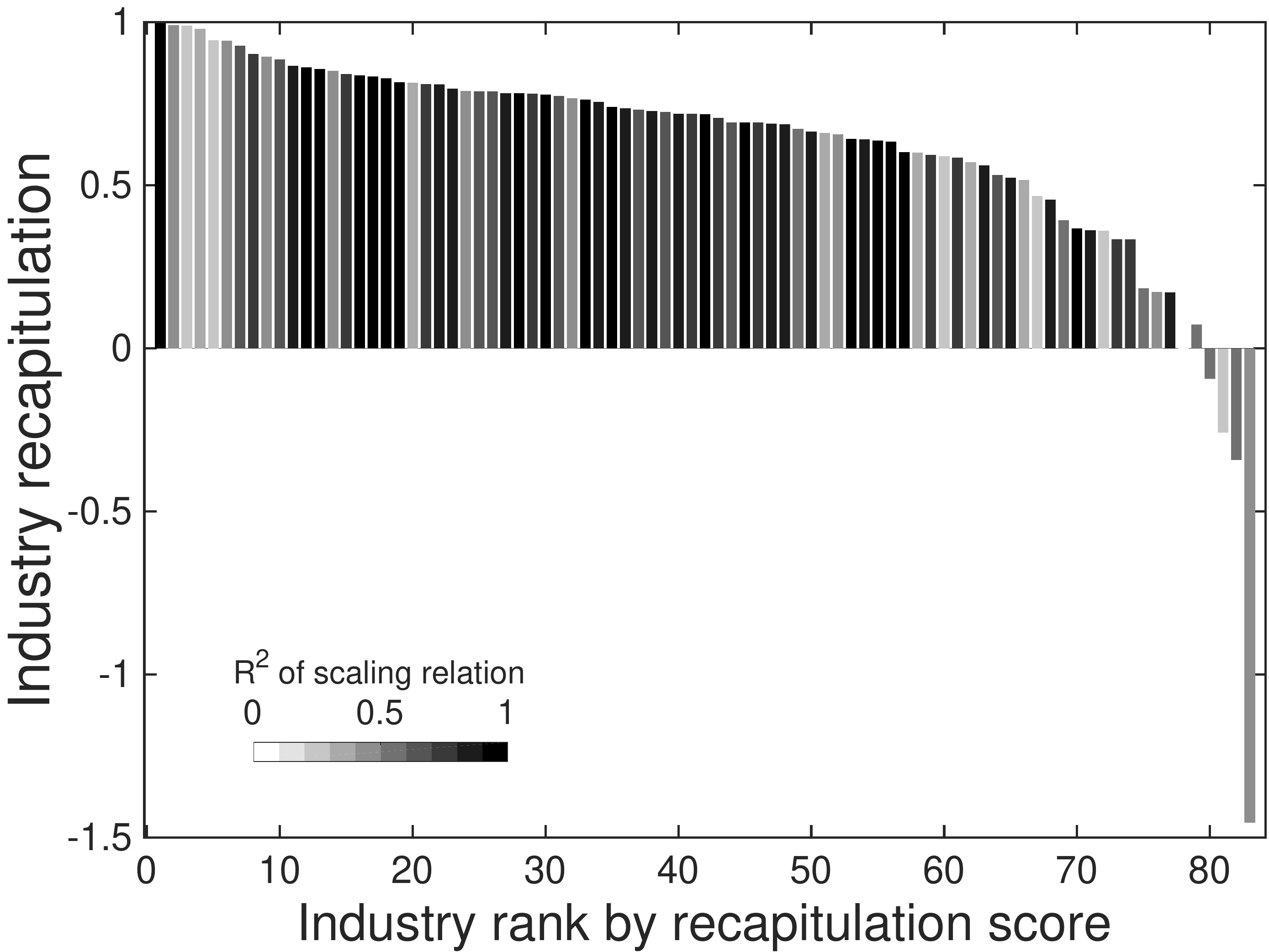}\label{fig:recap_score_N3}}
	\\
	\labelarial{E}
	\subfloat{\includegraphics[width=0.45\textwidth]{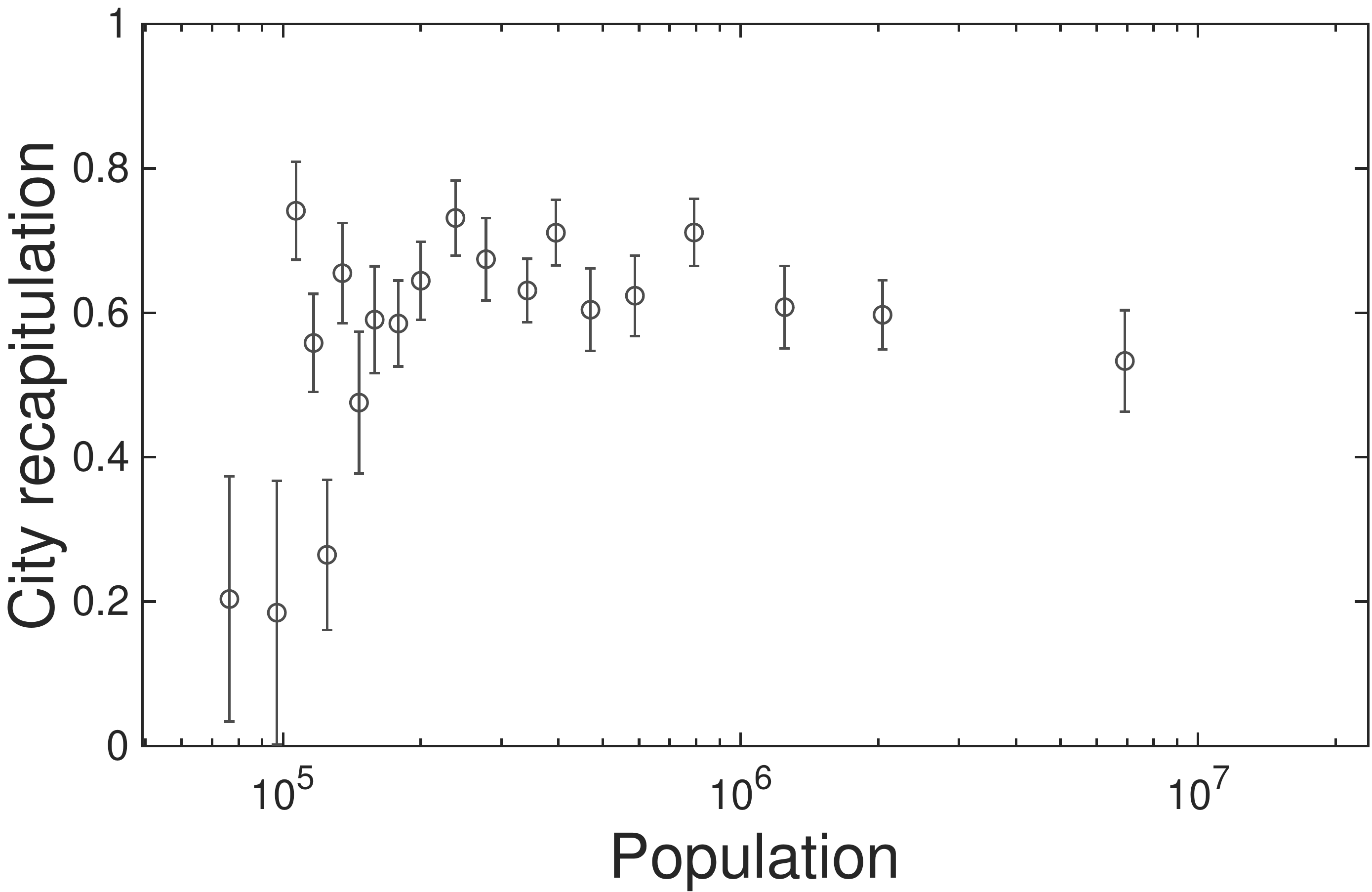}\label{fig:city_recap_score_N2}}\hspace{0.3cm}
	\labelarial{F}
	\subfloat{\includegraphics[width=0.45\textwidth]{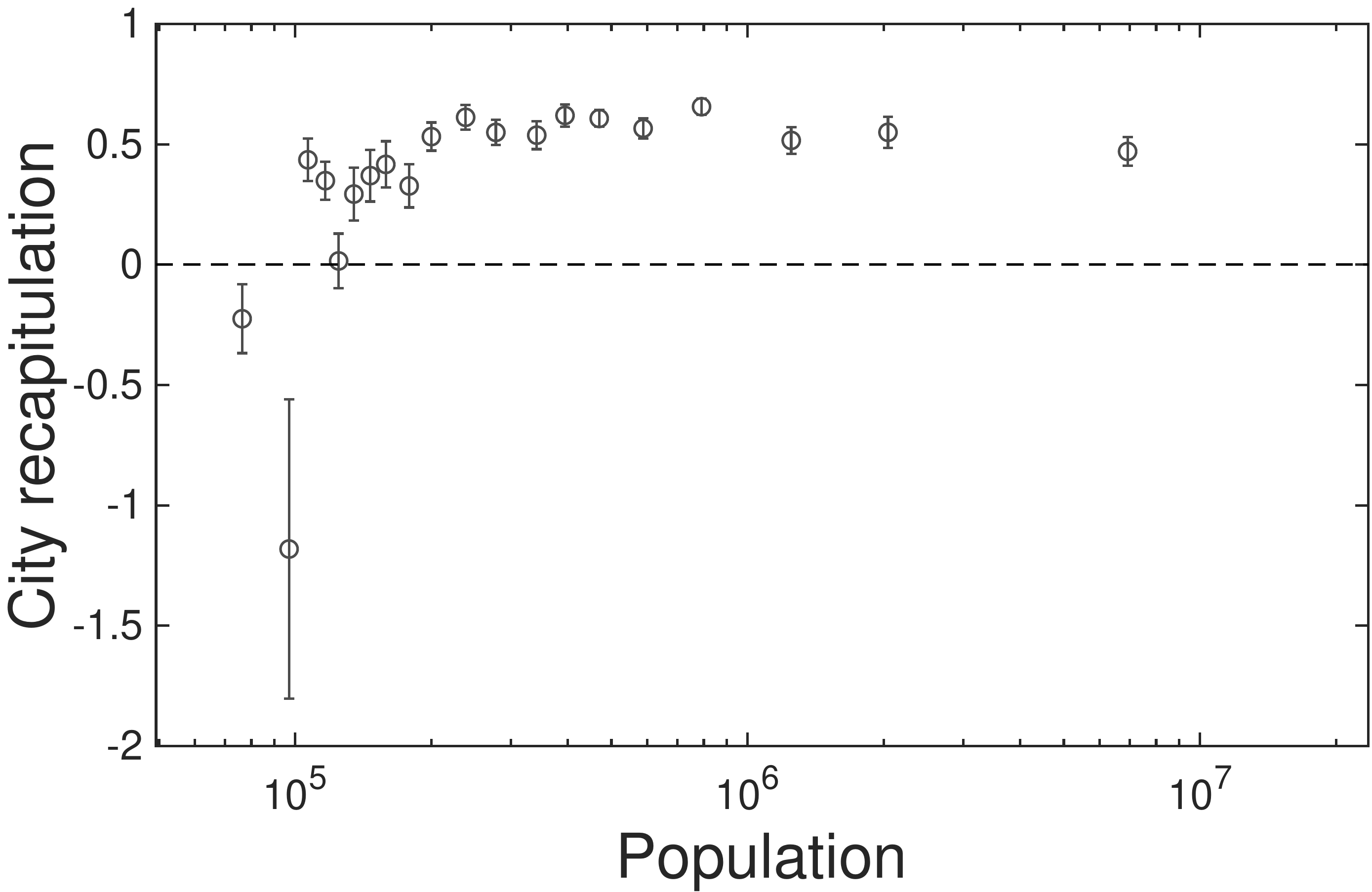}\label{fig:city_recap_score_N3}}
	\caption{
	(A) Change of logarithmic industry sizes versus on the change of logarithimic populations. Each dot in the figure means each city. For simplicity, we use the changed values as the differences between the first (1998) and last years (2013) in the data. Each industry in 2-digit NAICS classification is highlighted by colors. 
	(B) Comparison of the longitudinal size-dependencies and cross-sectional scaling exponents for 2-digit NAICS classification. The red dotted line is the prediction line by theoretical derivations.
	(C-D) Recapitulation scores for each industry in (C) 2-digit and (D) 3-digit NAICS classifications. We denoted the $R^2$ of scaling relations as greyscale. The average score is 0.70 when we exclude ``Agriculture'' and ``'Mining'' whose scaling relations are not significant with $R^2 < 0.65$.  
	(E-F) Recapitulation scores of cities measured for industries in (E) 2-digit and (F) 3-digit NAICS classifications. We group the cities into 20 groups by their population sizes, and calculate for the industries with $R^2 \geq 0.65$ of the scaling relation. 
	}
\end{figure*}

Now, let us consider that the relative changes of populations and industries are not uniform across cities. It can be described as a function of $\Delta\log{N}(c)$ and $\Delta\log{Y}(c,i)$ for city $c$. Fig.~\ref{fig:scaling-trend} shows their relations for NAICS 2-digit industries. It can be simplified to a linear relation,
\begin{equation}
	\Delta\log{Y}(c,i) = \Delta y(i) + k(i)\Delta\log{N}(c)
	\label{eq:log-growth_relation}
\end{equation}
where $\Delta y(i)$ is a global growth across all cities, and $k(i)$ is the coefficient of population-dependent growth. Using $\Delta\log{Y}={\Delta}Y/Y$, we can derive the time-dependent scaling relation.
\begin{flalign}
	d(\log{Y}(c,i)) &= dy(i) + k(i)d(\log{N}(c))\\
	\frac{Y(c,i,t_{2})}{Y(c,i,t_{1})} &= e^{\Delta y(i)}\left(\frac{N(c,t_{2})}{N(c,t_{1})}\right)^{k(i)}
\end{flalign}
where $t_{2} = t_{1}+\Delta t$. Since the scaling relation is given as $Y(c,i,t_{1})=Y_{0}(i,t_{1})N(c,t_{1})^{\beta(i,t_{1})}$, the equation can be expressed as a perturbed form.
\begin{equation}
	\frac{Y(c,i,t_{2})}{Y_{0}(c,i,t_{1})N(c,t_{1})^{\beta(i,t_{1})}} = e^{\Delta y(i)}\left(\frac{N(c,t_{2})}{N(c,t_{1})}\right)^{k(i)}
\end{equation}
\begin{flalign}
	Y(c,i,t_{2}) &= Y_{0}(i,t_{1})e^{\Delta y(i)}N(c,t_{2})^{k(i)}N(c,t_{1})^{\beta(i,t_1)-k(i)}\nonumber\\
		&= Y_{0}(i,t_{1})e^{\Delta y(i)}N(c,t_{2})^{\beta(i,t_1)+\epsilon(i)}N(c,t_{1})^{-\epsilon(i)}\nonumber\\
		&= Y_{0}(i,t_{1})e^{\Delta y(i)}N(c,t_{2})^{\beta(i,t_1)+\epsilon(i)}(N(c,t_{2})-\Delta N(c))^{-\epsilon(i)}\nonumber\\
		&= Y_{0}(i,t_{1})e^{\Delta y(i)}N(c,t_{2})^{\beta(i,t_1)}(1-\frac{\Delta N(c)}{N(c,t_{2})})^{-\epsilon(i)}\nonumber\\
		&\simeq Y_{0}(i,t_{1})e^{\Delta y(i)}N(c,t_{2})^{\beta(i,t_1)}(1+\epsilon(i)\frac{\Delta N(c)}{N(c,t_{2})})
\end{flalign}
where $\epsilon(i) = k(i) - \beta(i,t_1) \ll \beta(i,t_1)$ and $\Delta N(c) = N(c,t_2)-N(c,t_1) \ll N(c,t_2)$. Since $\epsilon(i)\Delta N(c)$ is the second order term, we can obtain an approximated scaling relation for $t=t_{2}$ in the case of small difference between $k(i)$ and $\beta(i,t_1)$.
\begin{equation}
	Y(c,i,t_{2}) \simeq Y_{0}(i,t_{2})N(c,t_{2})^{\beta(i,t_1)}		
\end{equation}
where $Y_{0}(i,t_{2}) = Y_{0}(i,t_{1})e^{\Delta y(i)}$ means a global exponential growth. From the above derivation, the global growth term $\Delta y(i)$ is related to the growth of $Y_{0}(i,t)$. It represents a global drift of every city in industry space. The population-dependence coefficient $k(i)$ is related to the scaling exponent $\beta(i)$. When there is little difference between $k(i)$ and $\beta(i, t_1)$ as in our data, the scaling exponent $\beta(i, t_2)$ at the later time $t_2$ becomes the same with $\beta(i,t_1)$. 



\begin{table*}[!th]
    \centering
    \begin{tabular}{l|c|c|c|c}
    \hline
    Industry & NAICS code & $\beta$ & $\hat{\beta}$ & $\Delta\log{\hat{Y}_{0}}$ \\
    \hline
    Agriculture, Forestry, Fishing and Hunting  	& 11 & 0.65 & 1.13 & -0.31 \\
    Mining, Quarrying, and Oil and Gas Extraction  	& 21 & 0.78 & 1.51 & -0.30 \\
    Manufacturing  									& 31 & 0.94 & 0.62 & -0.47 \\
    Retail Trade  									& 44 & 0.96 & 0.82 & -0.12 \\
    Health Care and Social Assistance  				& 62 & 0.96 & 0.66 & 0.15 \\
    Utilities  										& 22 & 0.99 & 1.20 & -0.30 \\
    Accommodation and Food Services  				& 72 & 1.00 & 0.69 & 0.10 \\
    Other Services (except Public Administration)  	& 81 & 1.02 & 0.76 & -0.15 \\
    Construction  									& 23 & 1.05 & 0.86 & -0.28 \\
    Arts, Entertainment, and Recreation  			& 71 & 1.09 & 0.77 & 0.66 \\
    Transportation and Warehousing  				& 48 & 1.11 & 0.72 & 0.14 \\
    Real Estate and Rental and Leasing  			& 53 & 1.12 & 0.85 & -0.09 \\
    Wholesale Trade  								& 42 & 1.12 & 0.88 & -0.18 \\
    Information  									& 51 & 1.16 & 0.67 & -0.23 \\
    Finance and Insurance  							& 52 & 1.17 & 0.59 & -0.07 \\
    \makecell[l]{Administrative and Support and \\ Waste Management and Remediation Services} &	56 &	 1.17 & 0.44 & -0.06 \\
    Educational Services  							& 61 & 1.21 & 1.09 & 0.18 \\
    Professional, Scientific, and Technical Services& 54 & 1.22 & 0.78 & 0.07 \\
    Management of Companies and Enterprises  		& 55 & 1.46 & 1.03 & 0.02\\
    \hline
    \end{tabular}
    \caption{
    Scaling exponents of industries in 2-digit NAICS classification. The scaling exponent is the time average of scaling exponents in 1998-2013. $\hat{\beta}$ and $\Delta\log{\hat{Y}_{0}}$ are measured for the difference between 1998 and 2013 following Eq.~\ref{eq:slope}}
    \label{table:growth_regression}
\end{table*}


\end{document}